\documentclass[sigconf,screen]{acmart}
\AtBeginDocument{%
  \providecommand\BibTeX{{%
    \normalfont B\kern-0.5em{\scshape i\kern-0.25em b}\kern-0.8em\TeX}}}

\copyrightyear{2022}
\acmYear{2022}
\setcopyright{rightsretained}
\acmConference[AIES'22]{Proceedings of the 2022 AAAI/ACM Conference
on AI, Ethics, and Society}{August 1--3, 2022}{Oxford, United Kingdom}
\acmBooktitle{Proceedings of the 2022 AAAI/ACM Conference on AI,
Ethics, and Society (AIES'22), August 1--3, 2022, Oxford, United
Kingdom}\acmDOI{10.1145/3514094.3534184}
\acmISBN{978-1-4503-9247-1/22/08}

\usepackage{multirow}
\usepackage{booktabs}
\usepackage{tabularx}
\usepackage{amsmath}
\usepackage{xr}
\usepackage{graphicx}
\usepackage{float}
\usepackage{dcolumn}
\usepackage{lettrine}
\usepackage{xcolor}
\usepackage{placeins}
\usepackage{subfig}
\usepackage{dirtytalk}

\newcolumntype{d}[1]{D{.}{.}{#1}}
\newcommand\mc[1]{\multicolumn{1}{c}{#1}} 

\newcommand{\tcr}{C}
\newcommand{\ir}{I}
\newcommand{\mbbE}{\mathbb{E}}
\newcommand*{\threefrac}[3]{%
  \begin{array}{@{\,}c@{\,}}%
    #1\\
    \hline
    #2\\
    \hline
    #3%
  \end{array}%
}

\newcommand{\beginsupplement}{%
        \setcounter{table}{0}
        \renewcommand{\thetable}{S\arabic{table}}%
        \setcounter{figure}{0}
        \renewcommand{\thefigure}{S\arabic{figure}}%
     }

\DeclareFontFamily{U}{mathx}{\hyphenchar\font45}
\DeclareFontShape{U}{mathx}{m}{n}{<-> mathx10}{}
\DeclareSymbolFont{mathx}{U}{mathx}{m}{n}
\DeclareMathAccent{\widebar}{0}{mathx}{"73}

\begin{document}

\title[Racial Disparities in the Enforcement of Marijuana Violations in the US]{Racial Disparities in the Enforcement\\of Marijuana Violations in the US}

\author{Bradley Butcher}
\authornote{Authors contributed equally to this research.}
\email{b.butcher@sussex.ac.uk}
\orcid{0000-0002-4711-7826}
\affiliation{%
  \institution{University of Sussex}
  \streetaddress{P.O. Box 1212}
  \city{Brighton}
  \country{UK}}
\affiliation{%
  \institution{The Alan Turing Institute}
  \city{London}
  \country{UK}}

\author{Chris Robinson}
\authornotemark[1]
\orcid{0000-0002-5296-7657}
\email{cfr20@sussex.ac.uk}
\affiliation{%
  \institution{University of Sussex}
  \streetaddress{P.O. Box 1212}
  \city{Brighton}
  \country{UK}
}

\author{Miri Zilka}
\authornotemark[1]
\orcid{0000-0001-9640-8139}
\email{mz477@cam.ac.uk}
\affiliation{%
  \institution{University of Cambridge}
  \city{Cambridge}
  \country{UK}
}
\affiliation{%
  \institution{The Alan Turing Institute}
  \city{London}
  \country{UK}}

\author{Riccardo Fogliato}
\orcid{0000-0002-8636-9639}
\affiliation{%
 \institution{Carnegie Mellon University}
 \state{Pennsylvania}
 \country{USA}}

\author{Carolyn Ashurst}
\affiliation{%
  \institution{The Alan Turing Institute}
  \city{London}
  \country{UK}}

\author{Adrian Weller}
\orcid{0000-0003-1915-7158}
\affiliation{%
  \institution{University of Cambridge}
  \city{Cambridge}
  \country{UK}}
\affiliation{%
  \institution{The Alan Turing Institute}
  \city{London}
  \country{UK}}

\renewcommand{\shortauthors}{Butcher, Robinson, Zilka et al.}

\begin{abstract}
        Racial disparities in US drug arrest rates have been observed for decades, but their causes and policy implications are still contested. 
        Some have argued that the disparities largely reflect differences in drug use between racial groups, while others have hypothesized that discriminatory enforcement policies and police practices play a significant role. In this work, we analyze racial disparities in the enforcement of marijuana violations in the US. Using data from the National Incident-Based Reporting System (NIBRS) and the National Survey on Drug Use and Health (NSDUH) programs, we investigate whether marijuana \textit{usage} and \textit{purchasing} behaviors can explain the racial composition of offenders in police records. We examine potential driving mechanisms behind these disparities and the extent to which county-level socioeconomic factors are associated with corresponding disparities. 
        Our results indicate that the significant racial disparities in reported incidents and arrests cannot be explained by differences in marijuana days-of-use alone. Variations in the location where marijuana is purchased and in the frequency of these purchases partially explain the observed disparities. 
        We observe an increase in racial disparities across most counties over the last decade, with the greatest increases in states that legalized the use of marijuana within this timeframe. 
        Income, high school graduation rate, and rate of employment positively correlate with larger racial disparities, while the rate of incarceration is negatively correlated. 
        We conclude with a discussion of the implications of the observed racial disparities in the context of algorithmic fairness.
    
\end{abstract}

\ccsdesc[500]{Social and professional topics}
\ccsdesc[500]{Social and professional topics~Governmental regulations}
\ccsdesc[500]{Social and professional topics~Race and ethnicity}
\ccsdesc[300]{Security and privacy~Privacy protections}
\ccsdesc[300]{Computing methodologies~Artificial intelligence}
\ccsdesc[300]{Computing methodologies~Machine learning algorithms}
\ccsdesc[300]{Computing methodologies~Machine learning approaches}
\ccsdesc[300]{Computing methodologies~Model verification and validation}
\ccsdesc[300]{Applied computing~Sociology}

\keywords{Racial Disparities, Law Enforcement, Marijuana}


\maketitle

\section{Introduction}

Racial disparities in incarceration and arrest rates have been observed in the US for decades \cite{Tonry2010}. In 2018, the rate of imprisonment of black males was 5.8 times higher than that of white males with respect to their share of the population  \cite{imprisonment_2018}. Being convicted of a criminal charge can have life-altering effects on employment prospects, the ability to get a loan, rent a home, and retain child custody. The consequences go beyond the convicted individual, impacting their children, families, neighborhoods, and communities 
\cite{martin2017hidden, PettitGutierrez2018}. 

Racial disparities in arrests for drug offenses have been studied extensively \cite{MitchellCaudy2015} and have been a major factor in decisions to legalize marijuana \cite{Clark2018, Vitiello2019}. In 2006, a study of racial and ethnic disparities in arrests for buying and possession of drugs in Seattle concluded that the large disparities could not be explained by race-neutral factors \cite{BecNyrPfi2006}. A follow-up study challenged these findings, suggesting that race-neutral police deployment strategies, which prioritized areas with higher crime rates, led to more encounters with minorities, resulting in the observed disparities \cite{EngSmiCull2012}. 

\begin{table*}[t]
\centering
\caption{Models used to estimate the baseline crime rate based on NSDUH data}
\label{Tab:Models}
\begin{tabular}{llll}
Model & Incidents & Proxy for criminal activity &  Additional information\\
\midrule
\texttt{Use}\textsubscript{Dmg+Pov} & all & days-of-use & poverty level \\
\texttt{Use}\textsubscript{Dmg} & all & days-of-use & - \\
\texttt{Use}\textsubscript{Dmg+Metro} & all & days-of-use & metropolitan area \\
\texttt{Use}\textsubscript{Dmg+Pov, Arrests}& arrests & days-of-use & poverty level \\
\texttt{Purchase}& all & days-of-purchase & poverty level\\
\texttt{Purchase}\textsubscript{Public}& all & days-of-purchase in public & poverty level\\
\bottomrule
\multicolumn{4}{l} {\parbox[t]{0.7\textwidth}{\footnotesize{All models consider sex, race, and age, in addition to the information listed in the table.
}}}
\end{tabular}
\end{table*}

Related works examining racial disparities in drug arrests have suggested several mechanisms that could explain the phenomenon. These include factors relating to consumer behavior, drug policy, and police practices. With respect to consumer behavior, more frequent use of marijuana has been shown to increase the risk of arrest \cite{Goode2002}. 
Differences in purchasing behaviors can also affect exposure to the police and the consequent likelihood of being arrested \cite{Goode2002}. 
For example, using and buying marijuana in public spaces is associated with a higher probability of arrest compared to using it in private spaces \cite{Goode2002}. Similarly, individuals that purchase drugs from strangers, and in smaller amounts but at a higher frequency, are more likely to be arrested \cite{Blumstein1993,Caulkins2006,Duster1993,Goode2002,JohPetWel1977,Sterling1997,RamPacIgu2006}. 
These behaviors may be more prevalent among individuals of lower socioeconomic status, as a result of overcrowded accommodation, or due to risks associated with consuming drugs within rented accommodation \cite{Bender2016}. 
More frequent encounters with the police due to engagement in other criminal activities \cite{VauSalRei2016} and outstanding warrants \cite{Gaston2019} may also increase exposure.

Differences in drug policies, be it at a national, state, or local level, may lead to disparities in arrest rates between racial groups. Legalization of marijuana has occurred in several states and its consequences are not well understood yet \cite{adinoff2019implementing, fischer2020considering}. While legalization has led to lower arrest rates for marijuana-specific crimes overall, it may not have impacted racial groups equally \cite{FirMahDil2019}. Legalization is not a discrete, absolute change: Only certain activities under specific circumstances become legal. Following legalization, marijuana violations can still lead to arrests in case of underage users or public consumption. The remaining types of arrests may exhibit particularly high racial disparities.

Disparities in arrest rates for marijuana-related crimes are also impacted by police practices. Policing strategies that target specific drugs or geographic areas may result in stricter enforcement for certain types of users or dealers \cite{BecNyrPfi2006}. Strategies informed by predictive policing have the potential for inadvertent feedback loops in disparate enforcement \cite{feedback18}. Similarly, profiling in traffic stops and stop-question-and-frisk (SQF) have been shown to affect the black community disproportionately \cite{Baumgartner2018, Gaston2019, GellerFagan2010, Gelman2007, GroggerRidgeway2006, Ridgeway2009, PieSimOve2020, goel2016precinct}. 

The records pertaining to illegal drug activity are inherently limited. Only a small fraction of offenses that are committed in reality are observed by law enforcement and then recorded \cite{morgan2019criminal, piquero2008disproportionate}. Crimes become known to law enforcement only if they are directly reported by victims or witnesses, or, in the case of drugs, through discovery efforts.
In addition, one needs to define what act constitutes an offense.
Unlike offenses such as burglaries, it is harder to define the number of times possession and use of illegal substances occurs. 


In this work, we conduct a large-scale, national analysis of racial disparities in recorded incidents of marijuana violations (often resulting in an arrest), while accounting for specific aspects of marijuana usage and purchasing habits. To estimate the number of crimes that are recorded by law enforcement, we use incident-level crime data from the National Incident Based Reporting System (NIBRS), which is part of the FBI’s Uniform Crime Reporting (UCR) program. Through NIBRS, participating law enforcement agencies submit detailed data on incidents, including characteristics of the offenses and involved parties. 

We estimate marijuana days-of-use, days-of-purchase, and days-of-purchase in public spaces, using data from the National Survey on Drug Use and Health (NSDUH), a yearly survey on substance use and mental health. These measures may not fully account for dimensions of use that may increase the risk of arrest or the extent of illegal activity.  
Leveraging both datasets, we can estimate the likelihood that a drug offense will become known to law enforcement for each racial group in each county.
We give an overview of our methodology in Section~\ref{sec:Approach}, with additional details deferred to Section~\ref{Sec:Method}.

We analyze several mechanisms that have been proposed by previous works as drivers of racial disparities in drug arrests, specifically focusing on how differences in use, consumer behavior, and legalization impact disparities. 
We then examine the relationship between arrest disparities and county-level socioeconomic factors, including income, education, employment and levels of incarceration. We also compare county-level disparities in marijuana violations to disparities in arrests for drunkenness and driving under the influence (DUI), to study the extent to which disparities are due to specific factors relating to marijuana, compared to illegal activity more generally. Finally, we briefly discuss the implications of the observed racial disparities on algorithmic fairness in the context of predictive tools deployed within the criminal justice system.

\begin{figure*}[t]
    \centering
    \includegraphics[width=1.0\textwidth]{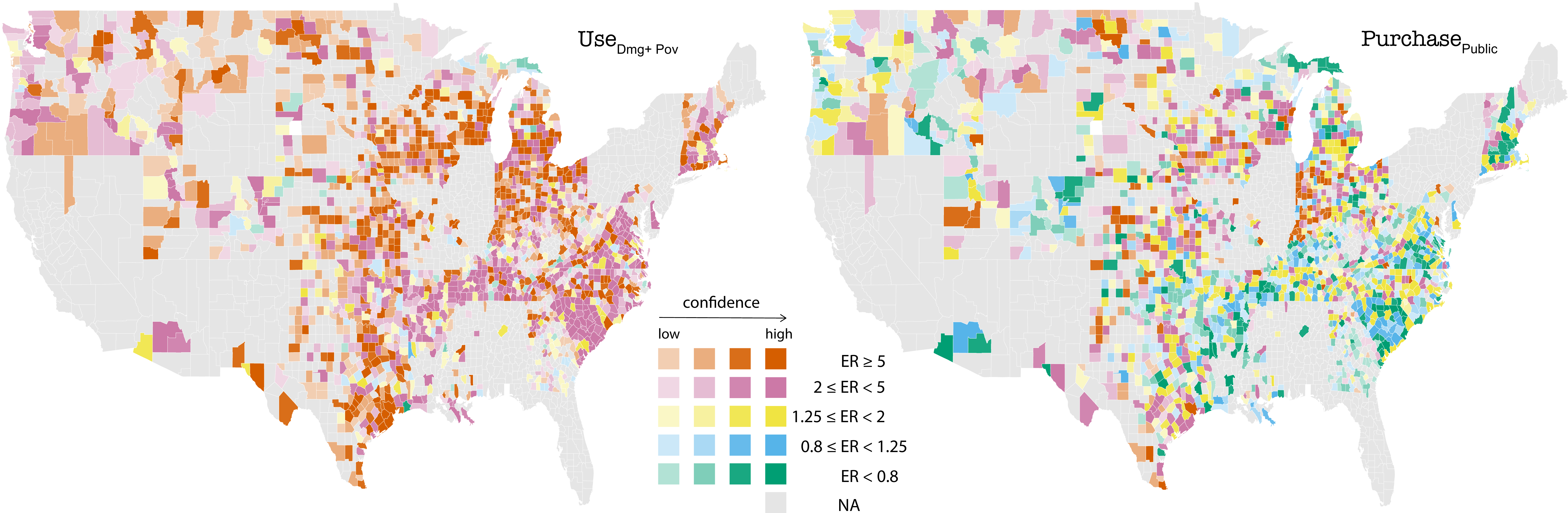}
    \caption{Heatmap of the enforcement ratio across counties in the United States, based on 2017--2019 NIBRS and NSDUH data. The colors indicate different levels of the ratio. Higher opacity indicates greater statistical confidence. 
    Grey counties are those with insufficient data available. 
    The left map shows ratios based on days-of-use, \texttt{Use}\textsubscript{Dmg+Pov}, with ratios above 1.25 for most counties. The right map shows ratios based on days-of-purchase in public spaces, \texttt{Purchase}\textsubscript{Public}, with ratios above 1.25 for just under half of the counties.}
    \label{fig:map}
\end{figure*}

\section{Approach}
\label{sec:Approach}

We denote with $C$ the \textit{baseline crime rate}, i.e., the rate of engagement in an illegal activity, whether by an individual, a subgroup of the population, or the population as a whole. We indicate with $I$ the rate at which incidents are recorded by law enforcement.
The main target of interest in this work is the \textit{enforcement rate}, being the proportion of crimes that are recorded by law enforcement agencies, which is defined as $E = I/C$. 

Due to data limitations, in this paper we consider just two racial groups: black and white individuals. We examine differences in county-level enforcement rates between black and white individuals. To examine how the enforcement rate of black individuals $E_b$ compares to that of white individuals $E_w$, we use the \textit{enforcement ratio} $ER = E_b/E_w$. An enforcement ratio of 1 indicates no disparity. A ratio above 1 indicates a higher enforcement rate for black individuals. Conversely, a ratio below 1 indicates a higher enforcement rate for white individuals.

\subsection{Measuring incidents known to law enforcement} 
We obtain the number of incidents recorded by law enforcement for each racial group $I_b(c)$ and $I_w(c)$ in every reporting county $c$ from NIBRS data. We consider all incidents involving offenses related to the use, possession, and buying of marijuana. Further details on NIBRS processing and incidents excluded from the analysis are provided in Section \ref{Sec:Method}.

\subsection{Estimating baseline criminal activity}
We estimate the baseline crime rate $C$ using three self-reported measures from the NSDUH: days-of-use, days-of-purchase, and days-of-purchase in public spaces. 
We use six separate models to generate estimates of $C$ that are summarized in Table~\ref{Tab:Models}. 
Our main model \texttt{Use}\textsubscript{Dmg+Pov} estimates the baseline crime rate $C_r(c)$ for county $c$ and racial group $r\in\{$black,white$\}$ as
$$C_{r}(c) = 365 \sum\limits_{d} \text{use}_{d, r} \times \text{pop}_{d, r}(c)$$
where $\text{use}_{d,r}$ is the probability of marijuana use in a given day for an individual of race $r$ and characteristics $d$, and $\text{pop}_{d,r}(c)$ is the number of individuals of race $r$ and characteristics $d$ residing in the county. 

The individual characteristics $d$ include information such as race, sex, age, and whether an individual lives above or below the poverty level. Specifically, $\text{use}_{d,r}$ is the average number of days 
that NSDUH respondents with characteristics $d$ and race $r$ reported using marijuana in the 30 days prior to the interview, divided by 30. We estimate $\text{pop}_{d,r}(c)$ by using data from the Census and American Community Survey (ACS). We compute the proportion of individuals living above the poverty threshold\footnote{Whether or not an individual lives below the poverty line depends on total family income as well as on the size and composition of the family \citep{poverty}.} for each demographic group from ACS data. We combine this information with the population estimates from the Census to infer $\text{pop}_{d,r}$. Additional models differ from the main model as follows:
\texttt{Use}\textsubscript{Dmg} accounts only for sex and age (not poverty level); \texttt{Use}\textsubscript{Dmg+Metro} accounts for sex, age and whether the individual resides in a metropolitan area; \texttt{Use}\textsubscript{Dmg+Pov, Arrests} only accounts for incidents that resulted in an arrest in the computation of $I$; in \texttt{Purchase}, we consider the average number of days that NSDUH respondents reported buying marijuana in the past 30 days, divided by 30; lastly, in \texttt{Purchase}\textsubscript{Public}, we only consider the proportion of days that NSDUH respondents reported buying marijuana in public. 
$I$ and $C$ are estimated for the same time frame, a single or several calendar years, depending on the analysis.

\section{Results}

\subsection{County-level racial disparities when accounting for days-of-use or days-of-purchase}

We calculate the enforcement ratio between black and white individuals $ER$ for every county in the US where law enforcement agencies have reported crime data through NIBRS in the years 2017--2019. We consider all incidents for which the only offenses are concerned with personal use, i.e., possession, use of buying of marijuana, or a combination thereof. County-level enforcement ratios based on two crime rate models are displayed in Figure~\ref{fig:map}. In the left panel, we consider \texttt{Use}\textsubscript{Dmg+Pov}, where we take the self-reported frequency of marijuana use, in days, as a proxy for the baseline crime rate. In the right panel, we consider \texttt{Purchase}\textsubscript{Public} where we take the frequency of purchasing marijuana in public spaces, in days, as a proxy for the crime rate. Figure~\ref{fig:map} shows that most counties across the country have a high $ER$ (above $1.25$). As shown in Table~\ref{Tab:count}, in about 70\% of the 2084 counties in the available data, the enforcement rate for black individuals is more than twice as high as that for whites when differences in days-of-use are accounted for.

The disparities are less striking under the model that accounts for days-of-purchase in public spaces, instead of use. One might think that buying habits should only be relevant for offenses for which the criminal activity consists of buying an illegal substance. In practice, it may be indicative of behaviors that increase the likelihood of encounters between users and law enforcement, such as using in public or frequently carrying marijuana on one's person or in vehicle. The decrease in the county-level enforcement ratios suggests that differences in drug-related habits, such as consumption in public spaces, may partially explain the large disparities we observe. We note that even with this model, a significant number of counties, 574, still retain an enforcement ratio higher than 2.

\begin{table}[t]
\caption{County-level enforcement ratios}
  \begin{tabular}{ccccccc}
    \toprule
    & & $<0.5$ & $<0.8$ & $>1.25$ & $>2$ & $>5$ \\
    \midrule 
     \multirow{2}{*}\texttt{Use}\textsubscript{Dmg+Pov} & Value & 13 & 38 & 1660 & 1494 & 734 \\
        & 95\% conf. & 1 & 3 & 1384 & 1118 & 352\\ 
     \multirow{2}{*}\texttt{Purchase}\textsubscript{Public} & Value & 137 & 372 & 994 & 574 & 173\\
        & 95\% conf.  & 29 & 101 & 563 & 255 & 55 \\ 
     \midrule
 \multicolumn{7}{l} {\parbox[t]{\columnwidth}{\footnotesize{The number of counties above or below a given enforcement ratio threshold, from a total of 2083 reporting counties. The 95\% conf. rows indicate the number of counties for which the upper bound of the 95\% confidence interval is below the threshold (for $ER<0.5$ and $ER<0.8$) or the lower bound of the 95\% confidence interval is above the threshold (for $ER>1.25$, $ER>2$ and $ER>5$). Enforcement ratios correspond to those displayed in Figure~\ref{fig:map}.}}}
  \end{tabular}
  \label{Tab:count}
 \end{table} 
 
\begin{figure}[h]
\centering
\includegraphics[width=\columnwidth]{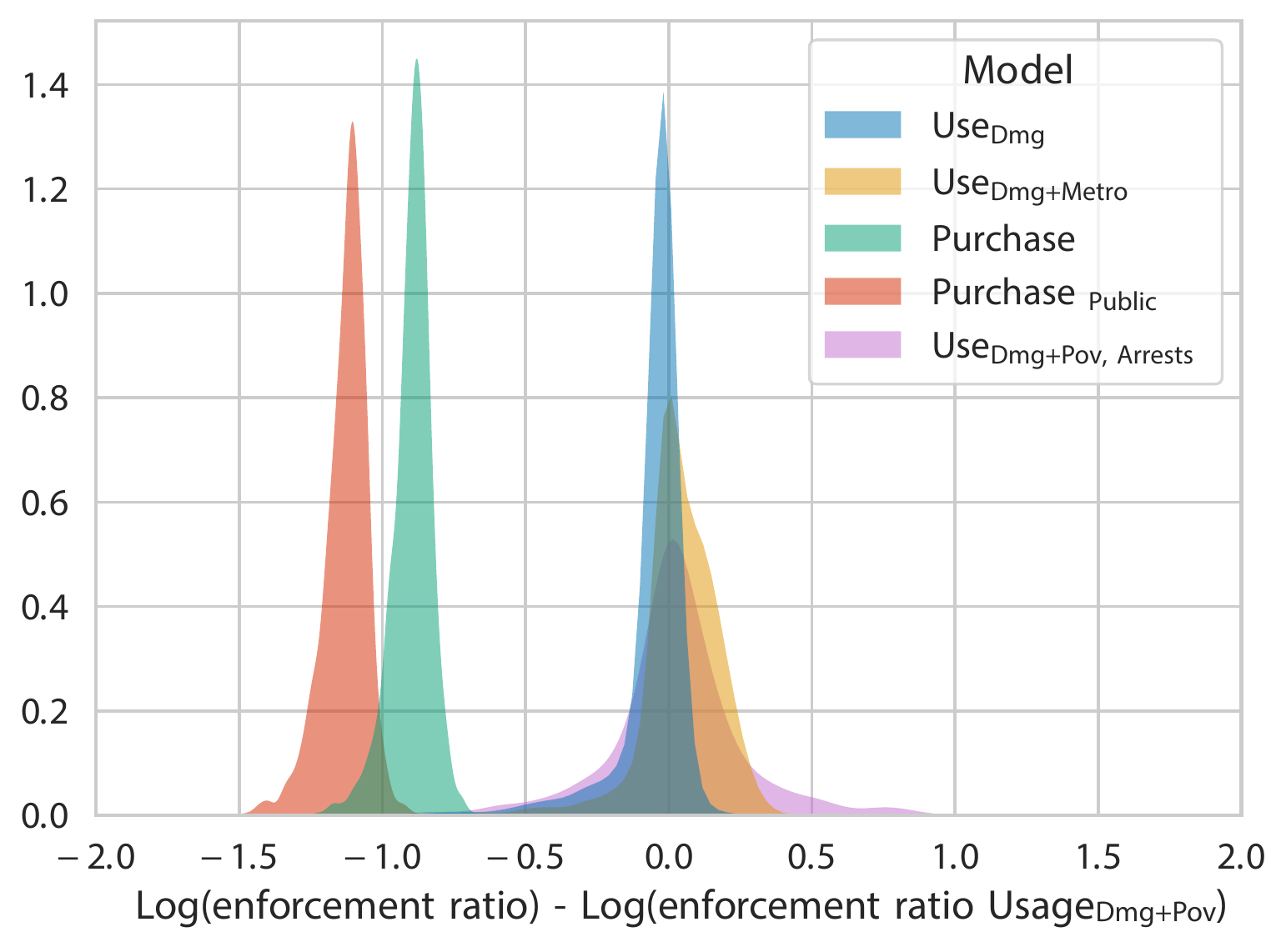}
\caption{Distributions of differences between logarithm of county-level enforcement ratios for each of the models listed in Table \ref{Tab:Models} versus the baseline model \texttt{Use}\textsubscript{Dmg+Pov} using aggregated 2017--2019 NIBRS and NSDUH data. Bandwidth of the kernels used for the density estimates was selected using Scott's rule \cite{scott2015multivariate}.}
\label{fig:diff}
\end{figure}

\subsection{Comparing racial disparities across models}
Figure~\ref{fig:diff} shows the differences in the estimates of enforcement ratios across the different estimation models compared to the baseline model \texttt{Use}\textsubscript{Dmg+Pov}. Models that use different demographic factors (metropolitan, poverty level) or that consider only incidents resulting in arrest do not yield substantially different distributions. However, both models which use days-of-purchase instead of days-of-use yield significantly lower enforcement ratios, with the largest difference occurring when accounting only for the frequency of marijuana purchases in public spaces. This suggests that differences in drug-related habits between black and white individuals may partly contribute to the observed disparities.

\subsection{Differences in marijuana buying habits across racial groups}

We now examine marijuana buying habits for black and white users from the NSDUH data. 
Table~\ref{Tab:buying} summarizes the self-reported buying habits of males between the ages of 18 and 25. We focus on this users' subgroup as it is the most overrepresented in our NIBRS data compared to their proportion in the US population, among both black and white individuals (see Tables \ref{SItab:NSDUH_metro} and \ref{SItab:NSDUH_pov} in the Appendix for other demographics). We observe that young black men buy marijuana more frequently than their white counterparts, and are also more likely to purchase it from a stranger and in public spaces. These findings persist even after we condition on the level of income. 

Our findings align with results from previous works that have concluded that racial differences in buying and usage culture may partially explain the observed disparity in incidents and arrests with drug offenses \cite{Blumstein1993,Caulkins2006,Duster1993,Goode2002,JohPetWel1977,Sterling1997,RamPacIgu2006}.

\begin{table}[h]
\caption{Self-reported marijuana buying habits of male users aged 18--25 based from NSDUH data}
  \centering
  \begin{tabular}{cccccc}
    \toprule
    & & \multicolumn{2}{c}{Above poverty} & \multicolumn{2}{c}{Below poverty} \\
    & & Black & White & Black & White \\
    \midrule 
    \multicolumn{2}{c}{Monthly purchases} &  10 (0) & 7 (0) & 12 (1) & 7 (0) \\ 
    \midrule \midrule 
     Quantity & $<$10 grams & 83\% (1) & 80\% (1) & 85\% (2) & 84\% (1) \\ 
     \midrule
     Price & $<\$$20 & 60\% (2) & 35\% (1) & 68\% (2) & 40\% (2) \\
    \midrule
    \multirow{2}{*}{Seller} & Friend & 66\% (2) & 81\% (1) & 59\% (2) & 80\% (1) \\ 
                            & Stranger & 30\% (2) & 16\% (1) & 37\% (2) & 17\% (1) \\
     \midrule
     \multirow{2}{*}{Location} & Residence & 37\% (2) & 60\% (1) & 38\% (2) & 68\% (1) \\ 
                & Public  & 35\% (2) & 20\% (1) & 39\% (2) & 15\% (1) \\ 
     \midrule
     \multicolumn{6}{l}{\parbox[t]{\columnwidth}{\small{Standard errors are reported within parentheses. Monthly purchases refers to the average number of days that NSDUH respondents reported buying marijuana in the 30 days prior to the interview. Quantity refers to the percentage of respondents that reported purchasing less then 10 grams the last time they purchased marijuana. For other demographics, see Appendix Tables~\ref{SItab:NIBRS_metro}--\ref{SItab:NIBRS_metro_all}.
}}}
  \end{tabular}
  \label{Tab:buying}
\end{table}

\subsection{Differences in locations and times of incidents across racial groups}
 
Table~\ref{Tab:NIBRS} displays summary statistics of incidents involving at least one marijuana violation for male offenders aged 18--25. We consider metropolitan and non-metropolitan areas separately. We find several differences between incidents involving young black and white offenders (see Tables \ref{SItab:NIBRS_metro} and \ref{SItab:NIBRS_metro_all} in the Appendix for other demographics). Incidents involving black offenders are more likely to involve other non-drug offenses in addition to the marijuana violations, particularly in metropolitan areas. 

\begin{figure*}[t]
\centering
\includegraphics[width=0.8\linewidth]{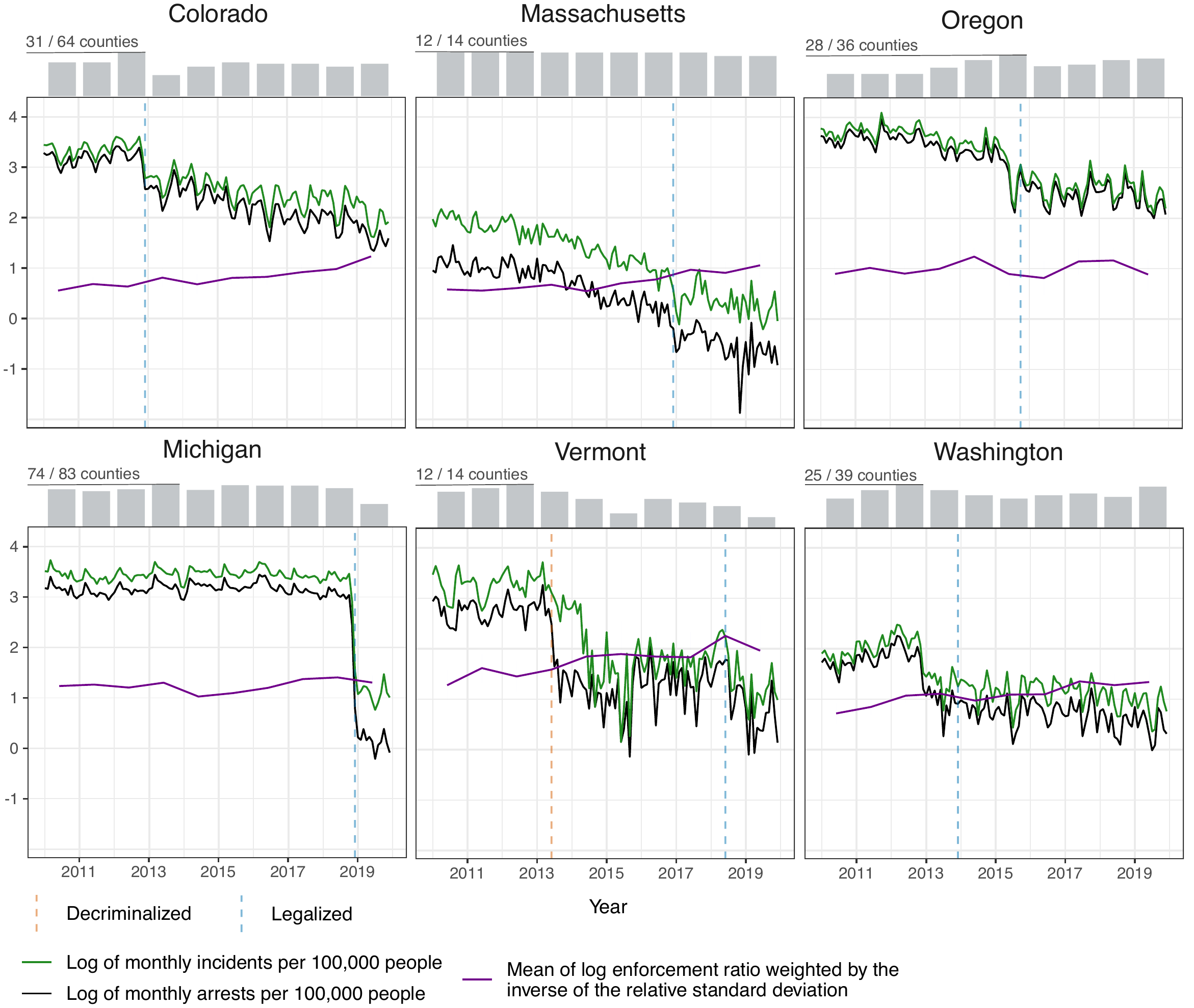}
\caption{Time trend of the rate of incidents and the enforcement ratio in the states where marijuana has been legalized between 2010--2019 and reported to NIBRS. The green line represents the log of the monthly number of incidents involving marijuana offenses per 100,000 people. The red line indicates the average of the log enforcement ratio computed for each year, weighted by the inverse of the relative standard deviation. For each year, only agencies that reported at least one incident for black individuals and one for whites in a given year were considered. The vertical lines correspond to decriminalization and legalization events. The histograms on the top of the panels represent the number of counties considered for the corresponding year.}
\label{fig:legal}
\end{figure*}

Incidents in the street and during daytime are slightly more frequent among black individuals. While in metropolitan areas incidents with white offenders are less likely to result in an arrest, arrest rates for whites tend to be higher in non-metropolitan areas. We note that the arrest rates decreased throughout the period 2012--2019 by approximately 5\% for all incidents involving at least one marijuana violation and by 7\% for incidents with only marijuana related charges (Figure~\ref{SIfig:arrest_rate} in the Appendix). Although the rates of marijuana incidents and arrests have decreased for all adult sex and age subgroups in that time window (Figure~\ref{SIfig:arrest_rate} in the Appendix), they have done so more rapidly for white individuals, resulting in an upwards trend in the enforcement ratio (rows 1--2 in Table~\ref{tab:correlate_table}).  

Previous analysis of police stops has observed racial disparities to be largest during daylight \cite{PieSimOve2020}. 
We assess whether an analogous pattern is present for marijuana violations by calculating separate enforcement ratios for incidents that occurred between dawn and dusk and vice versa. We find that the distributions of the county-level enforcement ratios are virtually identical (see Appendix Section~\ref{SISec:day}). 

To examine the role of the incident's location in the observed racial disparities, we assess how the enforcement ratio changes if we consider only incidents occurring in the street compared to those in a residence. Aggregating over all qualifying incidents in the years 2010--2019, the residence-only enforcement ratio is 2.5 while the street-only enforcement ratio is about 3.5 (see Figure~\ref{SIfig:location} in the Appendix). 

These results have implications for the development and deployment of predictive policing tools. Even when predictive tools are not used to proactively indicate where drug offenses will occur, their deployment may lead to increased police presence in certain neighborhoods, which may result in a disproportionate increase in drug-related arrests for black individuals.

\begin{table}[]
\caption{Summary statistics based on NIBRS 2010--2019 incidents involving marijuana violations for male offenders aged 18--25} 
\label{Tab:NIBRS}
  \centering
  \begin{tabular}{ccccc}
    \toprule
    & \multicolumn{2}{c}{Metropolitan} & \multicolumn{2}{c}{Non-Metropolitan} \\
    & Black & White & Black & White \\
    \midrule 
     Incidents & 426045 & 585236 & 49782 & 168178 \\
     \midrule
     Drugs only & 87\% & 95\% & 90\% & 96\% \\
     \midrule
     Possession & 81\% & 83\% & 77\% & 78\% \\
     Distributing & 13\% & 9\% & 15\% & 9\% \\
     \midrule
     One offender  & 58\% & 58\% & 58\% & 59\% \\
     \midrule
     Residence  & 15\% & 16\% & 18\% & 21\% \\ 
     Street & 61\% & 58\% & 58\% & 55\% \\
     \midrule
     6am -- 8 pm & 54\% & 47\% & 52\% & 50\% \\
     \midrule
     No arrest & 21\% & 23\% & 25\% & 23\% \\
     \bottomrule

  \end{tabular}
 \end{table}  

\subsection{Racial disparities following marijuana legalization}

There are six states where marijuana has been legalized and in which most counties have submitted their crime data through NIBRS over the past decade: Colorado, legal effective from December 2012; Massachusetts, decriminalized\footnote{Decriminalization means that the the acquisition, possession and consumption of a drug, under certain circumstances, is no longer considered a criminal offense. However, these activities are still illegal and may be reproached with a civil fine, mandated drug education or treatment.} in 2008, legal effective from December 2016; Michigan, legal effective from December 2018; Oregon, decriminalized from 1973, legal effective from October, 2015; Vermont, decriminalized from June 2013 legal effective from July 2018; and Washington, legal since December 2013 \cite{ColoradoAmendment64, MassachusettsAct, MichiganAct, OregonMeasure91, VermontAct86, WashingtonAct502}. Figure~\ref{fig:legal} shows the monthly rate of reported incidents per 100,000 people and the yearly enforcement ratio for the period 2010--2019 across the aforementioned states. The date of legalization is indicated by the vertical dotted line. We note, however, that the effects of legalization are not immediate but rather are often observed over the months or even years before and after the policy change. 

Nonetheless, we observe a decrease in the rate of reported incidents following legalization across all states. A similar trend can be observed for arrest rates, shown in black in Figure~\ref{fig:legal}. Concurrently, we observe a corresponding increase in the enforcement ratio (in purple) following legalization across all states other than Oregon and Michigan. 

To further assess the connection between legalization and the increase in disparity, we compare the change in the enforcement ratios between states where marijuana was illegal in the period 2010--2019 and the six reporting states where the legal status changed within that timeframe. When we only consider counties that reported to NIBRS throughout the entire period, the log of the enforcement ratio in these six states increased at a rate one and a half times higher than in the non-legalized states (rows 1-2 Table~\ref{tab:correlate_table}). When considering all counties, the rate of change in the log of the enforcement ratio is more than double in the states that legalized marijuana (Table~\ref{SItab:corr1}). This suggests that the enforcement ratio has increased at higher rates in states where the legal status of marijuana changed. Following legalization, incidents and arrests for marijuana violations include a narrower scope of activities, including, for example, use under the legal age or in public spaces. Racial disparities may be exacerbated if black users are more likely to engage in these activities. Our analysis of NSDUH data suggests this may be the case for use in public spaces.

\subsection{Investigating the association between county characteristics and racial disparities}
We examine the relationship between the county-level enforcement ratios and the
characteristics of the county (such as household income and employment rates)
through linear regression models. We regress via ordinary least squares the log of the enforcement ratio on each characteristic of the county, transformed into percentiles with respect to all reporting counties; see Section~\ref{Sec:Method}
for more details. The resulting estimates of the regression coefficients are
reported in Table~\ref{tab:correlate_table}. We find a positive association
between the enforcement ratio and the ratio of black to white populations in the
county, suggesting that the enforcement ratio is higher in areas where black
residents represent a smaller portion of the overall population. We find that
a higher levels of income and education in the county are associated with higher enforcement
ratios, while the county-level socioeconomic (income and education) disparities between black and white
individuals are not. For example, the association between the mean household
income and the enforcement ratio is positive, however, the ratio between the
mean household income of the black and white populations is only weakly
correlated with the enforcement ratio. This suggests that while the size of the
black population in the county is associated with the level of disparity, their
relative affluence, compared to the white population, is not.  



\begin{table}[h]
  \centering
    \caption{Association between county-level enforcement ratios and county characteristics}
    \label{tab:correlate_table}
    \begin{tabular}{lll}
    \toprule
     &  \mc{\texttt{Use}\textsubscript{Dmg+Pov}} &           \mc{\texttt{Purchase}\textsubscript{Public}} \\
    \midrule
    Time (legalized states) & 0.041 (0.019)* & 0.051 (0.019)** \\
    Time (non legalized) & 0.024 (0.004)*** & 0.030 (0.004)*** \\
    Population density & -0.268 (0.137) & -0.203 (0.139) \\
    Population\textsubscript{B/W} ratio & -0.756 (0.107)*** & -0.721 (0.105)*** \\
    Household income & 0.787 (0.114)*** & 0.768 (0.110)***\\
    Household income\textsubscript{B/W} & -0.028 (0.112) & -0.016 (0.113) \\
    Incarceration & -0.773 (0.109)*** & -0.732 (0.106)*** \\
    Incarceration\textsubscript{B/W} & 0.282 (0.113)* & 0.264 (0.112)* \\
    HS graduation & 0.595 (0.113)*** & 0.573 (0.113)*** \\
    HS graduation\textsubscript{B/W} & 0.049 (0.101) & 0.069 (0.100) \\
    Employment & 0.557 (0.156)*** & 0.505 (0.158)** \\
    Employment\textsubscript{B/W} & -0.298 (0.137)* & -0.236 (0.135) \\
    \bottomrule
    \end{tabular}
    \parbox[t]{\columnwidth}{\vspace{1em}\small{Regression coefficients with robust standard errors are reported within parenthesis. Asterisks denote significance levels for Wald tests to assess the null hypothesis that the coefficients are equal to 0. Regressions on time are calculated on data from the period 2010--2019. All regressions were performed using the natural log of the enforcement ratio. In ``Time (legalized states)'', we only take into account states where marijuana has been legalized during the period considered. In ``Time (non legalized)'', we only consider states where marijuana has not been legalized. For both, we include only counties with continuous reporting throughout the period.\\ \phantom{h} \hfill $^{*}$p$<$0.05; $^{**}$p$<$0.01; $^{***}$p$<$0.001}}
\end{table}

\subsection{Racial disparities in incidents involving DUI and drunkenness offenses}
To examine the extent to which the disparities we observe are specific to marijuana violations, we perform a similar analysis on incidents (i) involving driving under the influence (DUI), and (ii) drunkenness that resulted in arrests (Appendix \ref{SISec:dui}). We find the county-level DUI and drunkenness enforcement ratios have a significant positive association with the corresponding marijuana enforcement ratio. In $73\%$ of the counties for which data was available, the enforcement rate for DUI offenses was more than twice as high for black individuals than for white individuals. For drunkenness, the equivalent was true for $62\%$ of the counties. These are largely on par with what we observed for marijuana, with DUI disparities being slightly higher and drunkenness slightly lower.

\section{Implications for algorithmic fairness within criminal justice}
Data on criminal offenses and arrests are central to machine learning algorithms used in criminal justice. Hence, a careful understanding of this data and its potential biases is critical in order to make progress towards building fair and effective sociotechnical systems. Arrests for drug abuse violations make up over $45\%$ of arrests in the US \citep{crimeexp}. As a result, many training, test, and validation datasets containing arrest data will include a large portion of arrests due to drug violations. Our results have shown that the risk of arrest due to marijuana violations varies significantly depending on an individual's race and location. Thus, these arrests represent a weak and biased proxy for offending. Algorithmic predictions based on such arrest data may reflect past enforcement behavior more than the level of offending. This can heavily impact spatio-temporal crime predictions, which are increasingly used in policing.
 A statement from the predictive policing tool developer \textit{PredPol} (now Geolitica) said that the company's guidance suggests not to use the tool to predict ``event types without clear victimization that can include officer discretion, such as drug-related offenses'' \citep{sankin_2021_crime}. 
In practice, however, there is evidence that the tool is used for this purpose, at least in some jurisdictions \citep{sankin_2021_crime}. Consistent with prior work \citep{feedback18,LumWill2016}, our results suggest that predictions generated by PredPol can vary significantly with the demographics of the resident population. Even when predictive policing is not used to directly predict drug offenses, changes in police presence will likely impact the local extent of drug arrests in an unequal manner due to differences in drug-related behaviors between black and white individuals. 
 
Algorithmic tools  for predicting recidivism, commonly used as a decision aid at pre-trial and sentencing \citep{{desmarais2021predictive, stevenson2021algorithmic}}, are evaluated on data of defendants’ future re-arrests. However, even if these predictions are accurate they are not devoid of bias. This could lead to significant differential impact on the lives of people from different subpopulations. Differences in the probability of arrest given that a crime was committed are particularly high for drug violations and DUI offenses, which make up a large proportion of overall arrests, but also occur in violent crimes \citep{fogliato2021validity}. 

\section{Limitations}
Our study suffers from several limitations. First, crime data from NIBRS may not be fully representative of crime trends at the national level or at the state level \cite{pattavina2017assessing}. Not all counties report to NIBRS and even at reporting counties, not all agencies report. However, we find that the coverage for reporting counties, with respect to the population covered, is generally above 75\%, as shown in Figure~\ref{SIfig:coverage_map} in the Appendix. Even at the level of individual agencies, the recorded data may not be representative of all crimes that are known to the agency due to selective reporting \cite{richardson2019dirty}. This would be particularly problematic if the level of reporting was associated to the offenders' race. Despite the above, NIBRS is, to date, the most comprehensive data source for non-aggregated crime reporting, and participation in NIBRS has been increasing in recent years.  

Another limitation is that we could not exclude all of the Hispanic and Latino individuals from the sample. In NIBRS, race and ethnicity are recorded as separate variables. The ethnicity field was introduced only in 2012 as an optional element and is not used by all reporting agencies \cite{nibrs_manual21}. We use the race field and ignore entries unless marked as white or black. In NIBRS, most of the Hispanic individuals are included among whites. However, due to inconsistent reporting, when we exclude those marked as Hispanics from the analysis, we cannot be sure that all Hispanics are indeed excluded. Thus, differences in offending behavior between whites and Hispanics will not be captured by our analysis. If Hispanic individuals suffer from disparities similar to Black individuals, this may mean that we are underestimating the $ER$ in our analysis. 

Estimates in rural counties have a large variance due to a low number of reported incidents, which is the reason why we take into account 95\% confidence intervals in our analysis. 
Furthermore, we assume that incidents reported by an agency involve residents from the county in which the agency is located.
We attempt to overcome both issues by applying spatial smoothing on rural counties. This analysis is presented in the Appendix Section \ref{SISec:smoothing}.

Our models for estimating the baseline crime rate are limited to the information available in the NSDUH data. Ideally, we would like to examine differences in location and time of use as well as the frequency at which an individual carries marijuana on their person or in their vehicle. The NSDUH survey data is also subject to respondents' biases in self-reporting. Problematically, there is evidence that individuals in both racial groups underreport use, but it has been suggested the phenomenon may be more prevalent among black respondents \cite{FendrichJohnson2005}. However, the likely extent of the potential under-reporting is not large enough to explain the disparities we observe. 

The combination of separate data sources presents its own challenges.
Due to lack of information on an individual's location in NSDUH data, our models rely on the assumption that the probability of daily marijuana use is constant across counties conditional on the county's urbanization level and demographics. Similarly, since income information was not available in the Census data we used, our estimation of population sizes across poverty levels represents an approximation to the real quantities. In addition, while NIBRS data is collected consistently year-round, the NSDUH sample is evenly allocated to four calendar quarters and the majority of the interviews completed in the first 4 to 6 weeks of each quarter. A study prepared for the Substance Abuse and Mental Health Services Administration found no statistically significant seasonal differences for marijuana use within the NSDUH data \cite{seasonal}. 



\section{Discussion}

We have shown that there are significant, widespread racial disparities in recorded incidents and arrests for marijuana violations. These disparities cannot be explained by differences in days-of-use alone. If we consider days-of-purchase in public instead of days-of-use, the magnitude of racial disparities shrinks, as black individuals self-report buying in public more frequently than whites. This suggests that cultural drug-related differences in where crimes are committed may explain, at least in part, the observed disparities. Although legalization of marijuana resulted in a reduction of arrests for both black and white individuals, relative racial disparities increased following legalization. This increase in disparities might be accounted for by differences in use and buying habits. Indeed, the activities that remain illegal, such as consumption in public spaces, are likely more prevalent among black users. 

The vast majority of incidents with marijuana violations concern possession rather than use or purchase. Discovering drug possession requires a search. Unfortunately, we have no record of the circumstances that led to the search and therefore do not know how many marijuana incidents are the byproduct of other enforcement activities, such as traffic stops. The addition of this information to the recorded incidents within NIBRS could advance our understanding of the observed racial disparities. 

Racial disparities are not unique to marijuana violations. Our analysis revealed widespread disparities in arrests for driving under the influence (DUI) and drunkenness. We found positive associations between county-level marijuana enforcement ratios and each of DUI and drunkenness enforcement ratios. 
These are suggestive of either racially disparate enforcement, or behavior by black individuals which makes their offenses occur in a more visible way across all three criminal categories, or some combination of both. 

To help disentangle these factors and more effectively inform policy-makers, more contextual law enforcement data is required. In future work, it would be valuable to explore the extent to which disparities can be explained by additional factors which are not currently available in the data, such as: frequency of use in public spaces; frequency of carrying marijuana on one's person or in a vehicle; opportunities to keep the use discreet; general exposure to police scrutiny; and patterns of police enforcement within the local environment. 

A better understanding of the mechanisms that give rise to the observed racial disparities is also important in the context of data-driven predictive tools that are increasingly deployed within the criminal justice system \citep{feedback18, babuta2020data, jansen2018data, sprick2019predictive}. These tools are typically trained and validated on arrest data, including a large portion of arrests for drug abuse violations. As a result, predictions made by these tools may reflect or even exacerbate past racially disparate enforcement. Greater transparency will facilitate discussion about what we should aim for as a society \citep{coyle2020explaining}.

The datasets used in this work have significant limitations, highlighted above. Nonetheless, the analysis 
is geographically comprehensive and considers multiple mechanisms that may contribute to differences in enforcement of drug violations. The large enforcement ratios we find suggest that, at a minimum, the enforcement of marijuana violations focuses on activities which are more \emph{visible} rather than more \emph{illegal}, disproportionately affecting black individuals.

\section{Methods}
\label{Sec:Method}

\subsection{Datasets}

In this work, we use data from NIBRS, NSDUH, Census, American Community Survey (ACS), Urban Influence Codes (UIC), the Opportunity Atlas.

The US Census is a national-level population census which takes place every ten years by order of constitutional mandate. In this work, we use the county-level data for estimating baseline crime rates. Released in May 2020, this data contains population estimates for age, race, sex, and Hispanic origin. More details on this dataset are in the Appendix Section \ref{SISec:Census}. 

Data on socioeconomic characteristics across demographic groups are obtained from the ACS \cite{acs2015}. Similarly to the decennial Census, this survey provides a snapshot of the population's demographics and income level across the country.

The Opportunity Atlas is a release of social mobility data from a collaboration between researchers at the Census Bureau, Harvard University, and Brown University \cite{jones2018opportunity}. The data was compiled from several sources: the 2000 and 2010 short-form census; federal income tax returns dating from 1989 until 2015; and the long form 2000 Census and the 2005-2015 ACS.

\subsection{Data Processing}

\subsubsection*{NIBRS}

In our main analysis, we only consider incidents that satisfy the following criteria: those that include only drug offenses or drug equipment offenses; the criminal activity is consuming, possessing or buying; the only drug associated with the incident is marijuana; and the reported quantity of marijuana is above-zero.

We include incidents with multiple offenders, treating each offense as a separate event for the purpose of modelling. We allow multiple locations, and take only the `most serious' drug-related criminal act recorded for the incident; we consider the following list of acts to be in order of seriousness: consuming, possessing, buying, transporting, producing, distributing -- offenses including any of the latter three are removed unless otherwise stated.

In the \texttt{Use}\textsubscript{Dmg+Pov, Arrests} model we add the condition that the incident led to an arrest. 

In Table~\ref{SItab:NIBRS_metro} we include all incidents that include at least one drug offense or drug equipment offenses, and that have an above-zero quantity of marijuana.
   
Offenders are only included where the data has age, race -- black or white -- and sex. Ethnicity is not always specified, and only indicates whether the offender is of Hispanic or Latino ethnicity, or not. Where specified, we include only those not of Hispanic or Latino ethnicity. The inclusion of offenders of unspecified ethnicity likely leads to white offenses being overrepresented \citep{steffensmeier2011reassessing} -- thereby leading to an overestimated white enforcement rate and a lower enforcement ratio.

\subsubsection*{NSDUH} 

We code survey respondents' age, poverty level, and whether the individual resides in a metropolitan area into categories matching those from the other data sources. 
We make the simplifying assumption that data are missing completely at random and thus drop the observations that lack one or more values among the features that we consider. Invalid data (e.g., those coded as ``bad data'') are treated as missing. 
For outcomes, we consider the number of days, in the past 30 days, that the respondent reported (i) using marijuana (variable: MJDAY30A), (ii) buying marijuana (MMBT30DY), (iii) buying marijuana in a public area (but not at school, MMBT30DY and MMBPLACE). Then, for each outcome and each demographic group, we compute the mean proportion of days (out of 30), accounting for the survey weights.

\subsubsection*{Census}

The age categories in the census are transformed to align with the NSDUH dataset. Individuals younger than twelve are not included, as they are not included in the NSDUH.

Poverty information is joined to the census data from the county-level 2015-2019 ACS \cite{acs2015}. This data source contains poverty information for each combination of age, race and sex. Poverty status was converted into two categories, reflecting whether the individual lives below or above the poverty threshold.

Information about the urbanization of the county is derived from the Urban Influence Codes (UIC) \cite{uic2013}. Each county is classified as being either a metropolitan or a non-metropolitan area.

\subsection{Variance estimation}

In our variance estimates we only account for the uncertainty arising from NIBRS and NSDUH data. We do not consider variation in the Census and ACS, as the NSDUH data -- due to the limited sample size -- is likely the main driver of the variance in the denominator of $\tcr$. 
We construct confidence intervals via the Delta method and the bootstrap method. 

A standard application of the Delta method to a logarithmic transformation of the enforcement ratio yields the following expression for the variance:
\begin{align}\label{eq:delta_method_log_selectionratio}
\begin{split}
  V\left( \log E_b /E_w \right) &= 
   \frac{1}{(\mbbE \ir_b)^2} V(\ir_b) + \frac{1}{(\mbbE \ir_w)^2} V(\ir_w)  + \frac{1}{(\mbbE \tcr_b)^2} V(\tcr_b) \\  &+ \frac{1}{(\mbbE \tcr_w)^2} V(\tcr_w) - \frac{1}{\mbbE\ir_b\mbbE\ir_w} Cov(\ir_b, \ir_w) \\ &- 
   \frac{1}{\mbbE\tcr_b\mbbE\tcr_w} Cov(\tcr_b, \tcr_w)
\end{split}
\end{align}
where we used the fact that
$Cov(\ir_{b}, \tcr_{b}) =  Cov(\ir_{w}, \tcr_{w}) = \\
Cov(\ir_{b}, \tcr_{w}) = 
Cov(\ir_{w}, \tcr_{b})=0$
because the estimates of $I$ and $C$ come from
separate data sources. Under the assumption that $\log E_b /E_w$ has approximately a Gaussian
distribution, the $1-\alpha$ confidence intervals are given by $E_b/E_w
e^{\pm z_{1-\alpha/2}\sqrt{\hat V}}$ where $z_{1-\alpha/2}$ is
the $1-\alpha/2$ quantile of a standard normal and $\hat V$ is
the estimate of $V(\log E_{b} /E_{w})$, the variance of $\log (E_{b} /E_{w})$. We apply an adjustment to the observations 
for which $\ir=0$ by adding $1/2$ to $\ir$ and $1$ to $\tcr$. 

For the yearly estimates of the selection ratio at the county level, 
we derive the variances in the first, third, and fourth terms in \eqref{eq:delta_method_log_selectionratio} directly from the NSUDH using standard
variance estimation techniques for survey data \cite{sarndal2003model}. 
We estimate the variances in the first two terms via the bootstrap method, assuming
that $\tcr$ corresponds to the (true) total number of users. Since we lack
structural knowledge of the relationship between $\ir$ and $\tcr$, we treat
the covariance terms as being 0. Since these terms are likely positive, our confidence intervals will be conservative. 

\subsection{Regression analysis} 

Least squares regression is performed to identify significant correlations between the county-level enforcement ratio and the county's socioeconomic characteristics. The regression is weighted by the reciprocal of the variance of the enforcement ratio. Wald tests are used to assess whether the regression coefficients were significantly different from zero. County characteristic are grouped into percentiles with respect to all counties in the analysis, to allow for a meaningful  comparison between regression coefficients.

The regression analysis included eight county-level socioeconomic measures from The Opportunity Atlas. Six variables have separate estimates for black individuals and white individuals for each county: household income at age 35; employment rate at 35; incarceration rate; high-school graduation rate; college graduation rate; and teenage birth rate. Two additional variables are estimated on the entire population of the county: census response rate; and population density.

\FloatBarrier

\subsubsection*{Data and Code Availability.} All data processing and analysis code is publicly available on GitHub, including an interactive data visualization tool \cite{CodeRepo}. 

\begin{acks}
The authors thank Alexandra Chouldechova and Jonathan Caulkins for valuable comments and discussion. We thank the reviewers for their helpful comments and suggestions. This work is partially supported by the European Research Council (ERC), grant agreement No 851538. M.Z. acknowledges support from EPSRC grant EP/V025279/1, The Alan Turing Institute, and the Leverhulme Trust grant ECF-2021-429. R.F. acknowledges support from PwC through Carnegie Mellon University's Digital Transformation and Innovation Center. A.W. acknowledges support from a Turing AI Fellowship under grant EP/V025279/1, The Alan Turing Institute, and the Leverhulme Trust via CFI.
\end{acks}
\bibliographystyle{ACM-Reference-Format}
\bibliography{AIES}


\appendix
\beginsupplement

\onecolumn

\clearpage

\section{Analysis of NSDUH data}
\label{SI:buying}
\begin{table}[!htbp]
\caption{Summary statistics based on 2010--2019 NSDUH data by county's urbanization (metropolitan vs. non metropolitan areas).}

  \centering
  \rotatebox{90}{
  \resizebox{0.9\textheight}{!}{
  \begin{tabular}{cccccccccc}
    \toprule
    & & \multicolumn{4}{c}{Metropolitan} & \multicolumn{4}{c}{Non metropolitan}\\
    \cmidrule{3-10} 
  & & \multicolumn{2}{c}{Black} & \multicolumn{2}{c}{White} & \multicolumn{2}{c}{Black} & \multicolumn{2}{c}{White}\\
  \cmidrule{3-10} 
 & & All & 18-25 male & All & 18-25 male & All & 18-25 male & All & 18-25 male\\ 
 \midrule 
    \multirow{4}{*}{\shortstack{How did you \\get marijuana\\ last time?}} &
    Bought it & 55\% (1) & 57\% (1) & 48\% (0) & 56\% (1) & 54\% (3) & 58\% (4) & 48\% (1) & 55\% (1) \\ 
     & Got it for free & 42\% (1) & 40\% (1) & 49\% (0) & 42\% (1) & 43\% (3) & 38\% (4) & 46\% (1) & 42\% (1) \\ 
    & Grew it & 1\% (0) & 1\% (0) & 2\% (0) & 1\% (0) & 2\% (1) & 3\% (1) & 4\% (0) & 1\% (0) \\ 
     & Traded for it & 2\% (0) & 2\% (0) & 1\% (0) & 1\% (0) & 1\% (0) & 1\% (0) & 2\% (0) & 2\% (0) \\ 
     \midrule
     \multirow{2}{*}{\shortstack{How often did you buy \\it in the past 30
     days?}} & \shortstack{Mean number of \\days among buyers} & 9 (0) & 11 (0) & 5 (0) & 6 (0) & 10 (1) & 11 (1) & 5 (0) & 7 (0) \\ 
     & \% buyers & 7\% (0) & 18\% (1) & 5\% (0) & 17\% (0) & 4\% (0) & 16\% (2) & 3\% (0) & 12\% (0) \\ 
     \midrule
     \multirow{2}{*}{\shortstack{How much marijuana\\ did you get last time?}} & $<$10 grams & 84\% (1) & 84\% (1) & 77\% (1) & 81\% (1) & 77\% (4) & 84\% (3) & 74\% (1) & 79\% (1) \\ 
     & $>$10 grams & 16\% (1) & 16\% (1) & 23\% (1) & 19\% (1) & 23\% (4) & 16\% (3) & 26\% (1) & 21\% (1) \\ 
     \midrule
     \multirow{4}{*}{\shortstack{How much did \\you pay for marijuana?}} &
     $<$20\$ & 61\% (1) & 62\% (2) & 30\% (0) & 36\% (1) & 65\% (4) & 68\% (4) & 31\% (1) & 37\% (2) \\ 
    & 21-50\$ & 22\% (1) & 19\% (1) & 33\% (1) & 32\% (1) & 19\% (4) & 18\% (4) & 33\% (1) & 33\% (2) \\ 
     & 51-100\$ & 9\% (1) & 9\% (1) & 20\% (0) & 18\% (1) & 6\% (2) & 6\% (2) & 20\% (1) & 17\% (1) \\ 
    & $>$100\$ & 7\% (1) & 9\% (1) & 17\% (0) & 14\% (1) & 9\% (2) & 7\% (2) & 16\% (1) & 14\% (1) \\ 
    \midrule
    \multirow{3}{*}{\shortstack{Who sold you \\marijuana \\last time?}} & Friend
    & 65\% (1) & 63\% (2) & 78\% (0) & 81\% (1) & 70\% (3) & 72\% (4) & 81\% (1) & 80\% (1) \\ 
     & Relative & 7\% (1) & 4\% (1) & 6\% (0) & 3\% (0) & 8\% (2) & 6\% (2) & 6\% (1) & 4\% (1) \\ 
     & Stranger & 28\% (1) & 33\% (1) & 17\% (0) & 16\% (1) & 22\% (3) & 22\% (4) & 14\% (1) & 15\% (1) \\ 
     \midrule
     \multirow{4}{*}{\shortstack{Where did you \\buy marijuana \\last time?}} &
     At school & 2\% (0) & 3\% (1) & 1\% (0) & 2\% (0) & 1\% (0) & 2\% (1) & 1\% (0) & 1\% (0) \\ 
    & Inside a home & 41\% (1) & 37\% (2) & 58\% (1) & 62\% (1) & 50\% (4) & 45\% (5) & 60\% (1) & 63\% (2) \\ 
     & Inside public building & 10\% (1) & 10\% (1) & 12\% (0) & 7\% (0) & 7\% (2) & 4\% (1) & 8\% (1) & 6\% (1) \\ 
     & Other & 14\% (1) & 13\% (1) & 13\% (0) & 10\% (1) & 15\% (3) & 16\% (3) & 15\% (1) & 11\% (1) \\ 
     & Outside in public area & 33\% (1) & 37\% (1) & 16\% (0) & 19\% (1) & 27\% (3) & 34\% (5) & 16\% (1) & 19\% (1) \\ 
     \midrule
     \multirow{2}{*}{\shortstack{How near you when \\you bought marijuana last
     time?}} & Near home & 34\% (1) & 31\% (1) & 49\% (1) & 46\% (1) & 29\% (3) & 30\% (5) & 41\% (1) & 39\% (2) \\ 
     & Somewhere else & 66\% (1) & 69\% (1) & 51\% (1) & 54\% (1) & 71\% (3) & 70\% (5) & 59\% (1) & 61\% (2) \\ 
     \bottomrule
     \multicolumn{10}{l} {\parbox[t]{\textheight}{\footnotesize{ \textit{Notes:} Standard errors are reported within parentheses. All values present in the data that did not correspond to the items listed in the table were dropped before computing the summary statistics (e.g., answers coded as ``bad data''). Responses from the period 2015-2017 were omitted for most of the elements listed in the table because they were not recorded in the data. 
}} }\\
  \end{tabular}
  }}
  
\label{SItab:NSDUH_metro}
  \end{table}

\clearpage

\begin{table}[!htbp]
\caption{Summary statistics based on 2010--2019 NSDUH data by socioeconomic status (living above vs. below the poverty threshold).}
\label{SItab:NSDUH_pov}
  \centering
  \rotatebox{90}{
  \resizebox{0.95\textheight}{!}{
  \begin{tabular}{cccccccccc}
    \toprule
    & & \multicolumn{4}{c}{Above poverty threshold} & \multicolumn{4}{c}{Below poverty threshold}\\
    \cmidrule{3-10} 
  & & \multicolumn{2}{c}{Black} & \multicolumn{2}{c}{White} & \multicolumn{2}{c}{Black} & \multicolumn{2}{c}{White}\\
  \cmidrule{3-10} 
 & & All & 18-25 male & All & 18-25 male & All & 18-25 male & All & 18-25 male\\ 
 \midrule 
    \multirow{4}{*}{\shortstack{How did you \\get marijuana\\ last time?}} &
    Bought it & 55\% (1) & 58\% (2) & 48\% (0) & 57\% (1) & 56\% (1) & 57\% (2) & 48\% (1) & 53\% (1) \\ 
     & Got for free & 43\% (1) & 40\% (2) & 49\% (0) & 40\% (1) & 41\% (1) & 39\% (2) & 48\% (1) & 44\% (1) \\ 
    & Grew it &  1\% (0) & 1\% (0) & 2\% (0) & 1\% (0) & 1\% (0) & 2\% (1) & 2\% (0) & 1\% (0) \\ 
     & Traded for it & 2\% (0) & 2\% (0) & 1\% (0) & 1\% (0) & 2\% (0) & 2\% (1) & 2\% (0) & 2\% (0) \\ 
     \midrule
     \multirow{2}{*}{\shortstack{How often did you buy \\it in the past 30
     days?}} & \shortstack{Mean number of \\days among buyers} & 9 (0) & 10 (0) & 5 (0) & 7 (0) & 10 (0) & 12 (1) & 6 (0) & 7 (0) \\ 
     & \% buyers & 6\% (0) & 17\% (1) & 4\% (0) & 16\% (0) & 8\% (0) & 19\% (1) & 8\% (0) & 18\% (1) \\ 
     \midrule
     \multirow{2}{*}{\shortstack{How much marijuana\\ did you get last time?}} & $<$10 grams & 82\% (1) & 83\% (1) & 76\% (1) & 80\% (1) & 87\% (1) & 85\% (2) & 81\% (1) & 84\% (1) \\ 
     & $>$10 grams & 18\% (1) & 17\% (1) & 24\% (1) & 20\% (1) & 13\% (1) & 15\% (2) & 19\% (1) & 16\% (1) \\ 
     \midrule
     \multirow{4}{*}{\shortstack{How much did \\you pay for marijuana?}} &
     $<$20\$ & 55\% (1) & 60\% (2) & 28\% (0) & 35\% (1) & 72\% (1) & 68\% (2) & 44\% (1) & 40\% (2) \\ 
    & 21-50\$ & 26\% (1) & 22\% (2) & 33\% (1) & 33\% (1) & 16\% (1) & 15\% (2) & 30\% (1) & 31\% (2) \\
     & 51-100\$ & 11\% (1) & 10\% (1) & 21\% (1) & 18\% (1) & 6\% (1) & 8\% (1) & 15\% (1) & 17\% (1) \\ 
    & $>$100\$ & 8\% (1) & 9\% (1) & 18\% (1) & 14\% (1) & 6\% (1) & 10\% (2) & 11\% (1) & 12\% (1) \\ 
    \midrule
    \multirow{3}{*}{\shortstack{Who sold you \\marijuana \\last time?}} & Friend
    & 67\% (1) & 66\% (2) & 78\% (1) & 81\% (1) & 62\% (2) & 59\% (2) & 78\% (1) & 80\% (1) \\ 
     & Relative & 7\% (1) & 4\% (1) & 6\% (0) & 3\% (0) & 6\% (1) & 5\% (1) & 6\% (1) & 3\% (1) \\ 
     & Stranger & 25\% (1) & 30\% (2) & 16\% (0) & 16\% (1) & 32\% (1) & 37\% (2) & 16\% (1) & 17\% (1) \\
     \midrule
     \multirow{4}{*}{\shortstack{Where did you \\buy marijuana \\last time?}} &
     At school & 2\% (0) & 3\% (1) & 1\% (0) & 2\% (0) & 2\% (0) & 3\% (1) & 1\% (0) & 2\% (0) \\ 
    & Inside a home & 44\% (1) & 37\% (2) & 57\% (1) & 60\% (1) & 37\% (2) & 38\% (2) & 62\% (1) & 68\% (1) \\ 
     & Inside public building & 10\% (1) & 11\% (1) & 12\% (0) & 8\% (0) & 11\% (1) & 8\% (1) & 9\% (1) & 5\% (1) \\ 
     & Other & 14\% (1) & 14\% (1) & 14\% (0) & 10\% (1) & 15\% (1) & 11\% (1) & 12\% (1) & 10\% (1) \\ 
     & Outside in public area & 31\% (1) & 35\% (2) & 16\% (0) & 20\% (1) & 35\% (2) & 39\% (2) & 15\% (1) & 15\% (1) \\ 
     \midrule
     \multirow{2}{*}{\shortstack{How near you when \\you bought marijuana last
     time?}} & Near home & 33\% (1) & 32\% (2) & 48\% (1) & 45\% (1) & 34\% (2) & 29\% (2) & 45\% (1) & 47\% (2) \\ 
     & Somewhere else & 67\% (1) & 68\% (2) & 52\% (1) & 55\% (1) & 66\% (2) & 71\% (2) & 55\% (1) & 53\% (2) \\ 
     \bottomrule
     \multicolumn{10}{l} {\parbox[t]{\textheight}{\footnotesize{ \textit{Notes:} Standard errors are reported within parentheses. All values present in the data that did not correspond to the items listed in the table were dropped before computing the summary statistics (e.g., `answers coded as `bad data''). Responses from the period 2015-2017 were omitted for most of the elements listed in the table because they were not recorded in the data. 
}}} \\
  \end{tabular}
  }}
 \end{table} 

\begin{figure}[!htb]
    \centering
    \includegraphics[height=0.95\textheight]{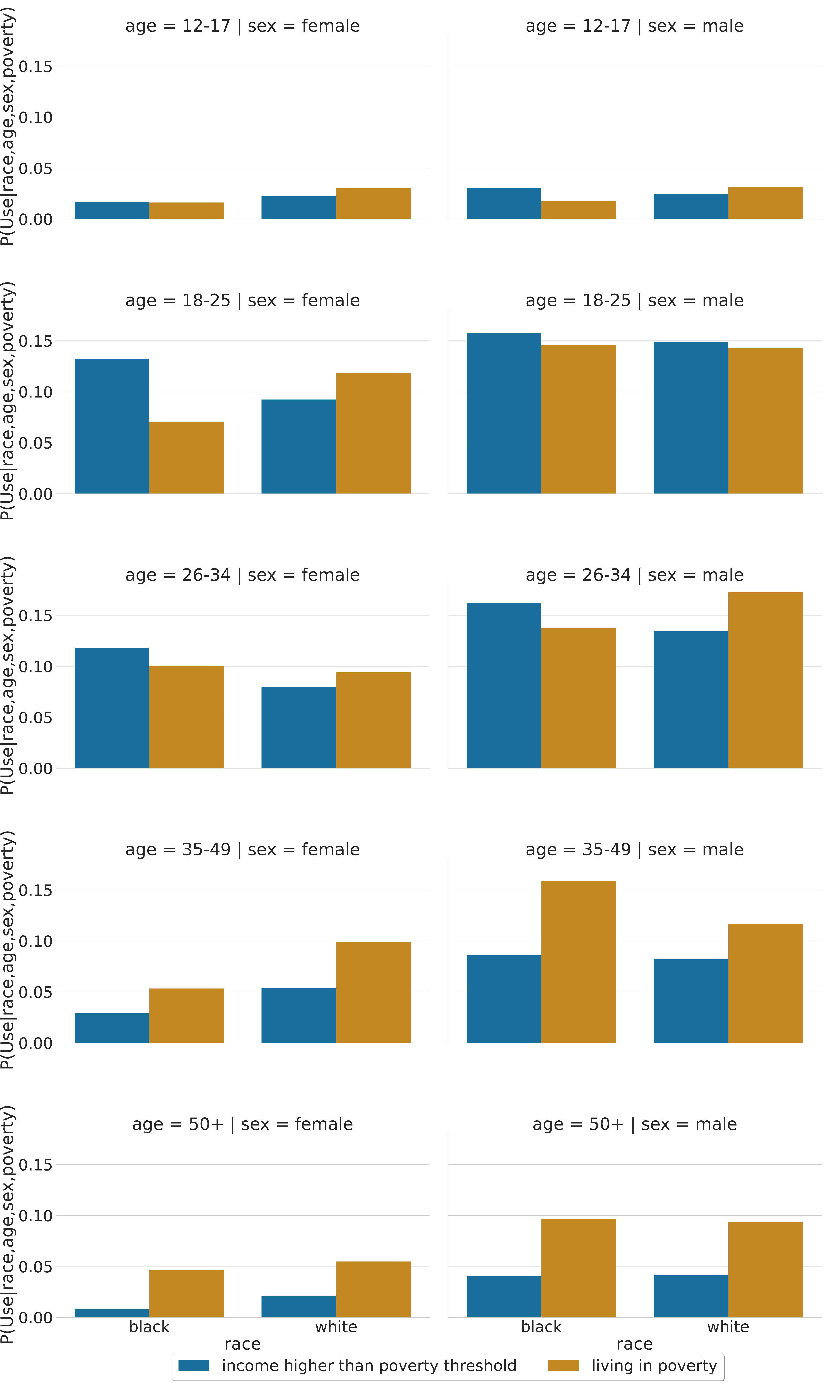}
    \caption{Usage probability by individual demographics and poverty status, based on self reported usage from the 2019 NSDUH survey.}
    \label{SIfig:dem_usage}
\end{figure}

\clearpage
  
\section{Analysis of NIBRS data}

\begin{table}[!htbp]
\caption{Summary statistics of incidents that include marijuana offenses based on 2010--2019 NIBRS data by county's urbanization (metropolitan vs. Non metropolitan areas).}
    \begin{minipage}{0.7\textheight}
  \centering
  \rotatebox{90}{
  \begin{tabular}{ccccccccc}
    \toprule
    & \multicolumn{4}{c}{Metropolitan} & \multicolumn{4}{c}{non metropolitan}\\
    \cmidrule{2-9} 
  & \multicolumn{2}{c}{Black} & \multicolumn{2}{c}{White} & \multicolumn{2}{c}{Black} & \multicolumn{2}{c}{White}\\
  \cmidrule{2-9} 
 & All & 18-25 male & All & 18-25 male & All & 18-25 male & All & 18-25 male\\ 
 \midrule 
 \# incidents & 1091087 & 426045 & 1670780 & 585236 & 125921 & 49782 & 530895 & 168178 \\ 
    \midrule
 \% drug only incidents & 88\% [88\%] & 87\% [86\%] & 94\% [94\%] & 95\% [95\%] & 91\% [90\%] & 90\% [89\%] & 95\% [94\%] & 96\% [96\%] \\ 
  \midrule
  \% possessing & 81\% [82\%] & 81\% [82\%] & 83\% [84\%] & 83\% [85\%] & 76\% [77\%] & 76\% [77\%] & 77\% [79\%] & 78\% [80\%] \\ 
  \% consuming & 2\% [2\%] & 2\% [2\%] & 4\% [3\%] & 4\% [3\%] & 4\% [4\%] & 5\% [4\%] & 6\% [5\%] & 6\% [6\%] \\ 
  \% buying & 1\% [1\%] & 1\% [1\%] & 1\% [1\%] & 1\% [1\%] & 1\% [1\%] & 1\% [1\%] & 2\% [2\%] & 1\% [1\%] \\ 
  \% distributing & 12\% [12\%] & 13\% [13\%] & 8\% [8\%] & 9\% [8\%] & 14\% [14\%] & 15\% [14\%] & 10\% [9\%] & 9\% [8\%] \\
    \midrule
 \% other drugs present & 17\% [17\%] & 13\% [14\%] & 18\% [18\%] & 13\% [14\%] & 14\% [15\%] & 11\% [12\%] & 19\% [20\%] & 13\% [13\%] \\  
    \midrule
  \% single offender & 60\% [63\%] & 58\% [61\%] & 57\% [60\%] & 58\% [61\%] & 61\% [64\%] & 58\% [61\%] & 59\% [62\%] & 59\% [61\%] \\ 
    \midrule
 \% residence & 17\% [16\%] & 15\% [14\%] & 19\% [17\%] & 16\% [15\%] & 20\% [18\%] & 18\% [16\%] & 26\% [23\%] & 21\% [19\%] \\ 
  \% hotel & 2\% [2\%] & 2\% [2\%] & 2\% [2\%] & 1\% [1\%] & 2\% [2\%] & 1\% [1\%] & 2\% [2\%] & 1\% [1\%] \\ 
  \% highway/road & 58\% [60\%] & 61\% [63\%] & 53\% [56\%] & 58\% [61\%] & 56\% [59\%] & 58\% [60\%] & 49\% [53\%] & 55\% [59\%] \\ 
  \% parking lot/garage & 9\% [9\%] & 10\% [10\%] & 8\% [8\%] & 9\% [9\%] & 6\% [6\%] & 7\% [7\%] & 6\% [6\%] & 6\% [6\%] \\ 
    \midrule
 \% during day (6-20) & 58\% [57\%] & 54\% [54\%] & 55\% [54\%] & 47\% [46\%] & 55\% [54\%] & 52\% [51\%] & 57\% [55\%] & 50\% [48\%] \\ 
    \midrule
 \% no arrest & 22\% & 21\% & 24\% & 23\% & 26\% & 25\% & 25\% & 23\% \\ 
  \% arrest: custody & 14\% & 14\% & 11\% & 11\% & 14\% & 13\% & 13\% & 13\% \\ 
  \% arrest: on view & 39\% & 39\% & 37\% & 36\% & 43\% & 43\% & 42\% & 42\% \\ 
  \% arrest: summoned/cited & 24\% & 26\% & 28\% & 30\% & 18\% & 19\% & 21\% & 23\% \\ 
     \bottomrule
     \multicolumn{9}{l} {\parbox[t]{\textwidth}{\footnotesize{\textit{Notes:} Summary statistics for arrests are reported within square brackets. Most of the standard errors are below 1\% and thus are omitted from the table.
}} }\\
  \end{tabular}
  }
  \end{minipage}
  \label{SItab:NIBRS_metro}
  \end{table}  

\clearpage

\begin{table}[!htbp]
\caption{Summary statistics of incidents that include only marijuana offenses based on 2010--2019 NIBRS data.}
  \centering
\rotatebox{90}{
  \begin{tabular}{ccccccccc}
    \toprule
    & \multicolumn{4}{c}{Metropolitan} & \multicolumn{4}{c}{Non metropolitan}\\
    \cmidrule{2-9} 
  & \multicolumn{2}{c}{Black} & \multicolumn{2}{c}{White} & \multicolumn{2}{c}{Black} & \multicolumn{2}{c}{White}\\
  \cmidrule{2-9} 
 & All & 18-25 male & All & 18-25 male & All & 18-25 male & All & 18-25 male\\ 
 \midrule 
\# incidents & 962283 & 369465 & 1575794 & 555872 & 114310 & 44640 & 502365 & 161045 \\ 
    \midrule
  \% possessing & 82\% [83\%] & 83\% [83\%] & 83\% [85\%] & 84\% [85\%] & 77\% [78\%] & 76\% [77\%] & 77\% [79\%] & 79\% [80\%] \\ 
  \% consuming & 3\% [2\%] & 3\% [2\%] & 4\% [3\%] & 4\% [3\%] & 5\% [5\%] & 5\% [5\%] & 6\% [6\%] & 6\% [6\%] \\ 
  \% buying & 1\% [1\%] & 1\% [1\%] & 1\% [1\%] & 1\% [1\%] & 1\% [1\%] & 1\% [1\%] & 2\% [1\%] & 1\% [1\%] \\ 
  \% distributing & 12\% [11\%] & 12\% [12\%] & 8\% [7\%] & 8\% [7\%] & 14\% [13\%] & 14\% [13\%] & 9\% [8\%] & 9\% [8\%] \\ 
    \midrule
  \% other drugs present & 15\% [16\%] & 12\% [12\%] & 16\% [17\%] & 12\% [13\%] & 13\% [14\%] & 10\% [10\%] & 18\% [19\%] & 12\% [13\%] \\ 
    \midrule
  \% single offender & 63\% [66\%] & 61\% [65\%] & 58\% [62\%] & 59\% [62\%] & 63\% [66\%] & 61\% [63\%] & 59\% [63\%] & 59\% [62\%] \\ 
    \midrule
 \% residence & 16\% [14\%] & 14\% [13\%] & 18\% [16\%] & 16\% [14\%] & 19\% [17\%] & 17\% [16\%] & 25\% [22\%] & 20\% [18\%] \\ 
  \% hotel & 2\% [2\%] & 2\% [2\%] & 2\% [2\%] & 1\% [1\%] & 2\% [2\%] & 1\% [1\%] & 2\% [2\%] & 1\% [1\%] \\ 
  \% highway/road & 58\% [60\%] & 62\% [63\%] & 53\% [56\%] & 58\% [61\%] & 57\% [60\%] & 58\% [60\%] & 50\% [54\%] & 56\% [59\%] \\ 
  \% parking lot/garage & 9\% [9\%] & 10\% [10\%] & 8\% [8\%] & 9\% [9\%] & 6\% [6\%] & 7\% [7\%] & 6\% [6\%] & 6\% [6\%] \\ 
    \midrule
  \% during day (6-20) & 58\% [57\%] & 54\% [54\%] & 55\% [54\%] & 47\% [46\%] & 55\% [54\%] & 52\% [51\%] & 57\% [55\%] & 49\% [48\%] \\ 
    \midrule
 \% no arrest & 23\% & 22\% & 24\% & 23\% & 26\% & 26\% & 25\% & 23\% \\ 
  \% arrest: custody & 13\% & 13\% & 11\% & 10\% & 13\% & 13\% & 13\% & 13\% \\ 
  \% arrest: on view & 37\% & 37\% & 36\% & 36\% & 41\% & 41\% & 41\% & 41\% \\ 
  \% arrest: summoned/cited & 26\% & 28\% & 29\% & 31\% & 19\% & 20\% & 21\% & 24\% \\ 
     \bottomrule
     \multicolumn{9}{l} {\parbox[t]{\textwidth}{\footnotesize{\textit{Notes:} Incidents that involve offenses other than marijuana-related violations are not considered in this table. Summary statistics relative to arrests are reported within square brackets. Most of the standard errors are below 1\% and thus are omitted from the table. 
}} }\\
  \end{tabular}
  }
  \label{SItab:NIBRS_metro_all}
  \end{table}  
  
\clearpage

\begin{figure}[!htbp]
    \centering
    \includegraphics[width=0.5\textwidth]{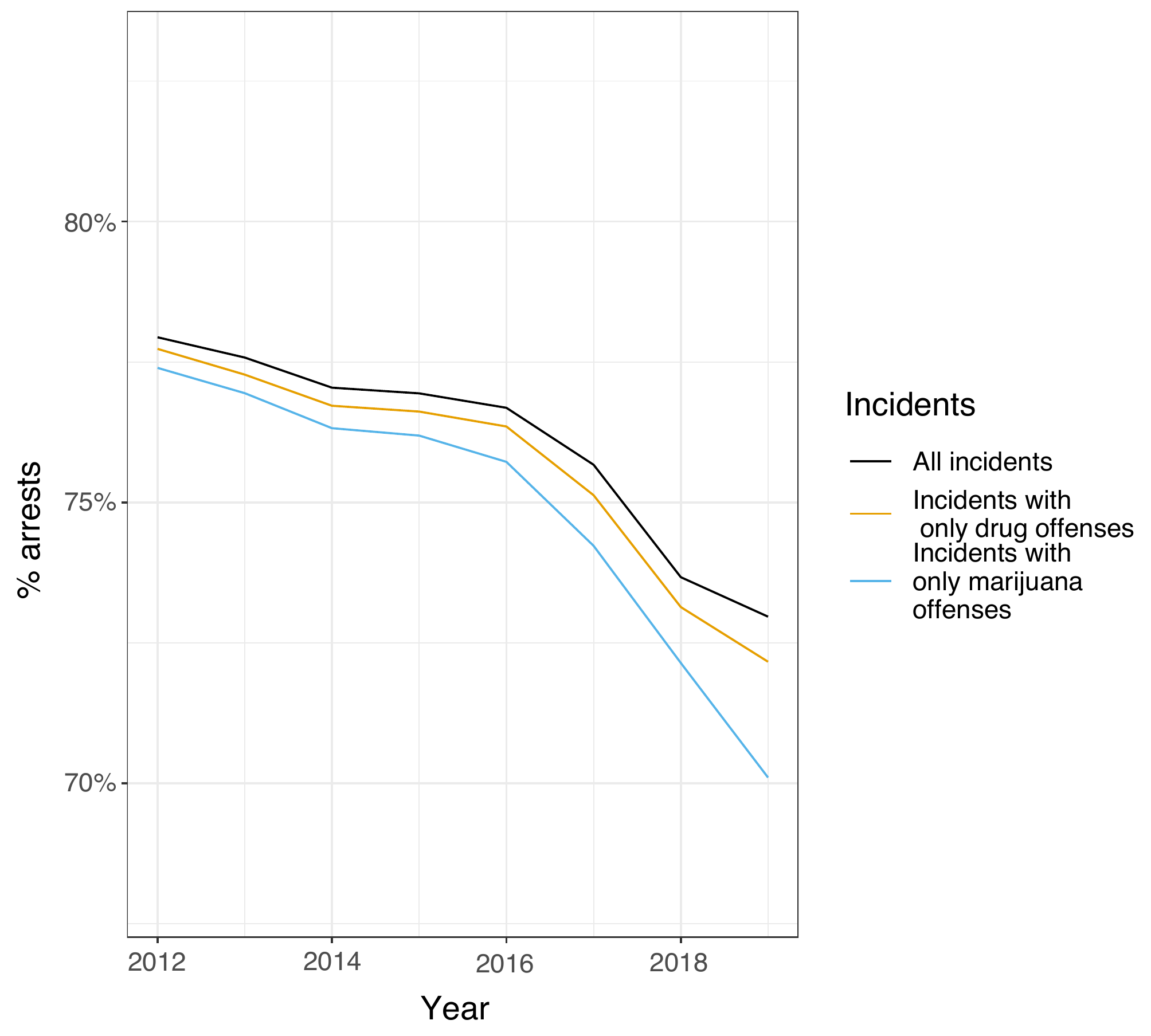}
    \caption{Arrest rate over time, by type of offense, out of all incidents with drug offenses, based on 2012--2019 NIBRS data.}
    \label{SIfig:arrest_rate}
\end{figure} 


\begin{figure}[!htb]
    \centering
    \includegraphics[width=0.5\textwidth]{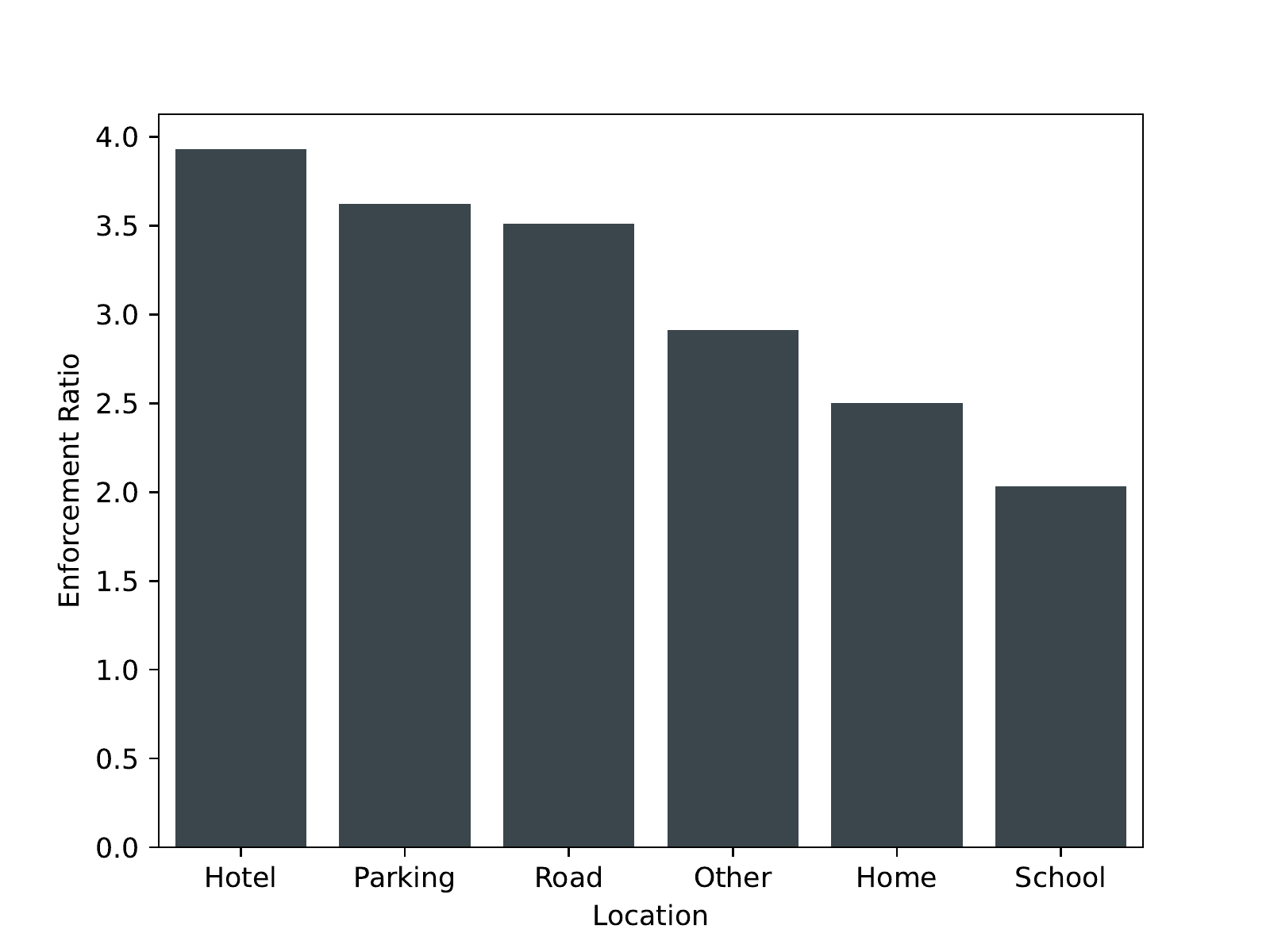}
    \caption{Enforcement ratio by location of incident for cannabis only incidents, where only a single location was recorded, for 2010-2019.}
    \label{SIfig:location}
\end{figure} 

\clearpage

\begin{figure}[!htb]
    \centering
    \includegraphics[width=0.85\textwidth]{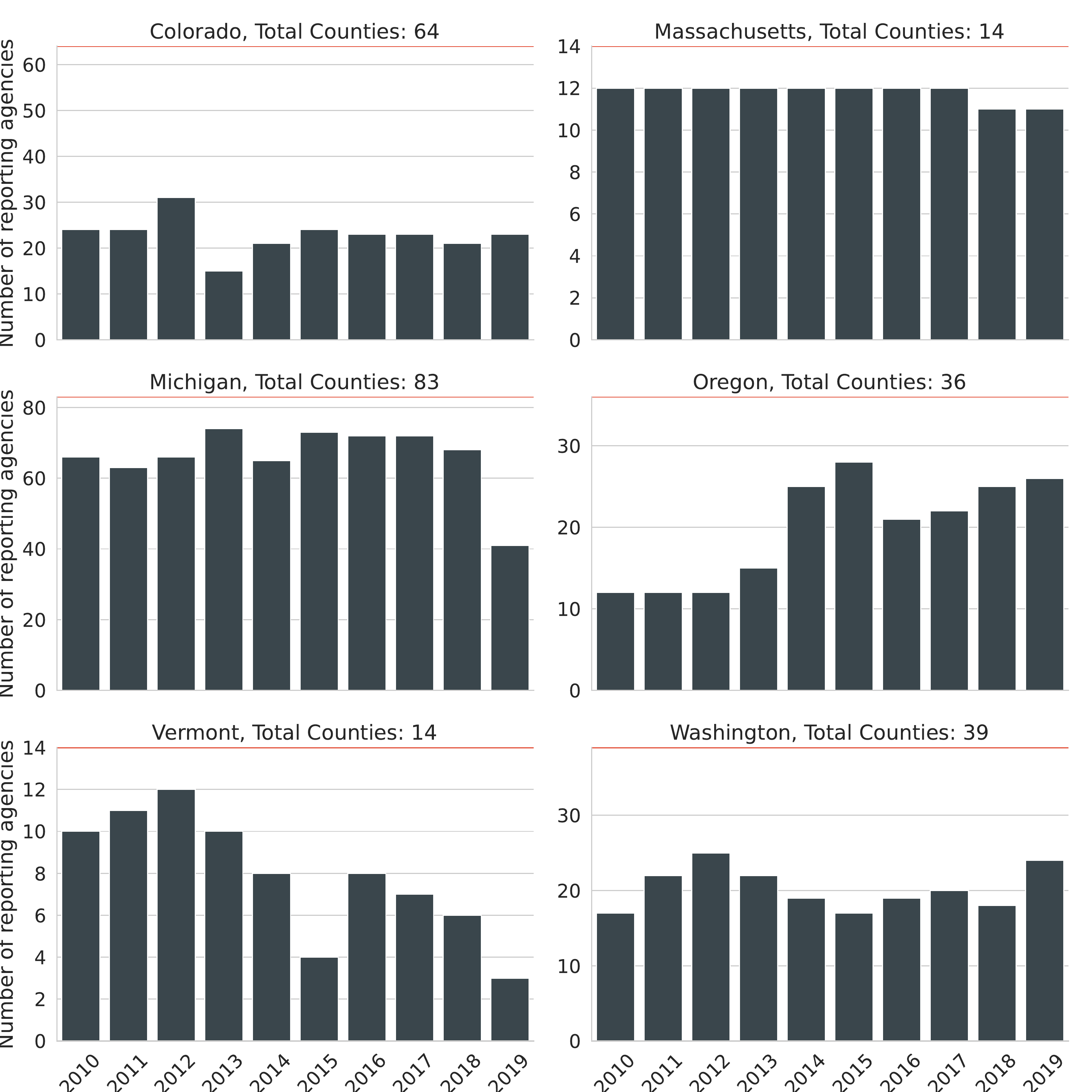}
    \caption{The number of counties reporting to the NIBRS per year for the legalized states in figure \ref{fig:legal}}
    \label{SIfig:legal_counties}
\end{figure} 

\clearpage

\begin{figure}[!htb]
    \centering
    \includegraphics[height=0.95\textheight]{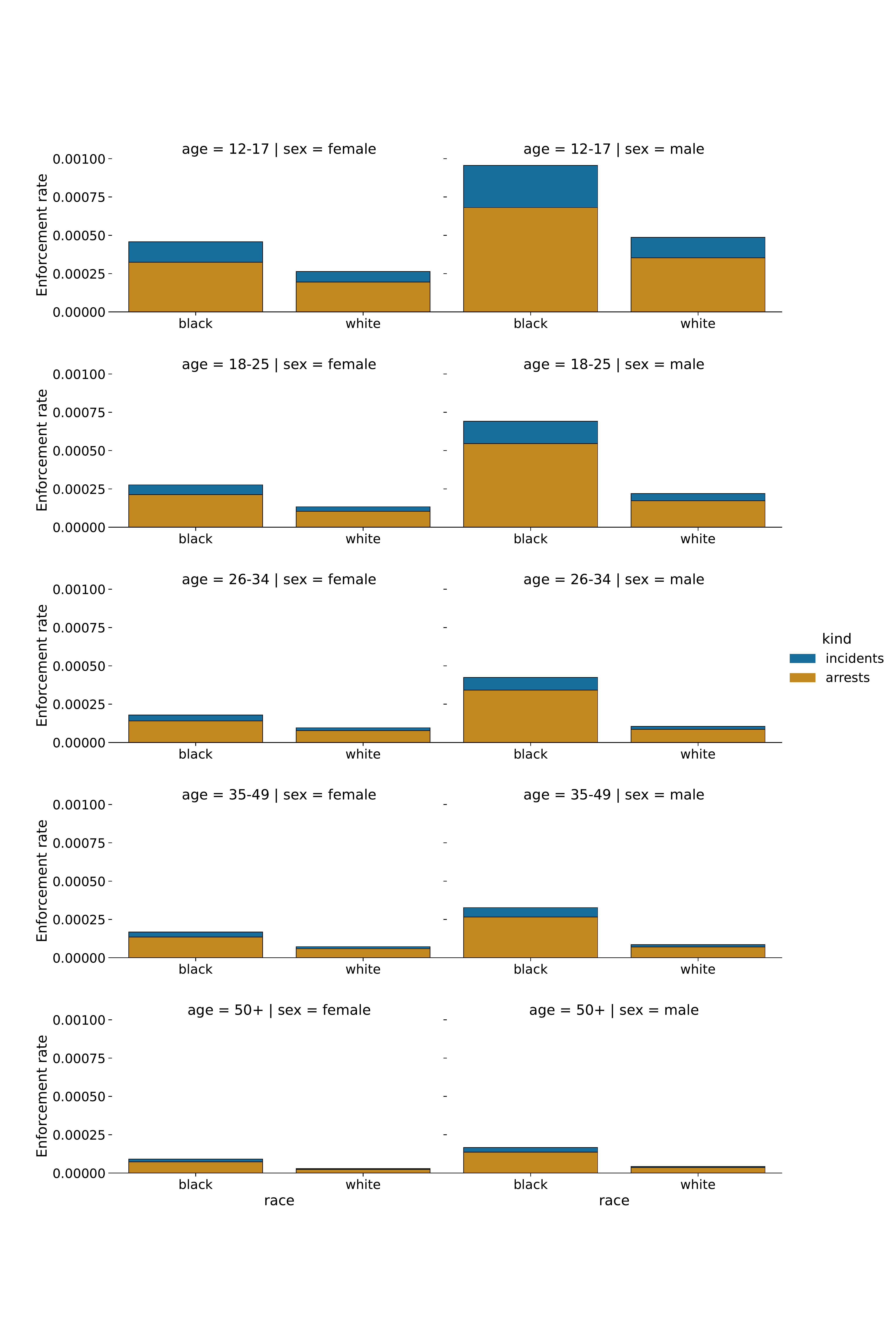}
    \caption{Enforcement rate per user, by demographic combination, averaged over the years 2012-2019.}
    \label{SIfig:dem_enforcement_bar}
\end{figure} 

\clearpage

\begin{figure}[!htb]
    \centering
    \includegraphics[height=0.95\textheight]{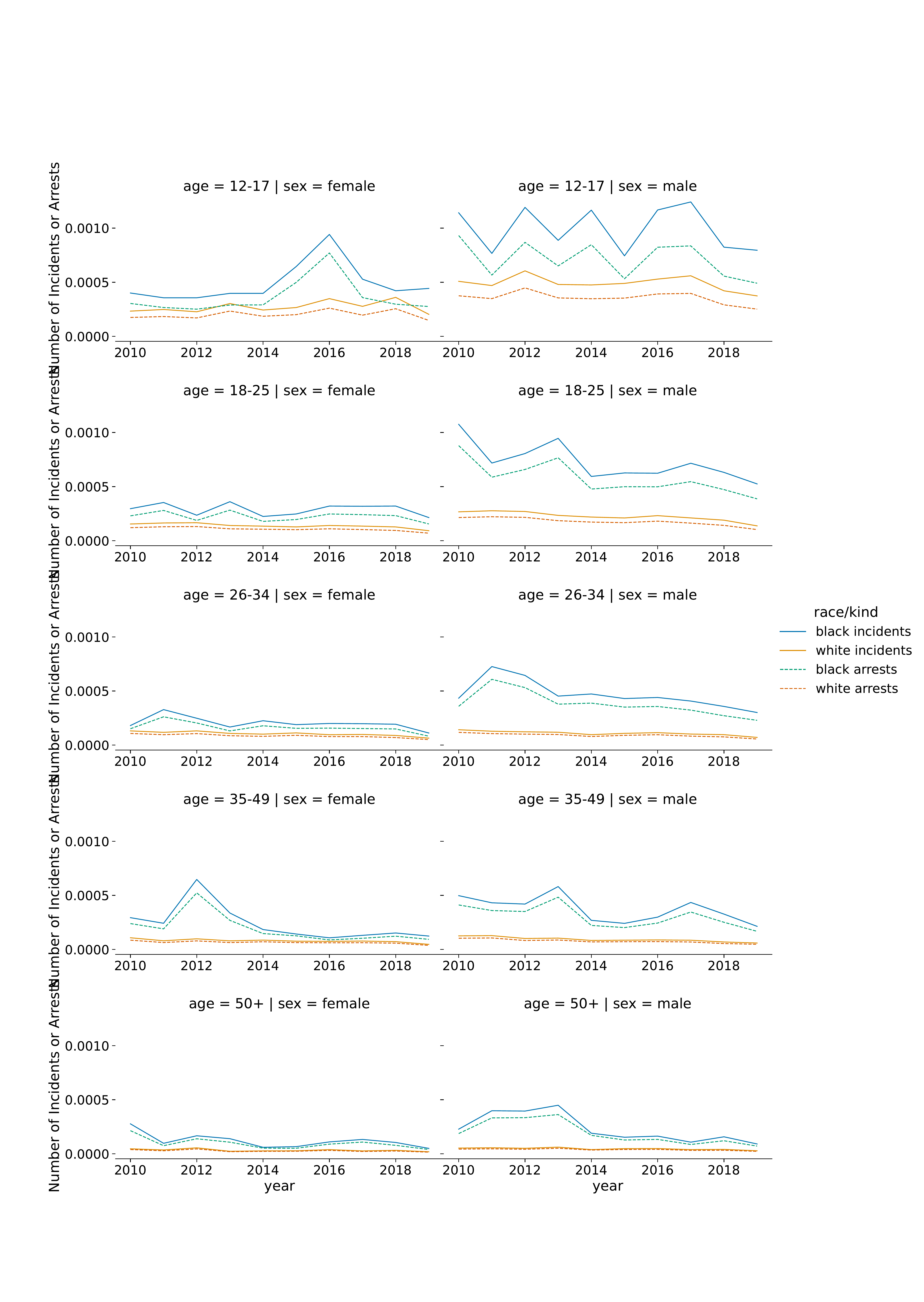}
    \caption{Yearly enforcement rate per user, by demographic combination, between 2012 and 2019.}
    \label{SIfig:dem_enforcement_line}
\end{figure} 

\clearpage


\section{Regression analysis}

\subsection{Uni-variate}

The regressions presented below were computed independently for each covariate. In the following section multi-variate regressions are presented.

\begin{table}[!htbp]
  \centering
      \caption{Regression coefficients estimates of enforcement ratio, based on different models on NUSDH data, on county-level covariates. }
          
  \rotatebox{90}{
  \resizebox{0.8\textheight}{!}{
    \begin{tabular}{l c * {6} {d{10}}}
    \toprule
     &  \mc{\texttt{Use}\textsubscript{Dmg}} & \mc{\texttt{Use}\textsubscript{Dmg+Pov}} &   \mc{\texttt{Use}\textsubscript{Dmg+Urb}} & \mc{\texttt{Purchase}} & \mc{\texttt{Purchase}\textsubscript{Public}} & \mc{\texttt{Arrests}\textsubscript{Dmg+Pov}} \\
    \midrule
    Time$^1$                                &  0.028 (0.004)***  &  0.024 (0.004)*** &  0.030 (0.004)*** &  0.018 (0.004)*** &  0.030 (0.004)*** &  0.024 (0.004)*** \\
    Time$^2$                                &  0.033 (0.004)***  &  0.029 (0.004)*** &  0.036 (0.004)*** &  0.023 (0.004)*** &  0.035 (0.004)*** &  0.028 (0.005)*** \\
    Time$^3$                                &    0.043 (0.020)*  &    0.041 (0.019)* &     0.039 (0.020) &    0.039 (0.020)* &   0.051 (0.019)** &    0.049 (0.023)* \\
    Time$^4$                                &  0.077 (0.021)***  &  0.074 (0.021)*** &  0.080 (0.021)*** &  0.075 (0.021)*** &  0.082 (0.020)*** &    0.065 (0.028)* \\
    B/W population ratio                  &  -0.742 (0.106)*** &  -0.756 (0.107)*** &  -0.764 (0.108)*** &  -0.725 (0.105)*** &  -0.721 (0.105)*** &  -0.756 (0.124)*** \\
    Income                                &   0.786 (0.114)*** &   0.787 (0.114)*** &   0.784 (0.116)*** &   0.759 (0.111)*** &   0.768 (0.110)*** &   0.765 (0.133)*** \\
    Income B/W ratio                      &     -0.065 (0.112) &     -0.028 (0.112) &     -0.038 (0.113) &     -0.027 (0.113) &     -0.016 (0.113) &     -0.038 (0.124) \\
    Incarceration                         &  -0.759 (0.108)*** &  -0.773 (0.109)*** &  -0.782 (0.111)*** &  -0.729 (0.107)*** &  -0.732 (0.106)*** &  -0.742 (0.124)*** \\
    Incarceration B/W ratio               &    0.308 (0.113)** &     0.282 (0.113)* &     0.275 (0.113)* &     0.269 (0.112)* &     0.264 (0.112)* &     0.291 (0.122)* \\
    Population density                    &     -0.190 (0.136) &     -0.268 (0.137) &   -0.384 (0.136)** &     -0.204 (0.138) &     -0.203 (0.139) &     -0.246 (0.144) \\
    High school graduation rate           &   0.594 (0.113)*** &   0.595 (0.113)*** &   0.586 (0.114)*** &   0.568 (0.113)*** &   0.573 (0.113)*** &   0.589 (0.125)*** \\
    High school graduation rate B/W ratio &      0.031 (0.101) &      0.049 (0.101) &      0.032 (0.101) &      0.062 (0.100) &      0.069 (0.100) &      0.046 (0.113) \\
    College graduation rate               &    0.270 (0.098)** &     0.239 (0.099)* &     0.198 (0.100)* &    0.271 (0.099)** &    0.275 (0.099)** &     0.258 (0.105)* \\
    College graduation rate B/W ratio     &     -0.172 (0.120) &     -0.145 (0.121) &     -0.160 (0.121) &     -0.124 (0.120) &     -0.105 (0.120) &     -0.184 (0.127) \\
    Employment rate at 35                 &   0.563 (0.156)*** &   0.557 (0.156)*** &   0.570 (0.158)*** &    0.502 (0.157)** &    0.505 (0.158)** &    0.569 (0.178)** \\
    Employment rate at 35 B/W ratio       &    -0.319 (0.136)* &    -0.298 (0.137)* &    -0.320 (0.138)* &     -0.255 (0.135) &     -0.236 (0.135) &     -0.280 (0.155) \\
    Teenage birth rate                    &  -0.733 (0.105)*** &  -0.736 (0.105)*** &  -0.725 (0.107)*** &  -0.706 (0.104)*** &  -0.713 (0.103)*** &  -0.727 (0.118)*** \\
    Teenage birth rate B/W ratio          &    0.313 (0.099)** &    0.280 (0.099)** &     0.253 (0.099)* &    0.282 (0.100)** &    0.287 (0.100)** &    0.323 (0.111)** \\
    Census Response rate                  &   0.847 (0.136)*** &   0.866 (0.136)*** &   0.885 (0.136)*** &   0.779 (0.138)*** &   0.785 (0.138)*** &   0.833 (0.150)*** \\
    \bottomrule
    \multicolumn{7}{l} {\parbox[t]{\textwidth}{\footnotesize{\textit{Notes:} Results of linear regression models fitted via ordinary least squares (OLS) with intercept and the individual terms indicated in the table as covariates. Each cell shows the results of a different regression model. Robust standard errors are reported within parenthesis. Asterisks following regression coefficients denote significance levels for Wald tests to assess the null hypothesis that the coefficients are equal to 0. Time regressions are calculated from years 2012-2019. Each feature is binned into it's percentile rank, with the exception of the Times. All regressions were performed using the natural log of the enforcement ratio.\\
    \phantom{h} \hfill $^{*}$p$<$0.05; $^{**}$p$<$0.01; $^{***}$p$<$0.001 \\
    $^{1}$  Only counties reporting throughout - excluding legalized states.\\
    $^{2}$ All counties - excluding legalized states.\\
    $^{3}$ Only counties reporting throughout - only legalized states.\\
    $^{4}$ All counties - only legalized states.}}}
    \end{tabular}}}
\label{SItab:corr3}
\end{table}

\clearpage

\subsection{Multi-variate}

Below are the multivariate regression results. Only a subset of counties report these covariates stratified by race, so two sets of results are presented. First, a table of multi-variate regressions where all counties have been included, not including / white ratio covariates. Second, a multi-variate regression including black / white ratio covariates, but only on the subset of counties that report this information.

\begin{table}[!htbp]
  \centering
      \caption{Multi-variate regression coefficients estimates of enforcement ratio, based on different models on NUSDH data, on county-level covariates. }
          
    \resizebox{\textwidth}{!}{
    \begin{tabular}{lllllll}
    \toprule
     &  \mc{\texttt{Use}\textsubscript{Dmg}} & \mc{\texttt{Use}\textsubscript{Dmg+Pov}} &   \mc{\texttt{Use}\textsubscript{Dmg+Urb}} & \mc{\texttt{Purchase}} & \mc{\texttt{Purchase}\textsubscript{Public}} & \mc{\texttt{Arrests}\textsubscript{Dmg+Pov}} \\
    \midrule
    B/W population ratio        &  -0.427 (0.143)** &  -0.418 (0.143)** &  -0.431 (0.145)** &  -0.426 (0.143)** &  -0.402 (0.141)** &  -0.501 (0.162)** \\
    Income                      &    0.782 (0.331)* &    0.736 (0.330)* &    0.795 (0.336)* &    0.766 (0.331)* &    0.797 (0.332)* &    0.695 (0.318)* \\
    Incarceration               &    -0.106 (0.151) &    -0.102 (0.150) &    -0.112 (0.150) &    -0.081 (0.152) &    -0.062 (0.152) &     0.013 (0.173) \\
    \% Republican Vote Share     &    -0.054 (0.392) &    -0.002 (0.392) &    -0.104 (0.398) &    -0.044 (0.400) &    -0.004 (0.400) &    -0.187 (0.443) \\
    Population density          &    -0.055 (0.170) &    -0.137 (0.171) &    -0.179 (0.174) &    -0.138 (0.172) &    -0.135 (0.173) &    -0.261 (0.191) \\
    High school graduation rate &    -0.210 (0.143) &    -0.218 (0.144) &    -0.242 (0.145) &    -0.197 (0.146) &    -0.207 (0.145) &    -0.209 (0.153) \\
    College graduation rate     &   0.448 (0.137)** &   0.444 (0.138)** &   0.445 (0.139)** &   0.458 (0.141)** &   0.461 (0.141)** &   0.422 (0.148)** \\
    Employment rate at 35       &    -0.253 (0.192) &    -0.256 (0.193) &    -0.259 (0.195) &    -0.310 (0.197) &    -0.311 (0.199) &    -0.181 (0.212) \\
    Teenage birth rate          &     0.329 (0.275) &     0.257 (0.274) &     0.314 (0.278) &     0.282 (0.275) &     0.254 (0.276) &     0.183 (0.288) \\
    Census Response rate        &  0.712 (0.132)*** &  0.704 (0.132)*** &  0.698 (0.131)*** &  0.658 (0.133)*** &  0.665 (0.133)*** &  0.683 (0.143)*** \\
    \bottomrule
    \multicolumn{7}{l}{\parbox[t]{\textwidth}{\footnotesize{\textit{Notes:} Results of linear regression models fitted via ordinary least squares (OLS) with intercept and the individual terms indicated in the table as covariates. Each cell shows the results of a different regression model. Robust standard errors are reported within parenthesis. Asterisks following regression coefficients denote significance levels for Wald tests to assess the null hypothesis that the coefficients are equal to 0. Time regressions are calculated from years 2012-2019. Each feature is binned into it's percentile rank, with the exception of the Times. All regressions were performed using the natural log of the enforcement ratio.\\
    \phantom{h} \hfill $^{*}$p$<$0.05; $^{**}$p$<$0.01; $^{***}$p$<$0.001}}}
    \end{tabular}}
\label{SItab:corr1}
\end{table}

\begin{table}[!htbp]
  \centering
      \caption{Multi-variate regression coefficients estimates of enforcement ratio, based on different models on NUSDH data, on black / white ratio county-level covariates. }
          
    \resizebox{\textwidth}{!}{
    \begin{tabular}{lllllll}
    \toprule
     &  \mc{\texttt{Use}\textsubscript{Dmg}} & \mc{\texttt{Use}\textsubscript{Dmg+Pov}} &   \mc{\texttt{Use}\textsubscript{Dmg+Urb}} & \mc{\texttt{Purchase}} & \mc{\texttt{Purchase}\textsubscript{Public}} & \mc{\texttt{Arrests}\textsubscript{Dmg+Pov}} \\
    \midrule
    Income B/W ratio                      &   0.083 (0.224) &   0.103 (0.223) &   0.043 (0.222) &   0.093 (0.223) &   0.126 (0.223) &   0.161 (0.224) \\
    Incarceration B/W ratio               &   0.217 (0.188) &   0.225 (0.188) &   0.223 (0.188) &   0.230 (0.183) &   0.223 (0.184) &   0.175 (0.198) \\
    High school graduation rate B/W ratio &   0.179 (0.150) &   0.178 (0.149) &   0.176 (0.150) &   0.191 (0.149) &   0.187 (0.150) &   0.173 (0.164) \\
    College graduation rate B/W ratio     &  -0.061 (0.150) &  -0.051 (0.150) &  -0.049 (0.150) &  -0.038 (0.146) &  -0.027 (0.147) &  -0.130 (0.159) \\
    Employment rate at 35 B/W ratio       &  -0.110 (0.178) &  -0.099 (0.177) &  -0.103 (0.179) &  -0.074 (0.175) &  -0.069 (0.175) &  -0.063 (0.185) \\
    Teenage birth rate B/W ratio          &   0.308 (0.172) &   0.289 (0.171) &   0.250 (0.172) &   0.277 (0.170) &   0.303 (0.170) &   0.368 (0.200) \\
    \bottomrule
    \multicolumn{7}{l} {\parbox[t]{\textwidth}{\footnotesize{\textit{Notes:} Results of linear regression models fitted via ordinary least squares (OLS) with intercept and the individual terms indicated in the table as covariates. Each cell shows the results of a different regression model. Robust standard errors are reported within parenthesis. Asterisks following regression coefficients denote significance levels for Wald tests to assess the null hypothesis that the coefficients are equal to 0. Time regressions are calculated from years 2012-2019. Each feature is binned into it's percentile rank, with the exception of the Times. All regressions were performed using the natural log of the enforcement ratio.\\
    \phantom{h} \hfill $^{*}$p$<$0.05; $^{**}$p$<$0.01; $^{***}$p$<$0.001}}}
    \end{tabular}}
\label{SItab:corr2}
\end{table}

\clearpage



\section{Choropleth Maps}

The choropleth maps presented throughout the paper are generated based on data from 2017--2019. 
We only considered data from a limited time range in order to avoid conflating disparities with time trends. 
Colors in the map correspond to different thresholds on the enforcement ratio. The opacity of the maps are binned as a quartile on the inverse relative standard error for the enforcement ratio on each county. The relative error is calculated as: $|\frac{\sqrt{\text{var}(\log ER)}}{\log ER}|$. In addition, we removed counties that reported no incidents for either white or black individuals, e.g., those for which no incident with black offenders was recorded. 

\subsection{Wilson score interval}

As we assume the enforcement rate is binomial, we can use the Wilson score interval to calculate the variance:

$$
\sigma(E)=z \sqrt{\frac{E(1-E)}{C}} \quad \sigma(ER)=|ER| \cdot \sqrt{\left(\frac{\sigma\left(E_{b}\right)}{E_{b}}\right)^{2}+\left(\frac{\sigma \left(E_{w}\right)}{E_{w}}\right)^{2}}
$$

\begin{align*}
\sigma(ER) &= |ER| \cdot \sqrt{z^{2} \left(\threefrac{E_{b}(1 - E_{b})}{C_{b}}{E_{b}^2}\right)  + z^{2} \left(\threefrac{E_{w}(1 - E_{w})}{C_{w}}{E_{b}^2}\right)} \\
 &= |ER| \cdot z \cdot \sqrt{\frac{E_b(1-E_b)}{E_{b}^{2} \cdot C_{b}} + \frac{E_w(1-E_w)}{E_{w}^{2} \cdot C_{w}}} \\
 &= |ER| \cdot z \cdot \sqrt{\frac{1 - E_{b}}{E_{b} \cdot C_{b}} + \frac{1 - E_{w}}{E_{w} \cdot C_{w}}} = |ER| \cdot z \cdot \sqrt{\frac{1}{E_{b} \cdot C_{b}} - \frac{1}{C_{b}} + \frac{1}{E_{w} \cdot C_{w}} - \frac{1}{C_{w}}} \\
 &= |ER| \cdot z \cdot \sqrt{\frac{1}{\frac{I_{b}}{C_b} \cdot C_{b}} - \frac{1}{C_{b}} + \frac{1}{\frac{I_{w}}{C_w} \cdot C_{w}} - \frac{1}{C_{w}}} = |ER| \cdot z \cdot \sqrt{\frac{1}{I_b} + \frac{1}{I_w} - \frac{1}{C_b} - \frac{1}{C_w}}
\end{align*}

The variance of the enforcement ratio is proportional to the sum of the reciprocal of the number of recorded incidents, which is the desired behavior.

\subsection{Spatial smoothing}
\label{SISec:smoothing}

Smoothing is performed independently for each state. 
It was determined smoothing over state lines may not be appropriate due to differing laws and policies around marijuana. 
Counties that were defined as metropolitan areas according to the UIC codes (equal to 1 or 2) were not considered in the computation of the smoothing.

The smoothing works as follows. 
First, an adjacency matrix $A$ of counties is computed for each state. Two counties are considered to be adjacent if they share boundaries. This matrix is referred to as \lq queen adjacency\rq. Note that only NIBRS-reporting counties are considered. The shortest-distance matrix $D$ is found from the adjacency matrix using the Floyd–Warshall algorithm, the weighted-distance matrix $W$ is then calculated by applying the following function $d$ to $D$: $W = d(D_{ij})$:
\[
    d(x)= 
\begin{cases}
    1, & \text{if } x=0\\
    \frac{1}{(x+1)^p}, & \text{if } m > x > 0 \\
    0,              & \text{if } x \geq m
\end{cases}
\]

where $p$ controls the smoothing drop-off and $m$ controls the maximum distance a county can be from another whilst influencing it. Both are user-specified hyperparameters affecting the degree of smoothing.
The smoothing matrix is then applied to each component of the enforcement rate independently. The resulting smoothed enforcement ratio $\widebar{ER}$ for a state $S$ is then given by
$$
\widebar{ER(S)} = \frac{\frac{WI_{b}(c \in S)}{WC_{b}(c \in S)}}{\frac{WI_{w}(c \in S)}{WC_{w}(c \in S)}}
$$
where $I_{r}(c \in S)$ is a vector of the number of incidents for each (non-metro) county $c$ in state $S$, with race $r$. $C_{r}(c \in S)$ is equivalently a vector of the baseline crime rate for county $c$ in state $S$ with race $r$. $\widebar{ER(E)}$ is thus a vector of smoothed enforcement rates for state $E$.

\clearpage

 \begin{figure}[!ht]
    \centering
    \includegraphics[width=\textwidth]{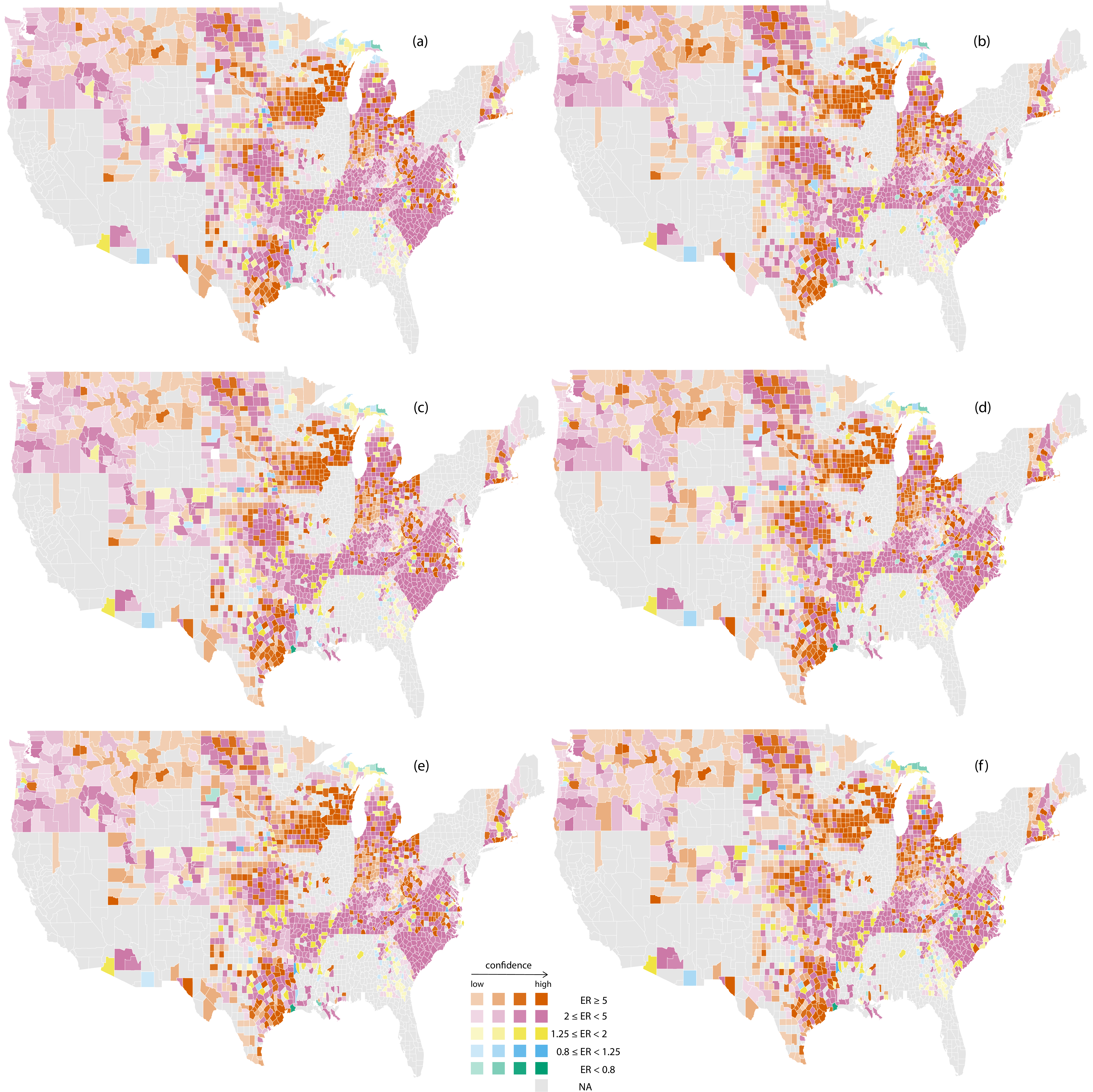}
    \caption{Six smoothed choropleths of enforcement ratio across counties in the United States, based on 2017--2019 NIBRS and NSDUH data. The colors indicate different levels of the ratio, while the opacity level is inversely proportional to the relative standard error. Null values, which correspond to counties in which the agencies did not report data for the period considered, are colored in grey. Smoothing, as described in section \ref{SISec:smoothing} has been performed with the following parameters:
    \\ \textbf{(a)} $m = 2$, $p = 1$
    \\ \textbf{(b)} $m = 3$, $p = 1$
    \\ \textbf{(c)} $m = 2$, $p = 1.5$
    \\ \textbf{(d)} $m = 3$, $p = 1.5$
    \\ \textbf{(e)} $m = 2$, $p = 2$
    \\ \textbf{(f)} $m = 3$, $p = 2$
    }
    \label{SIfig:smooth_maps}
\end{figure}

\clearpage
\section{Analysis UCR SRS data}

The summary reporting system (SRS) is an FBI program that collects summary statistics of crime incidents from law enforcement agencies across the US. NIBRS was developed as a more detailed incident-level alternative in the 1980s, more faithfully representing the diversity and complexity of crime in the USA. The SRS contains aggregated agency-level arrest data. Through the SRS, agencies report the number of offense arrests by demographic and time period.  In order to contrast SRS to the NIBRS we calculated the enforcement ratio for marijuana possession arrests using the data from the SRS. Additionally, we use the SRS data to calculate the enforcement ratio for driving under the influence and drunkenness, as these offenses aren't captured in NIBRS. The baseline crime rate in this analysis was based \texttt{Use}\textsubscript{Dmg+Pov} model.
\clearpage

\subsection{DUI Arrests}
\label{SISec:dui}
\phantom{This is dummy text.}

\begin{figure}[H]

\begin{minipage}{.5\linewidth}
\centering
\subfloat{\label{dui:a}\includegraphics[scale=.5]{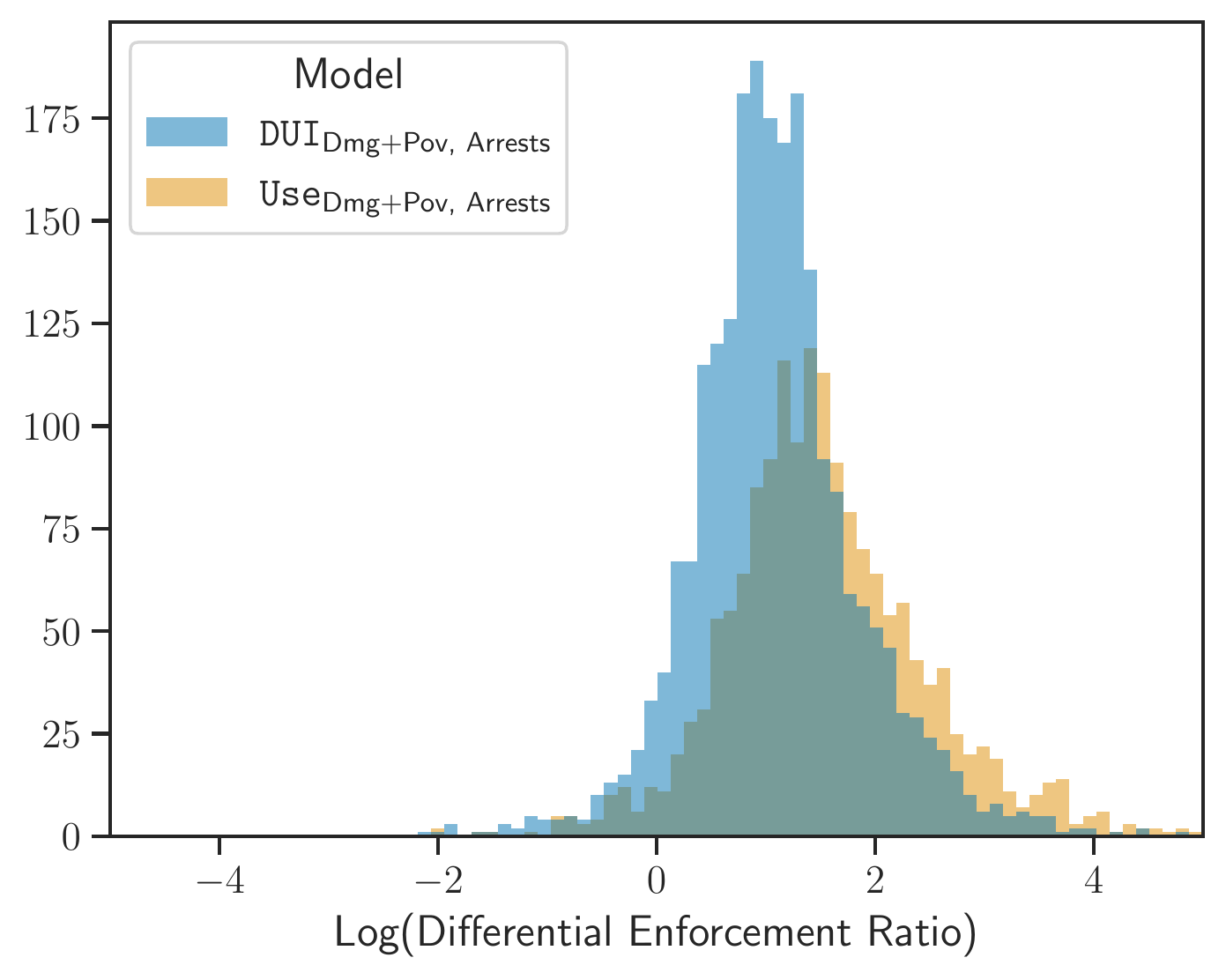}}
\end{minipage}%
\begin{minipage}{.5\linewidth}
\centering
\subfloat{\label{dui:b}\includegraphics[scale=.5]{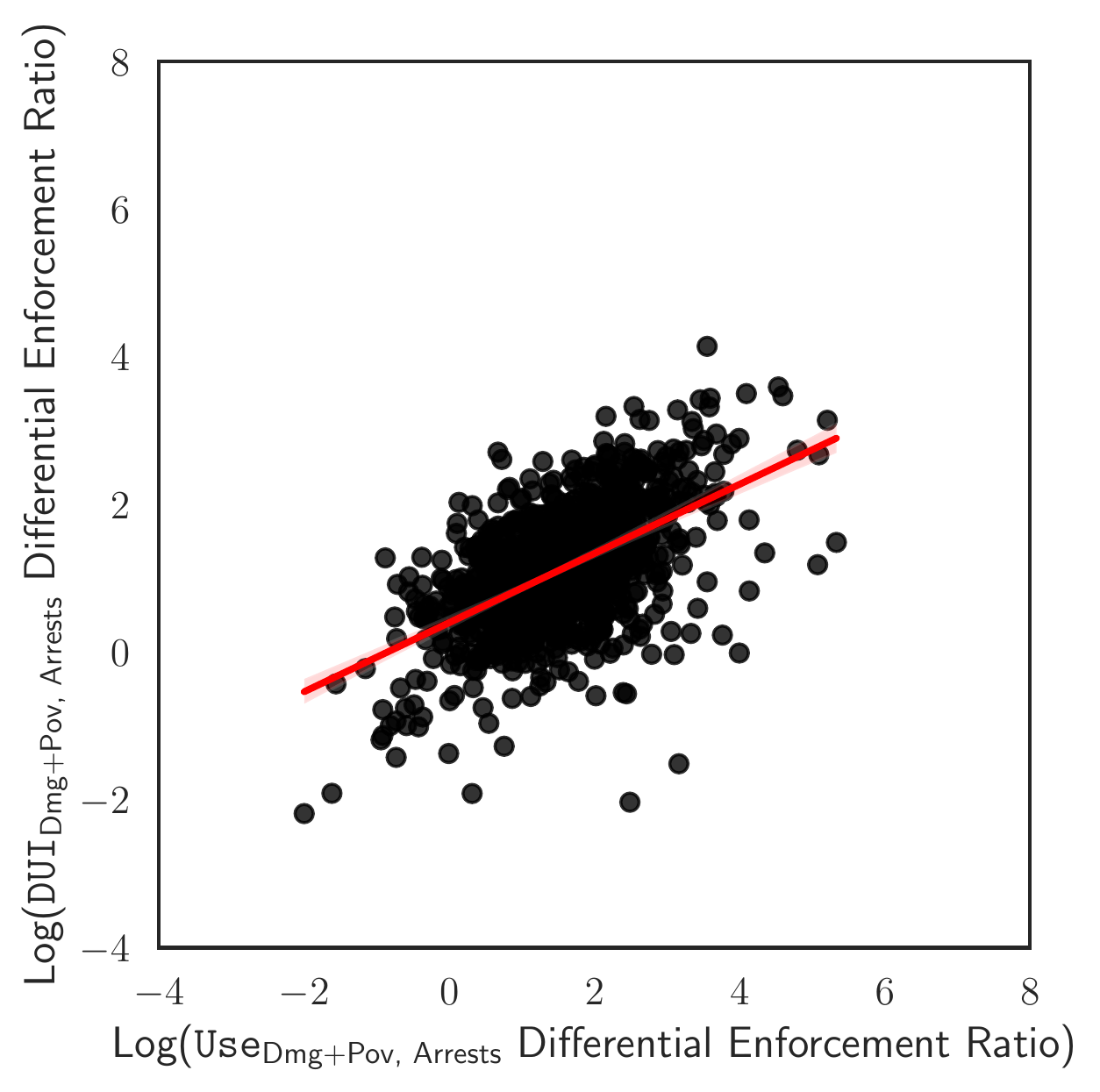}}
\end{minipage}\par\medskip
\centering
\subfloat{\label{dui:c}\includegraphics[scale=.5]{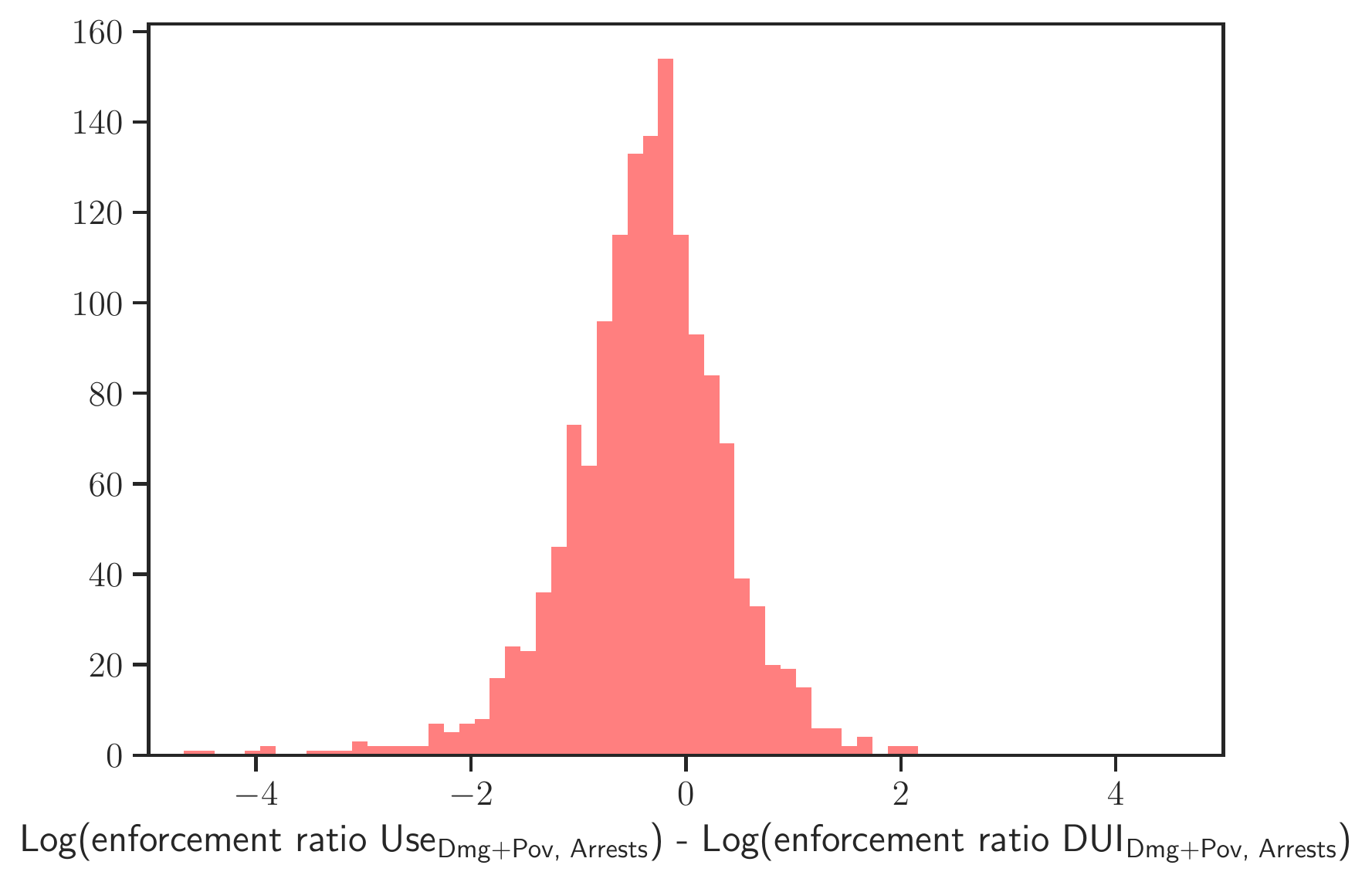}}

\caption{\\\textbf{(a)} Visualization of the distribution of enforcement ratio for arrests due to marijuana use (\texttt{Use}\textsubscript{Dmg+Pov, Arrests}) and driving under the influence (\texttt{DUI}\textsubscript{Dmg+Pov, Arrests}). \\ \textbf{(b)} Regression plot of \texttt{Use}\textsubscript{Dmg+Pov, Arrests} vs. \texttt{DUI}\textsubscript{Dmg+Pov, Arrests}. The regression produced the equation: \texttt{Use}\textsubscript{Dmg+Pov, Arrests} = (0.6)\texttt{DUI}\textsubscript{Dmg+Pov, Arrests} + 0.26. Significance with Wald tests found a significant relationship (p < 0.001). \\ \textbf{(c)} Visualization of the difference in distribution of enforcement ratio between for arrests driving under the influence (\texttt{DUI}\textsubscript{Dmg+Pov, Arrests} and for arrests due to marijuana use (\texttt{Use}\textsubscript{Dmg+Pov, Arrests}).}
\label{SIFig:dui}
\end{figure}

\begin{figure}[H]
    \centering
    \includegraphics[width=\textwidth]{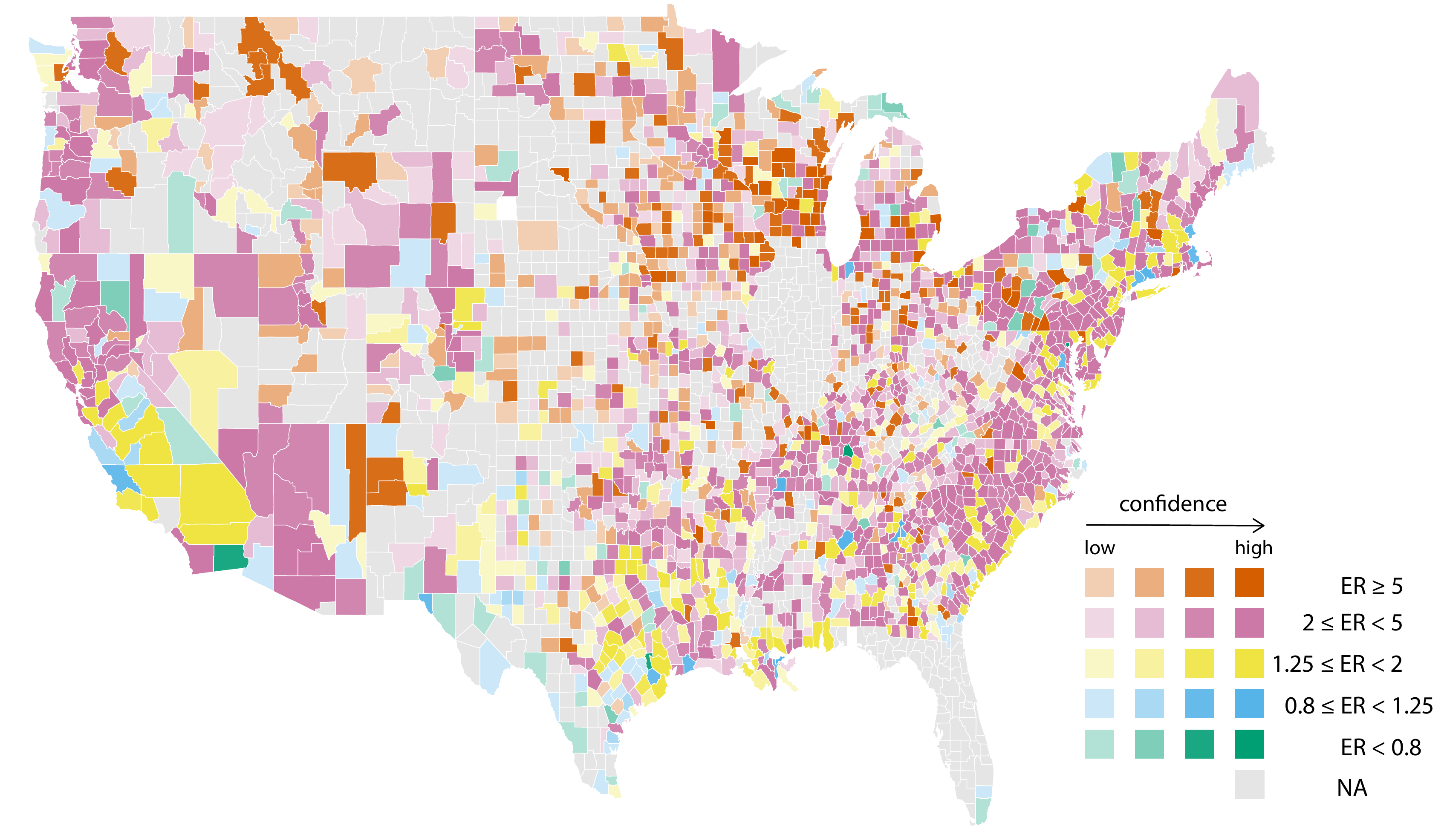}
    \caption{Maps of enforcement ratio for DUI across counties in the United States, using UCR and NSDUH data. The colors indicate different levels of the ratio, while the opacity level is inversely proportional to the relative standard error. Null values, which correspond to counties in which the agencies did not report data for the period considered, are colored in grey.}
    \label{SIfig:dui_map}
\end{figure}

\begin{table}[t]
\caption{County-level enforcement ratios}
  \begin{tabular}{ccccccc}
    \toprule
    & & $<0.5$ & $<0.8$ & $>1.25$ & $>2$ & $>5$ \\
    \midrule 
     \multirow{2}{*}\texttt{DUI}\textsubscript{Dmg+Pov} & Value & 31 & 73 & 2037 & 1639 & 454 \\
        & 95\% conf. & 1 & 389 & 1565 & 963 & 113\\ 
     \midrule
 \multicolumn{7}{l} {\parbox[t]{0.5\textwidth}{\footnotesize{The number of counties above or below a given enforcement ratio threshold, from a total of 2257 reporting counties. The 95\% conf. rows indicate the number of counties for which the upper bound of the 95\% confidence interval is below the threshold (for $ER<0.5$ and $ER<0.8$) or the lower bound of the 95\% confidence interval is above the threshold (for $ER>1.25$, $ER>2$ and $ER>5$). Enforcement ratios correspond to those displayed in Figure~\ref{SIfig:dui_map}.}}}
  \end{tabular}
  \label{si_tab:dui_count}
 \end{table} 
 
\clearpage

\subsection{Drunkenness Arrests}
\label{SISec:drunkenness}
\phantom{This is dummy text.}

\begin{figure}[!htb]

\begin{minipage}{.5\linewidth}
\centering
\subfloat{\label{drunkeness:a}\includegraphics[scale=.5]{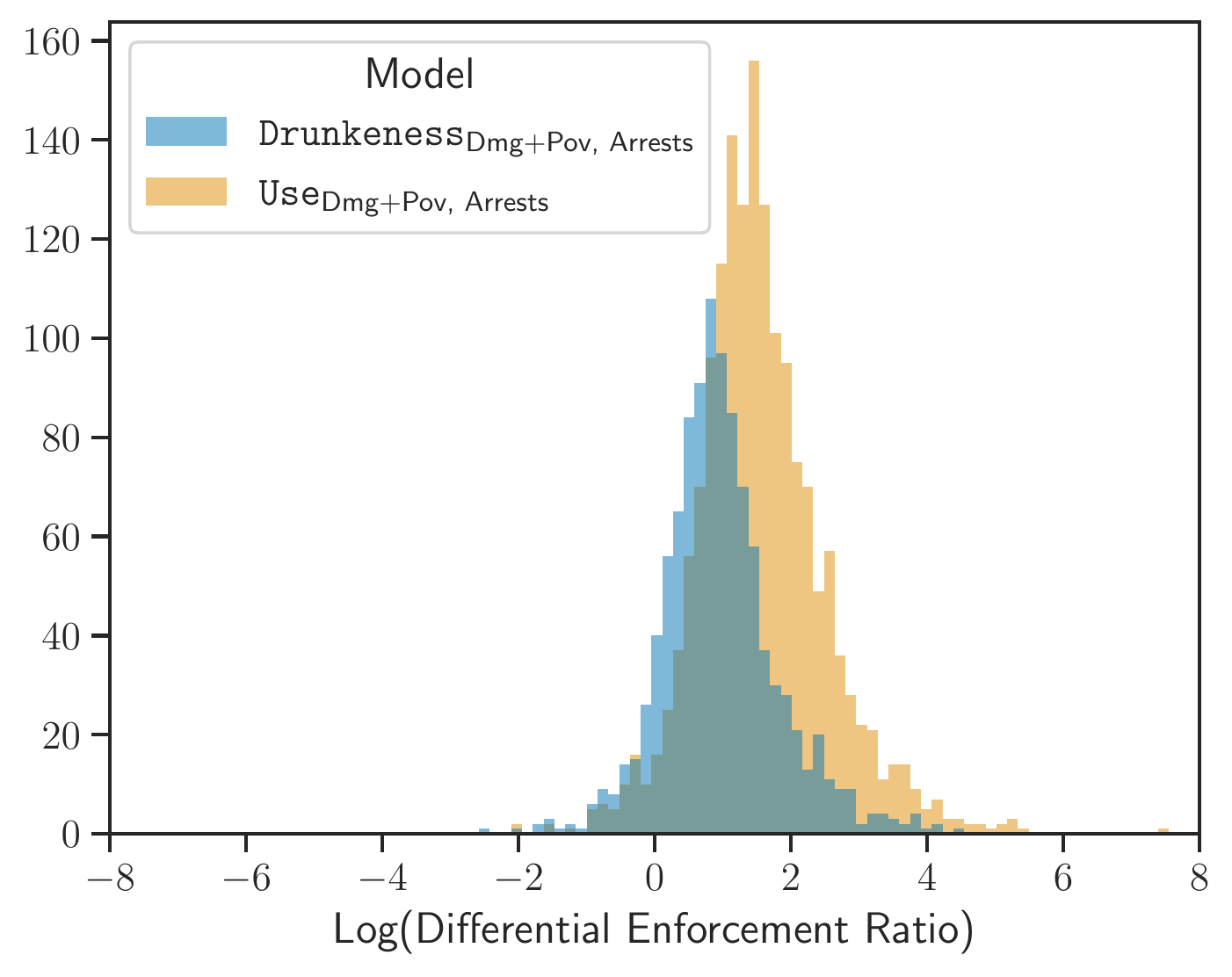}}
\end{minipage}%
\begin{minipage}{.5\linewidth}
\centering
\subfloat{\label{drunkeness:b}\includegraphics[scale=.5]{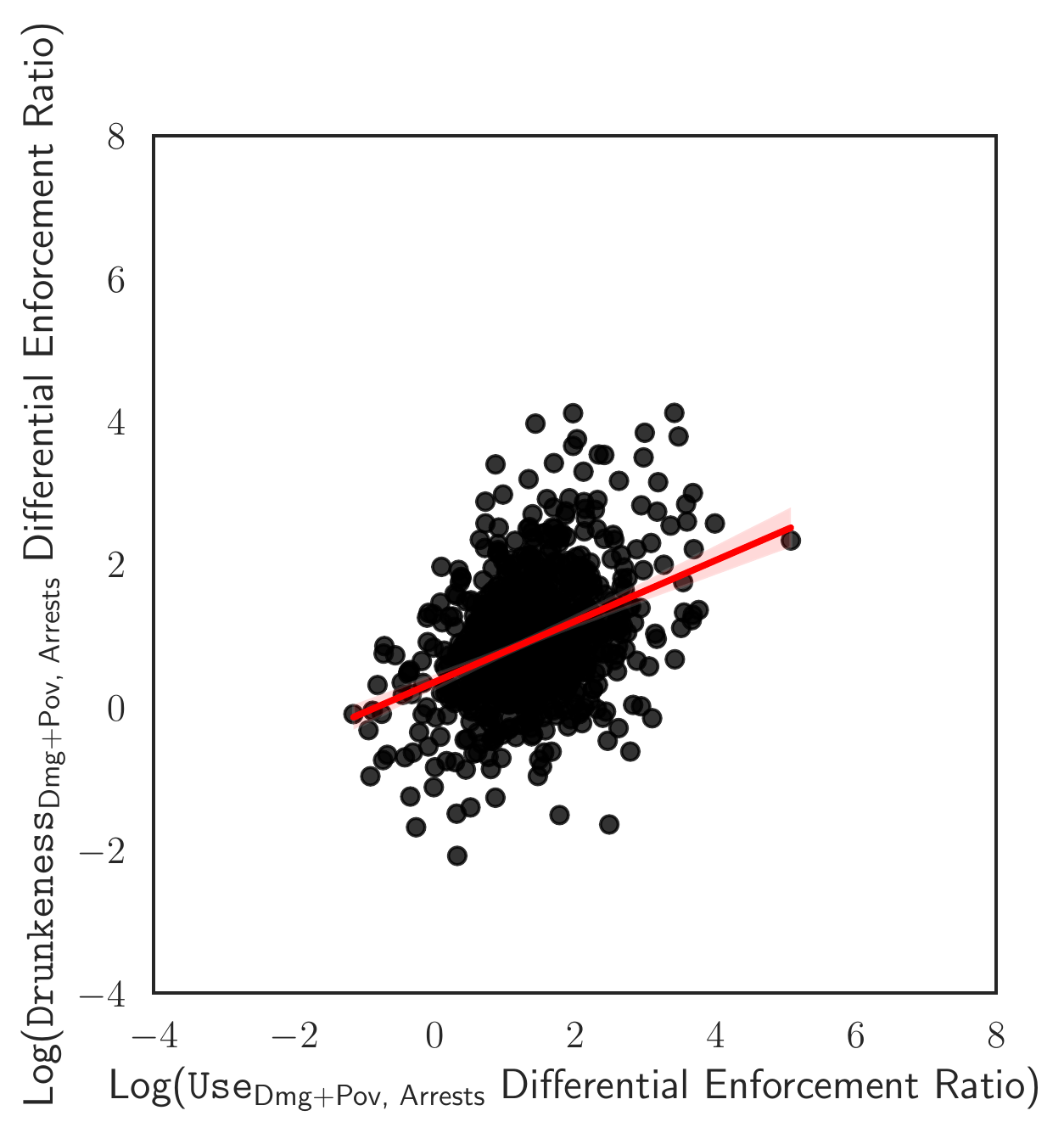}}
\end{minipage}\par\medskip
\centering
\subfloat{\label{drunkeness:c}\includegraphics[scale=.5]{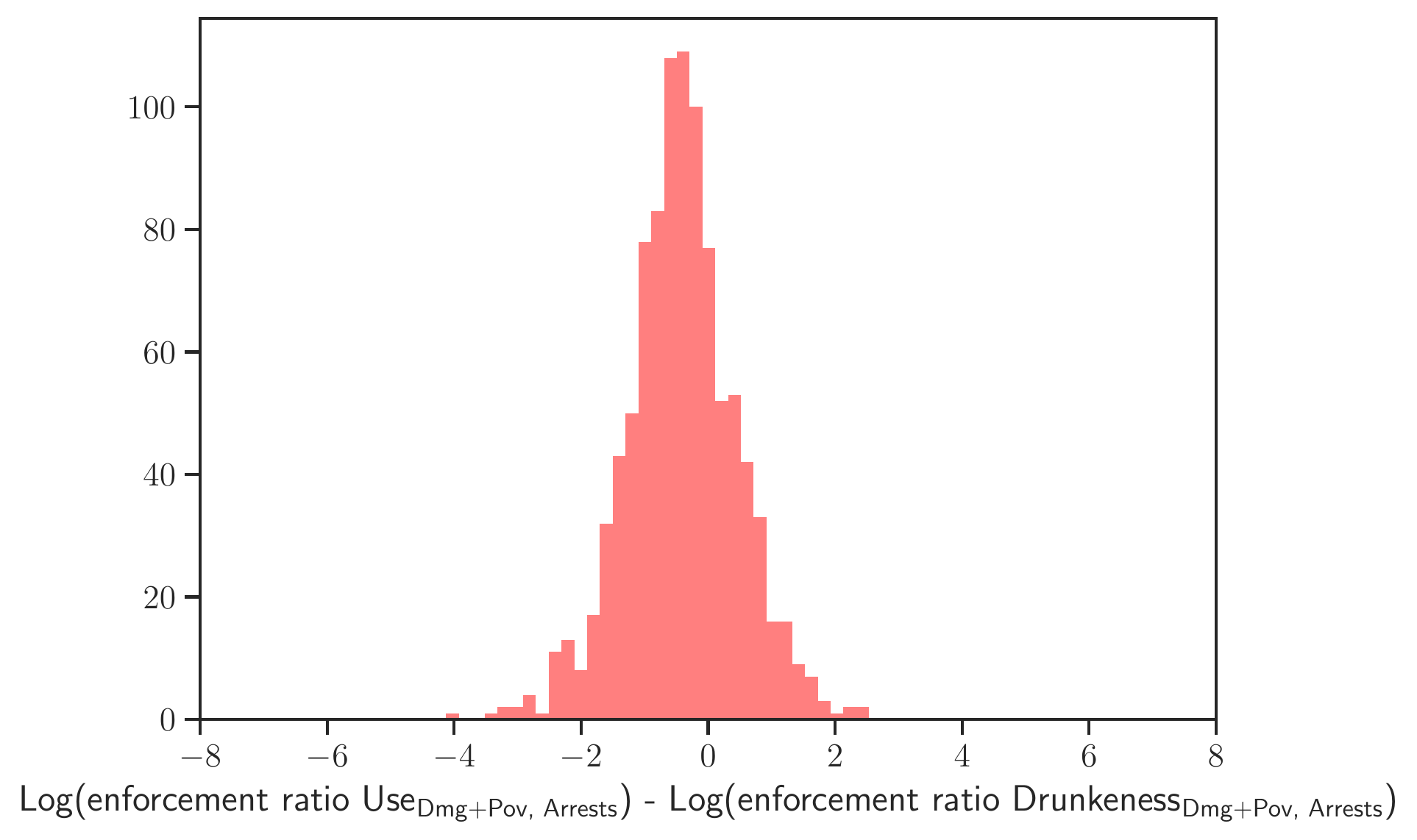}}

\caption{\\\textbf{(a)} Visualization of the distribution of enforcement ratio for arrests due to marijuana use (\texttt{Use}\textsubscript{Dmg+Pov, Arrests}) and drunkenness (\texttt{Drunkenness}\textsubscript{Dmg+Pov, Arrests}). \\ \textbf{(b)} Regression plot of \texttt{Use}\textsubscript{Dmg+Pov, Arrests} vs. \texttt{Drunkenness}\textsubscript{Dmg+Pov, Arrests}. The regression produced the equation: \texttt{Use}\textsubscript{Dmg+Pov, Arrests} = (0.43)\texttt{Drunkenness}\textsubscript{Dmg+Pov, Arrests} + 0.35. Significance with Wald tests found a significant relationship (p < 0.001). \\ \textbf{(c)} Visualization of the difference in distribution of enforcement ratio between arrests for drunkenness (\texttt{Drunkenness}\textsubscript{Dmg+Pov, Arrests} and for arrests due to marijuana use (\texttt{Use}\textsubscript{Dmg+Pov, Arrests}).}
\label{SIFig:drunkeness}
\end{figure}

\begin{figure}[!htbp]
    \centering
    \includegraphics[width=\textwidth]{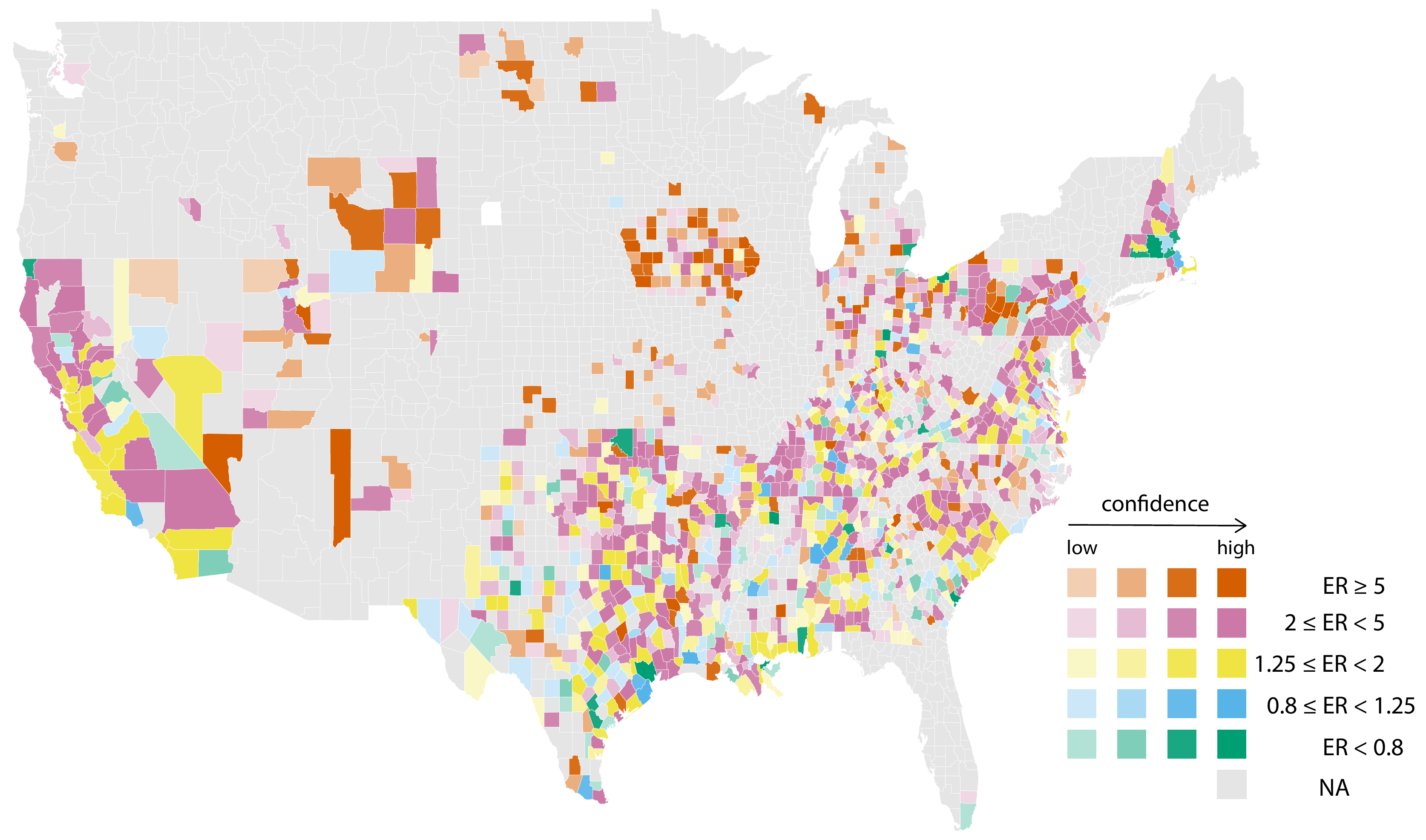}
    \caption{
 Maps of enforcement ratio for drunkenness across counties in the United States, using UCR and NSDUH data. The colors indicate different levels of the ratio, while the opacity level is inversely proportional to the relative standard error. Null values, which correspond to counties in which the agencies did not report data for the period considered, are colored in grey.}
    \label{SIfig:drunkenness_map}
\end{figure}

\begin{table}[t]
\caption{County-level enforcement ratios}
  \begin{tabular}{ccccccc}
    \toprule
    & & $<0.5$ & $<0.8$ & $>1.25$ & $>2$ & $>5$ \\
    \midrule 
     \multirow{2}{*}\texttt{Drunkenness}\textsubscript{Dmg+Pov} & Value & 36 & 89 & 1132 & 832 & 240 \\
        & 95\% conf. & 5 & 388 & 703 & 367 & 63 \\ 
     \midrule
 \multicolumn{7}{l} {\parbox[t]{0.5\textwidth}{\footnotesize{The number of counties above or below a given enforcement ratio threshold, from a total of 1351 reporting counties. The 95\% conf. rows indicate the number of counties for which the upper bound of the 95\% confidence interval is below the threshold (for $ER<0.5$ and $ER<0.8$) or the lower bound of the 95\% confidence interval is above the threshold (for $ER>1.25$, $ER>2$ and $ER>5$). Enforcement ratios correspond to those displayed in Figure~\ref{SIfig:drunkenness_map}.}}}
  \end{tabular}
  \label{si_tab:drunkenness_count}
 \end{table} 
 
\clearpage

\subsection{Marijuana Possession}

\phantom{This is dummy text.}




\begin{figure}[!htb]

\begin{minipage}{.5\linewidth}
\centering
\subfloat{\label{srscannabis:a}\includegraphics[scale=.5]{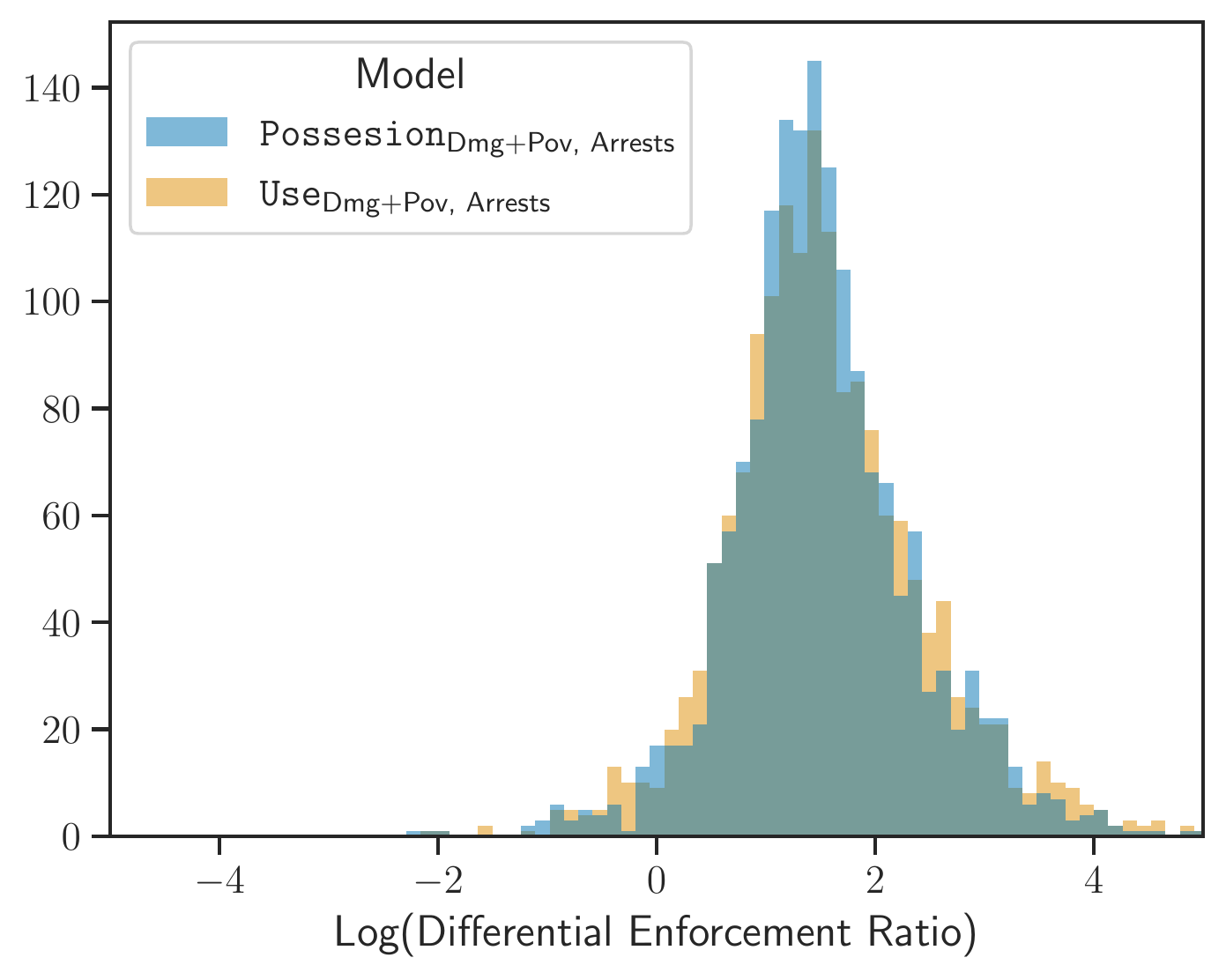}}
\end{minipage}%
\begin{minipage}{.5\linewidth}
\centering
\subfloat{\label{srscannabis:b}\includegraphics[scale=.5]{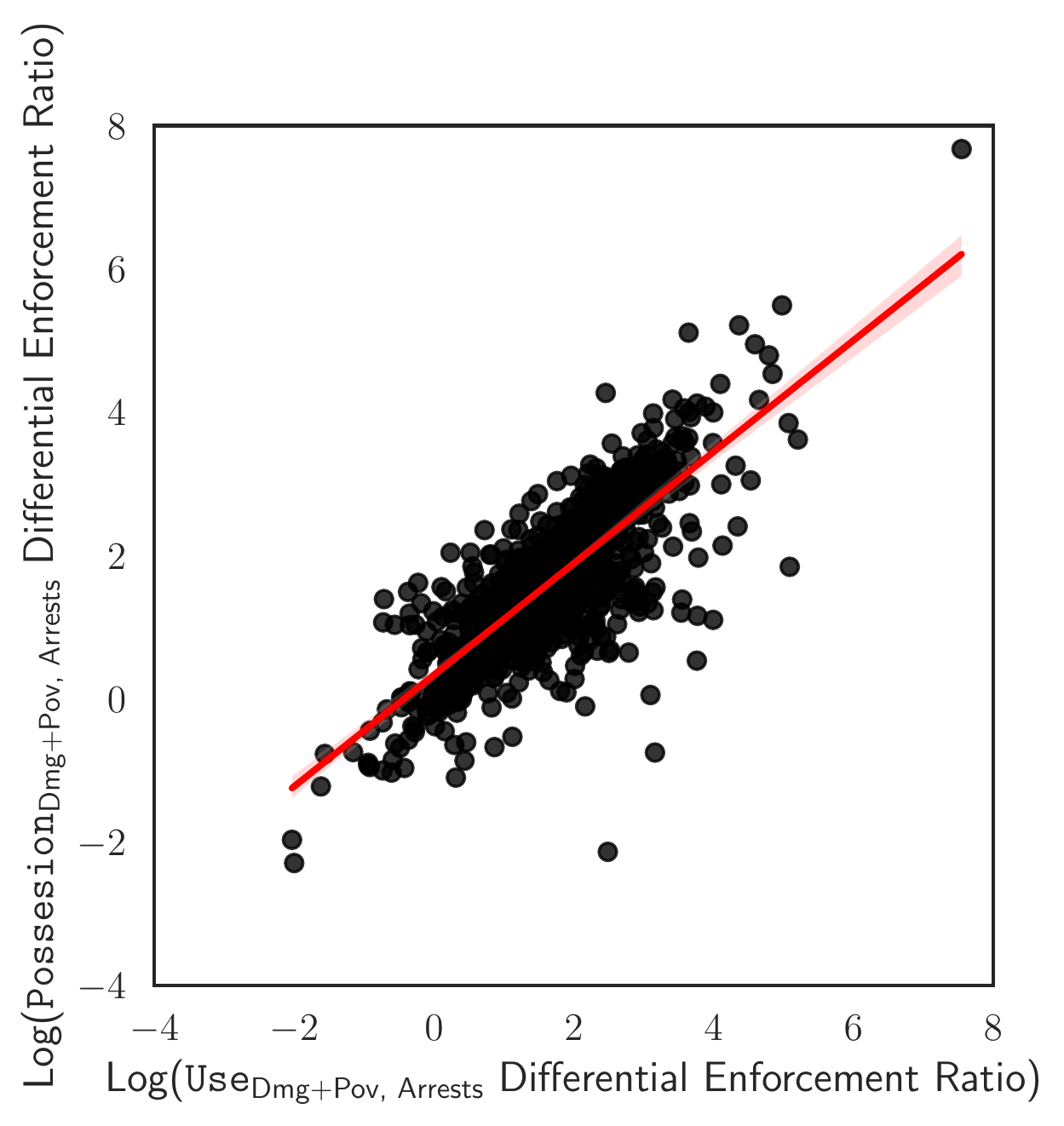}}
\end{minipage}\par\medskip
\centering
\subfloat{\label{srscannabis:c}\includegraphics[scale=.5]{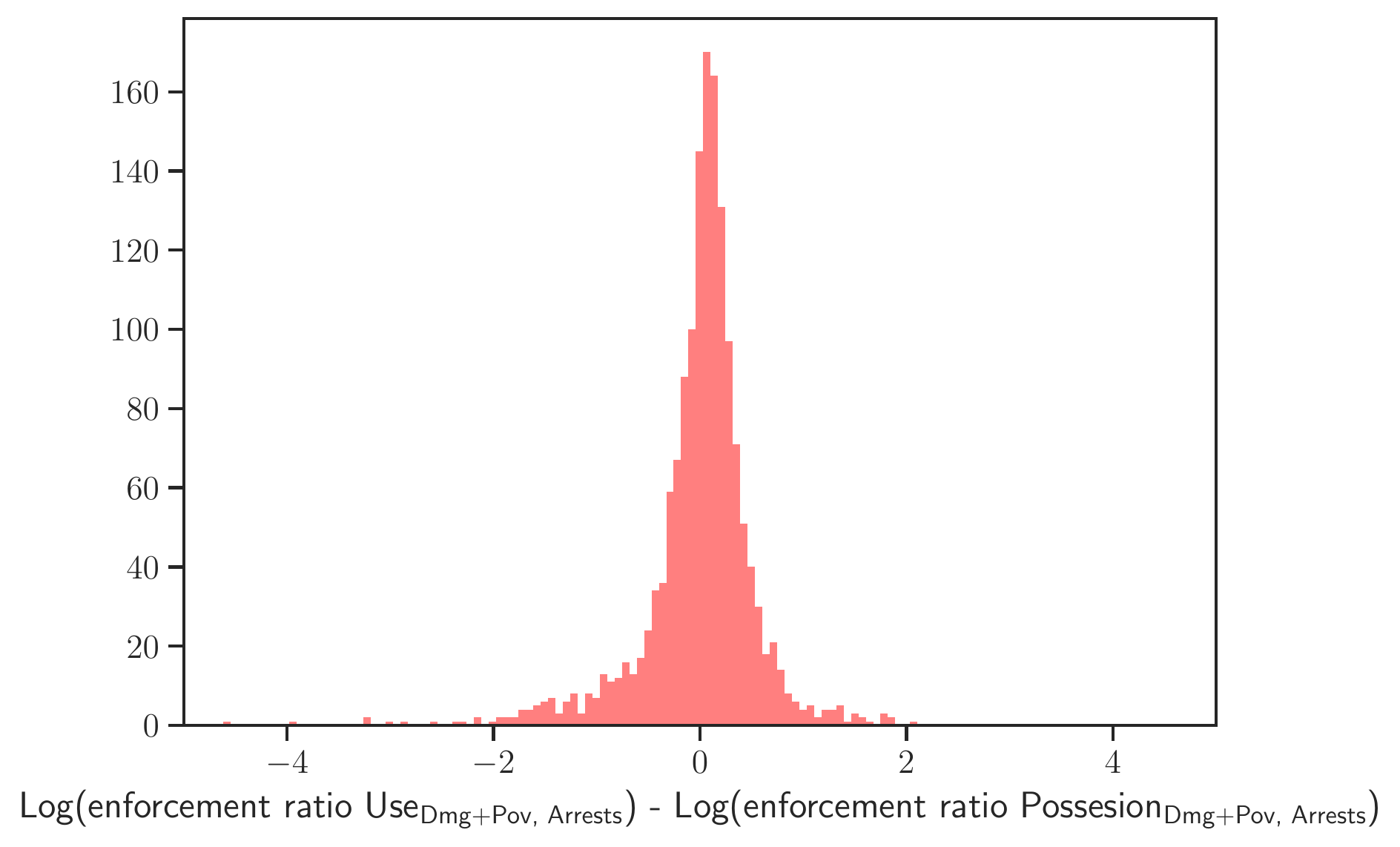}}

\caption{\\\textbf{(a)} Histogram of enforcement ratio for marijuana use (\texttt{Use}\textsubscript{Dmg+Pov, Arrests}) and (\texttt{Use}\textsubscript{Dmg+Pov, SRS}). \\ \textbf{(b)} The regression produced the equation: \texttt{Use}\textsubscript{Dmg+Pov, Arrests} = (0.78)\texttt{Use}\textsubscript{Dmg+Pov, SRS} + 0.33. Significance with Wald tests found a significant relationship (p < 0.001). \\ \textbf{(c)} Histogram of the difference in distribution of enforcement ratio between all Marijuana incidents (\texttt{Drunkenness}\textsubscript{Dmg+Pov, Arrests} and Marijuana incidents where non-isolated Marijuana incidents are excluded (\texttt{Use}\textsubscript{Dmg+Pov}).}
\label{fig:srscannabis}
\end{figure}

\begin{figure}[!htbp]
    \centering
    \includegraphics[width=\textwidth]{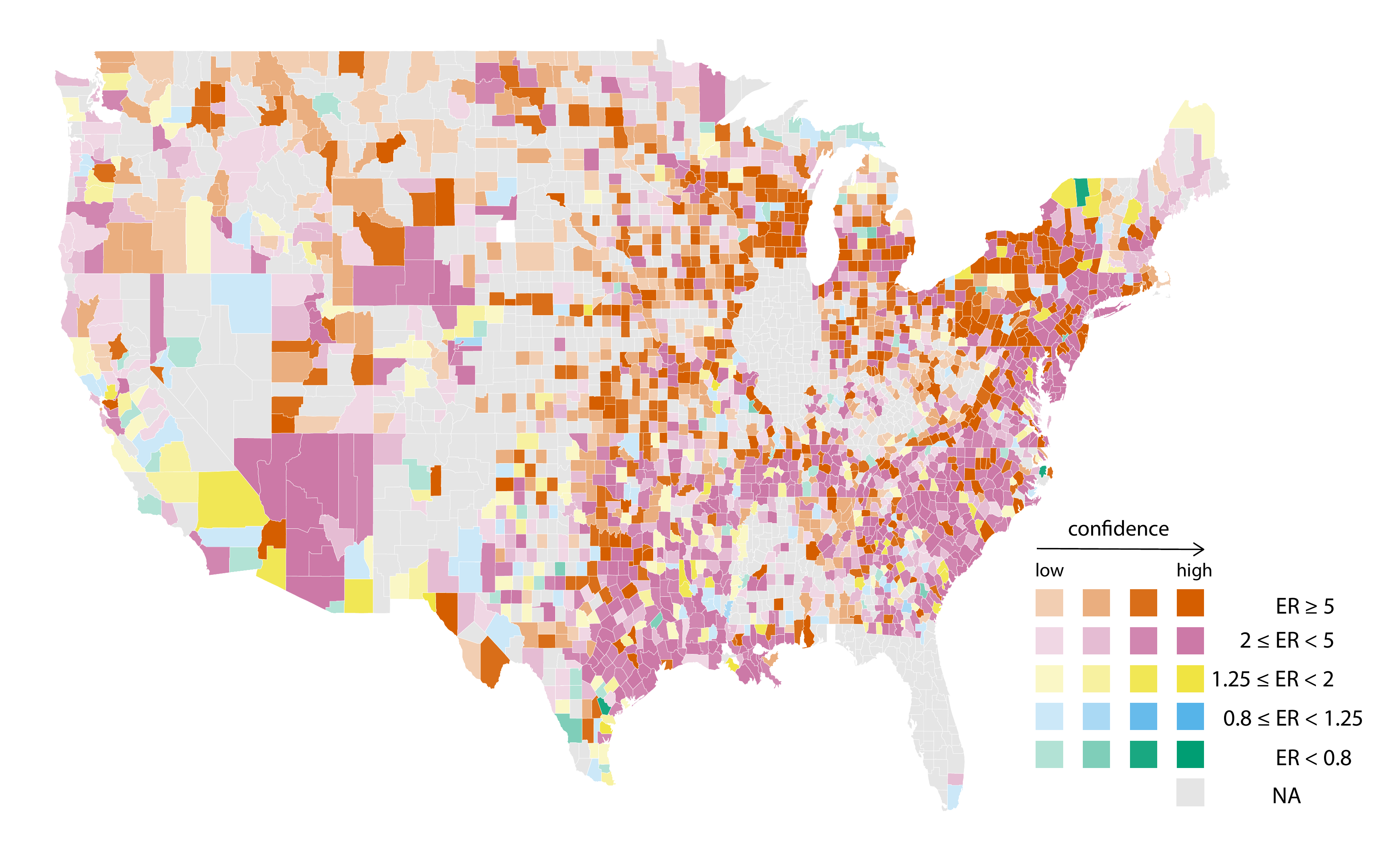}
    \caption{
 Maps of enforcement ratio for Marijuana possession across counties in the United States, using SRS and NSDUH data. The colors indicate different levels of the ratio, while the opacity level is inversely proportional to the relative standard error. Null values, which correspond to counties in which the agencies did not report data for the period considered, are colored in grey.}
    \label{SIfig:ucr_possesion_correlation}
\end{figure}

\clearpage

\section{NIBRS Ablation Study}

\subsection{All Incidents with Marijuana involvement}

The main analysis is performed solely on Marijuana incidents, where incidents involving other offences have been removed. This section contains the comparison between all incidents \textbf{involving} a marijuana offense against all incidents \textbf{only} containing a marijuana offense.

\begin{figure}[!htb]

\begin{minipage}{.5\linewidth}
\centering
\subfloat{\label{allincidents:a}\includegraphics[scale=.5]{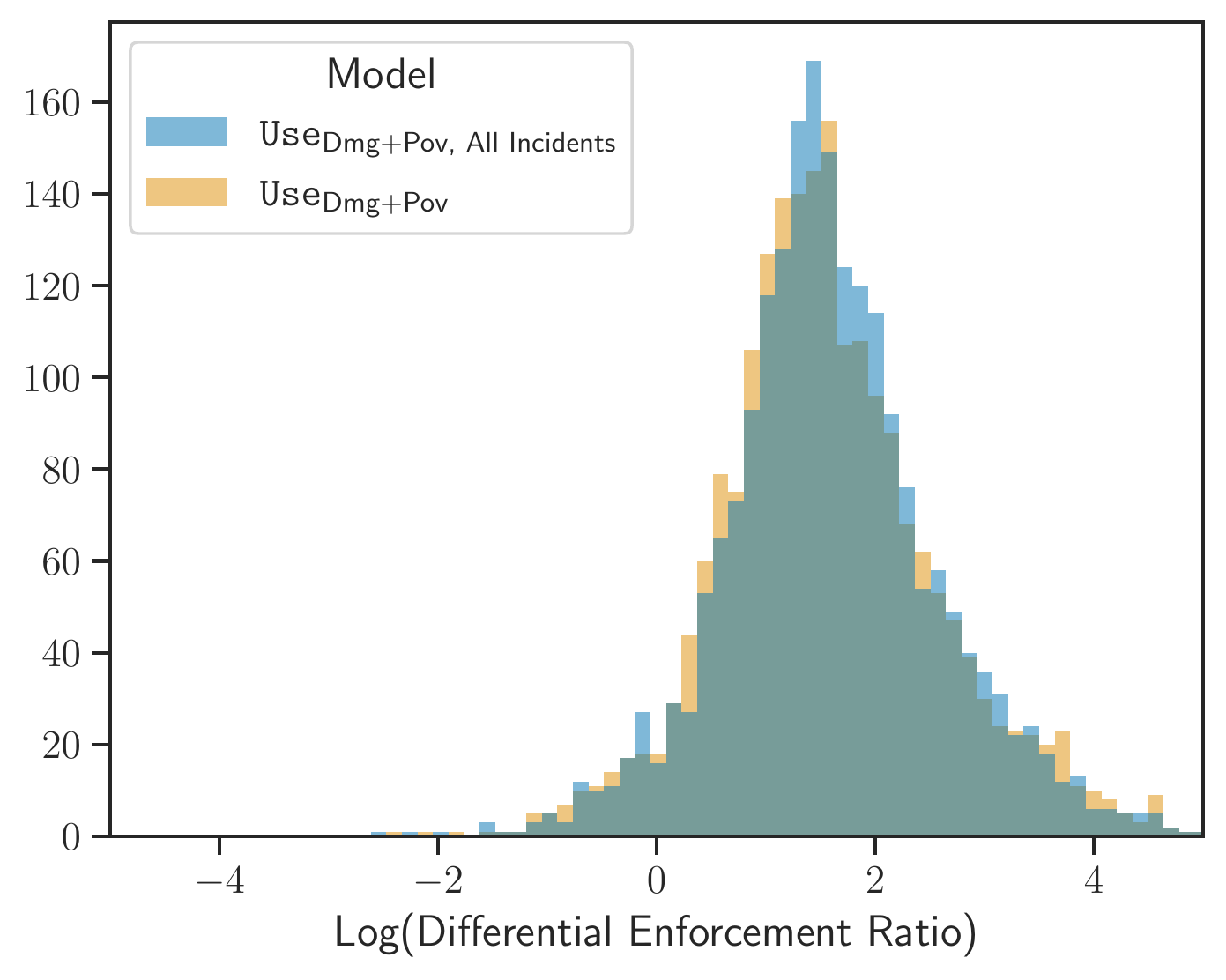}}
\end{minipage}%
\begin{minipage}{.5\linewidth}
\centering
\subfloat{\label{allincidents:b}\includegraphics[scale=.5]{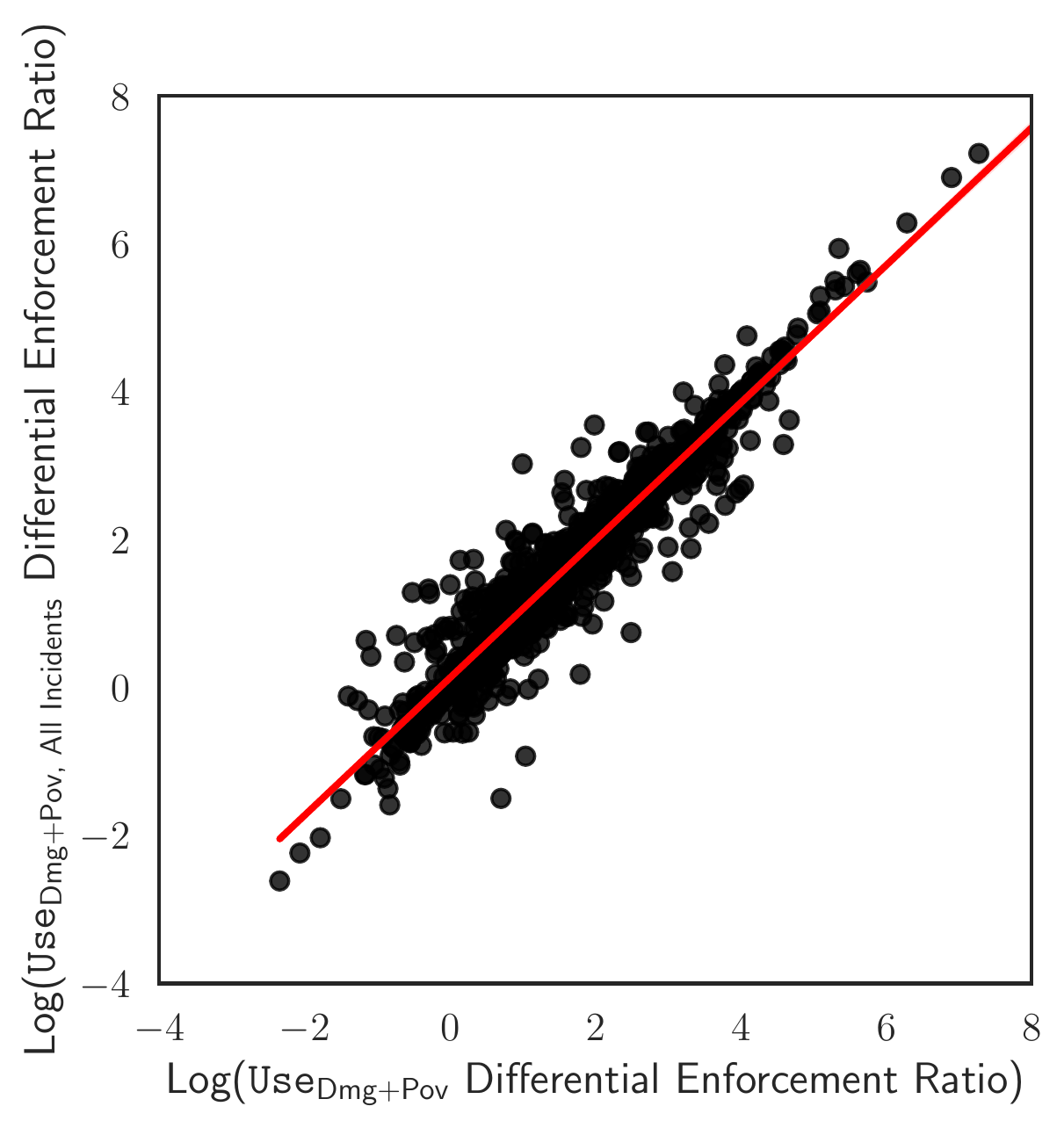}}
\end{minipage}\par\medskip
\centering
\subfloat{\label{allincidents:c}\includegraphics[scale=.5]{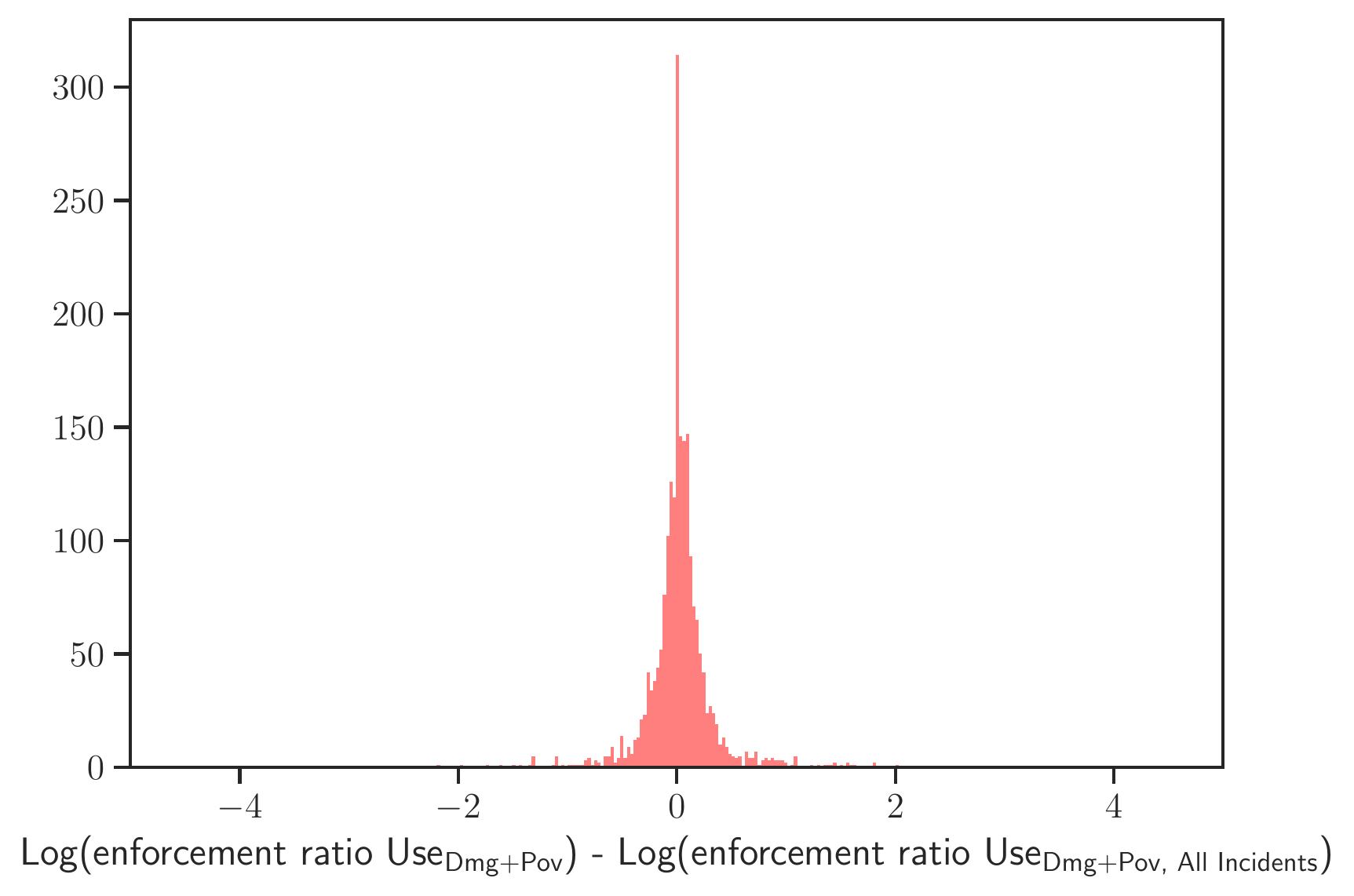}}

\caption{\\\textbf{(a)} Visualization of the distribution of enforcement ratio for marijuana use (\texttt{Use}\textsubscript{Dmg+Pov, Arrests}) and (\texttt{Use}\textsubscript{Dmg+Pov, All Incidents}). \\ \textbf{(b)} Regression plot of \texttt{Use}\textsubscript{Dmg+Pov} vs. \texttt{Use}\textsubscript{Dmg+Pov, All Incidents}. \\ \textbf{(c)} Visualization of the difference in the distribution of the enforcement ratio between all marijuana incidents (\texttt{Use}\textsubscript{Dmg+Pov, Arrests} and marijuana incidents where non-isolated marijuana incidents are excluded (\texttt{Use}\textsubscript{Dmg+Pov}).}
\label{fig:allincidents}
\end{figure}

\begin{figure}[!htbp]
    \centering
    \includegraphics[width=\textwidth]{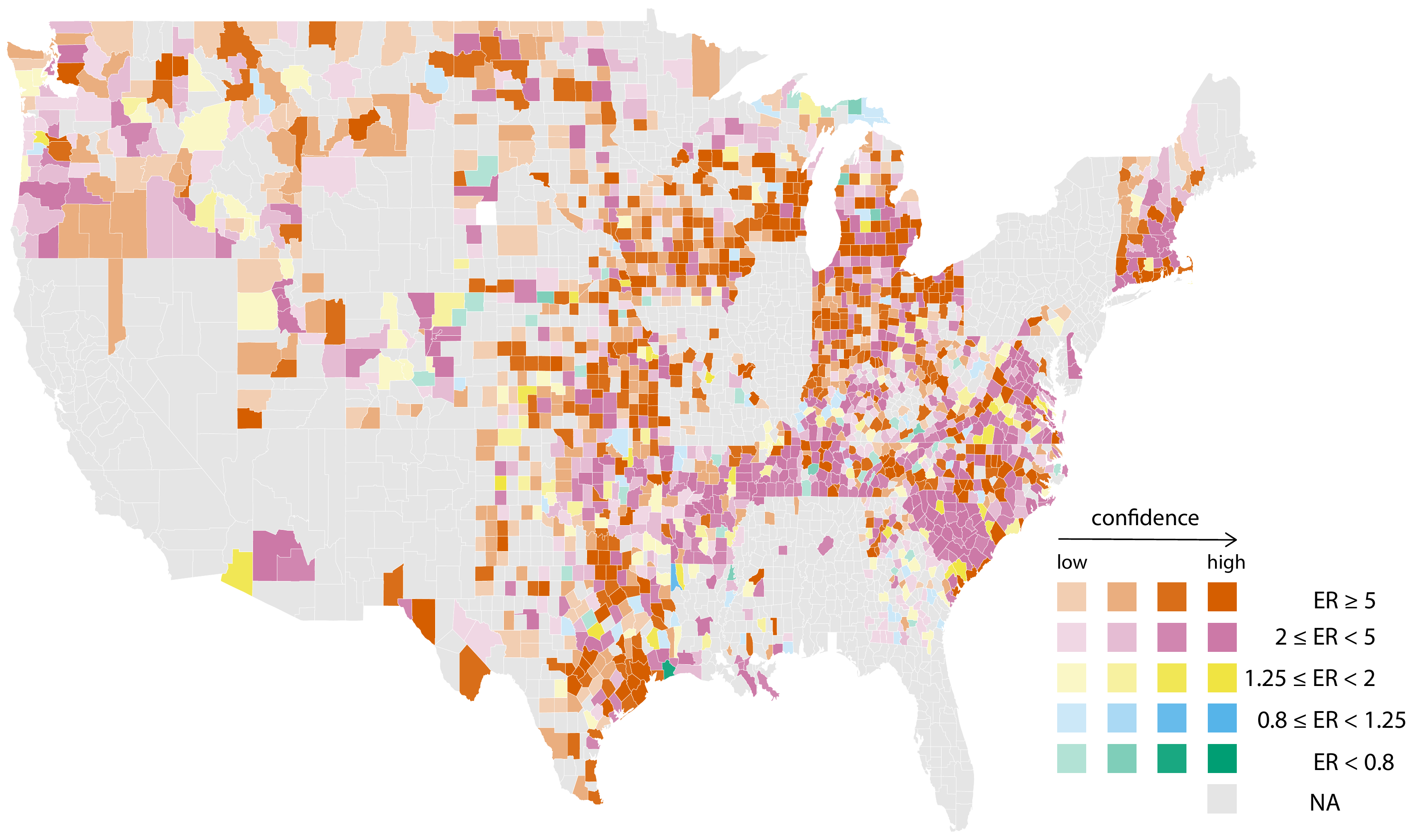}
    \caption{
 Maps of enforcement ratio for all incidents involving Marijuana across counties in the United States, using UCR and NSDUH data. The colors indicate different levels of the ratio, while the opacity level is inversely proportional to the relative standard error. Null values, which correspond to counties in which the agencies did not report data for the period considered, are colored in grey.}
    \label{SIfig:all_incidents_maps}
\end{figure}



\clearpage

\subsection{Including Hispanic Individuals}

Currently, Hispanic individuals are removed from the NIBRS dataset before the enforcement ratio is calculated. In this ablation, we remove this filtering process and observe the differences.




\begin{figure}[!htb]

\begin{minipage}{.5\linewidth}
\centering
\subfloat{\label{Hispanic:a}\includegraphics[scale=.5]{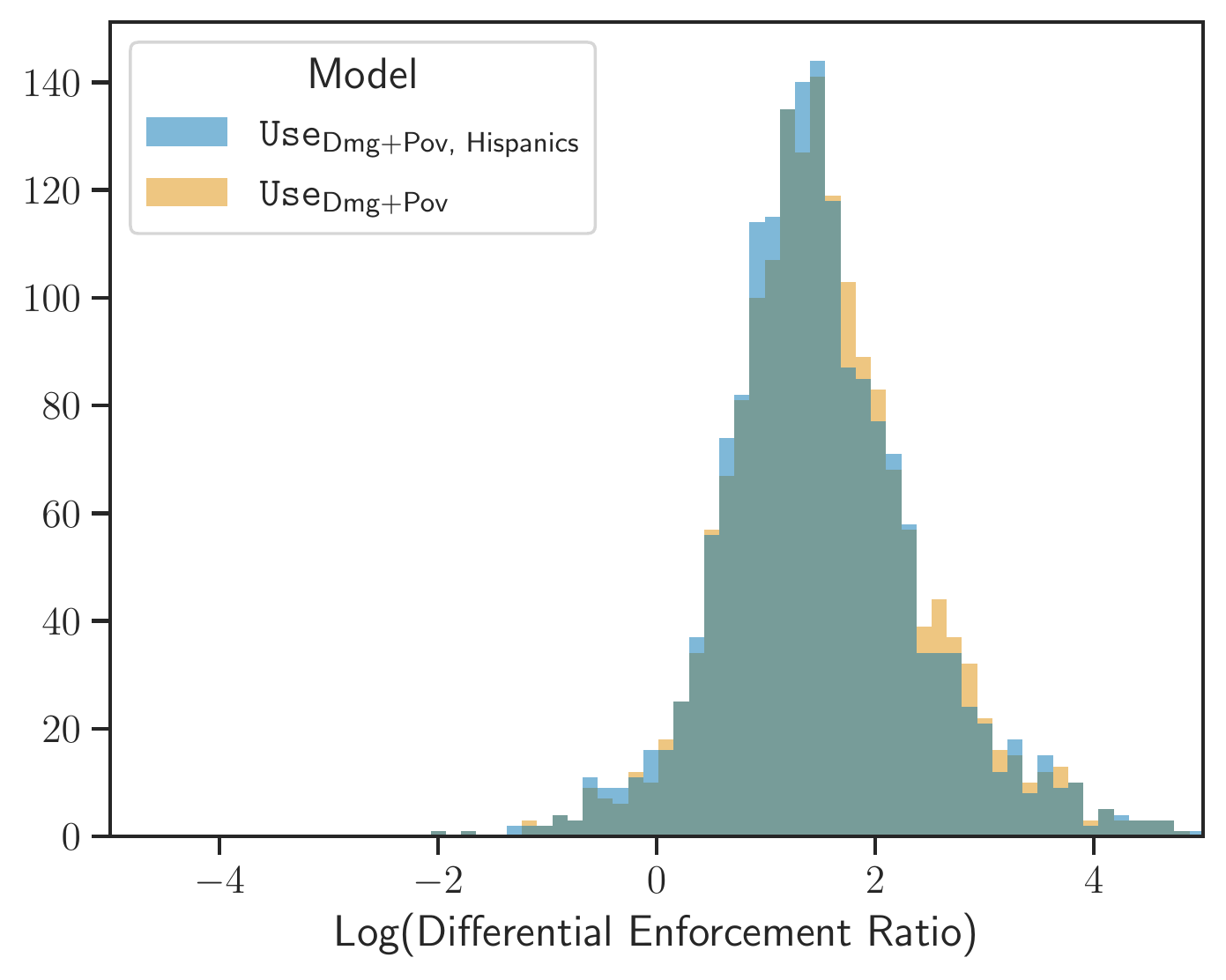}}
\end{minipage}%
\begin{minipage}{.5\linewidth}
\centering
\subfloat{\label{Hispanic:b}\includegraphics[scale=.5]{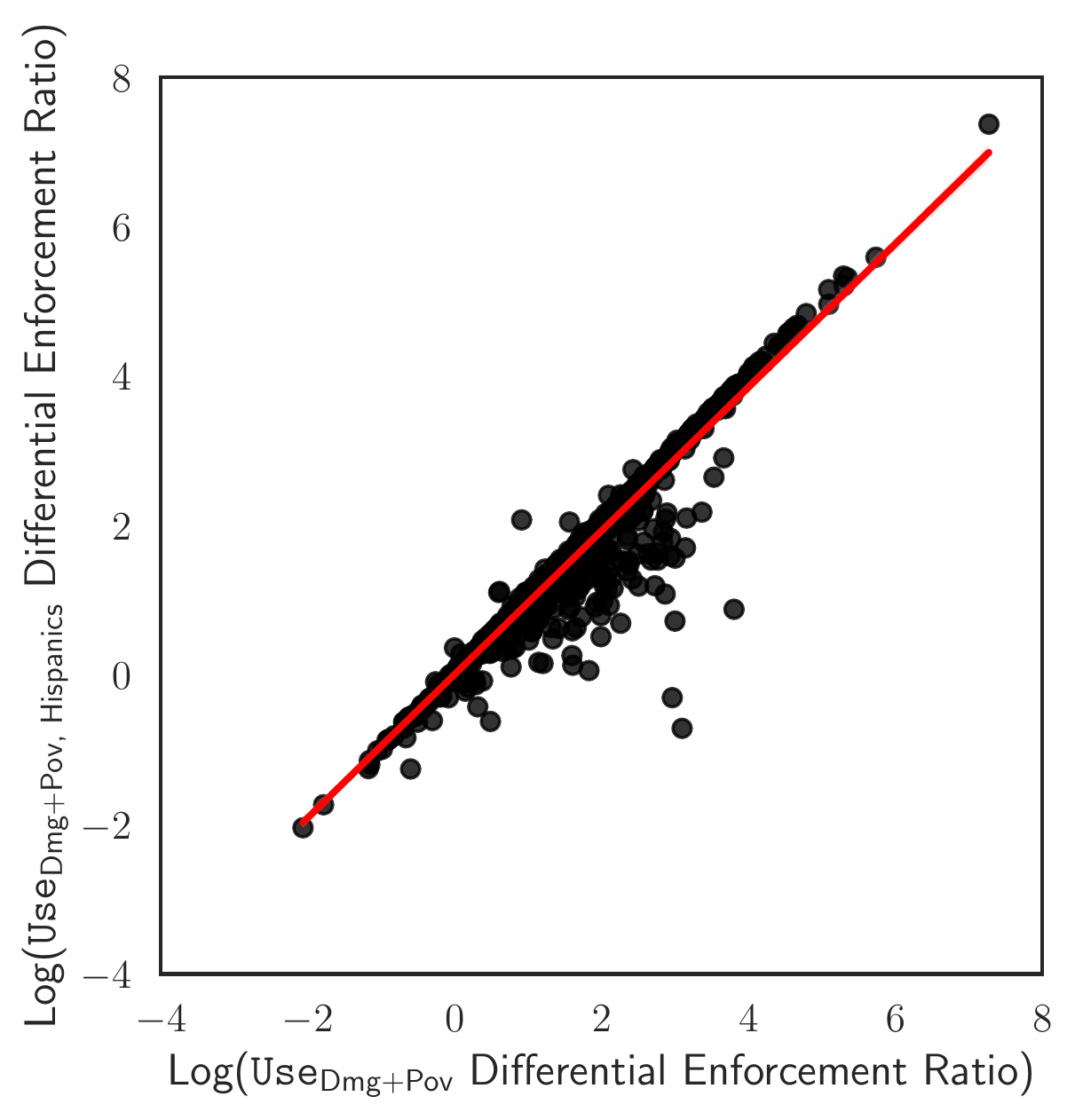}}
\end{minipage}\par\medskip
\centering
\subfloat{\label{Hispanic:c}\includegraphics[scale=.5]{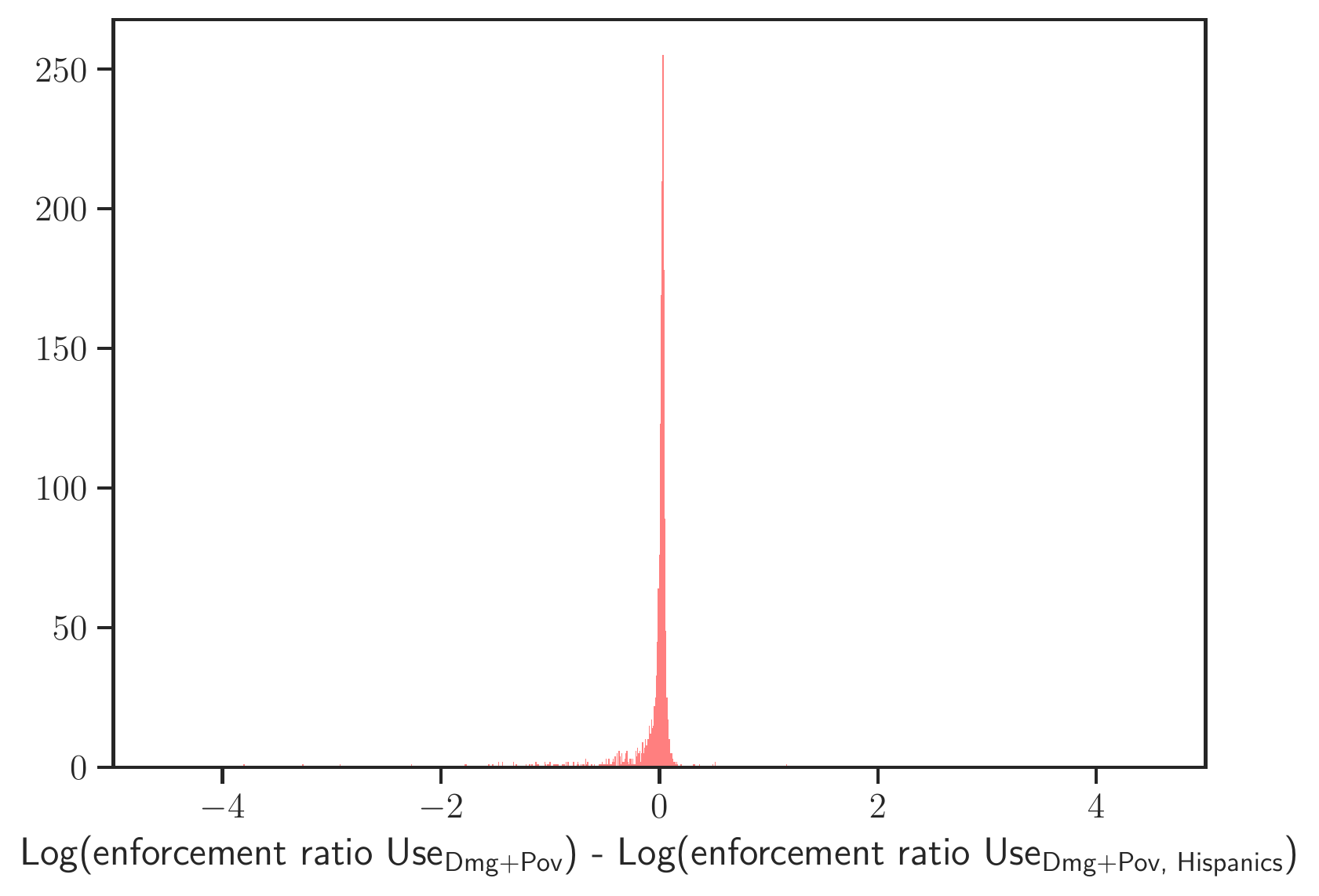}}

\caption{\\\textbf{(a)} Visualization of the distribution of enforcement ratio for marijuana use (\texttt{Use}\textsubscript{Dmg+Pov}) and (\texttt{Use}\textsubscript{Dmg+Pov, Hispanic}). \\ \textbf{(b)} Regression plot of \texttt{Use}\textsubscript{Dmg+Pov} vs. \texttt{Use}\textsubscript{Dmg+Pov, Hispanic}. \\ \textbf{(c)} Visualization of the difference in the distribution of the enforcement ratio between all marijuana incidents (\texttt{Use}\textsubscript{Dmg+Pov} and marijuana incidents in which Hispanic individuals are included (\texttt{Use}\textsubscript{Dmg+Pov, Hispanic}).}
\label{fig:Hispanic}
\end{figure}

\begin{figure}[!htbp]
    \centering
    \includegraphics[width=\textwidth]{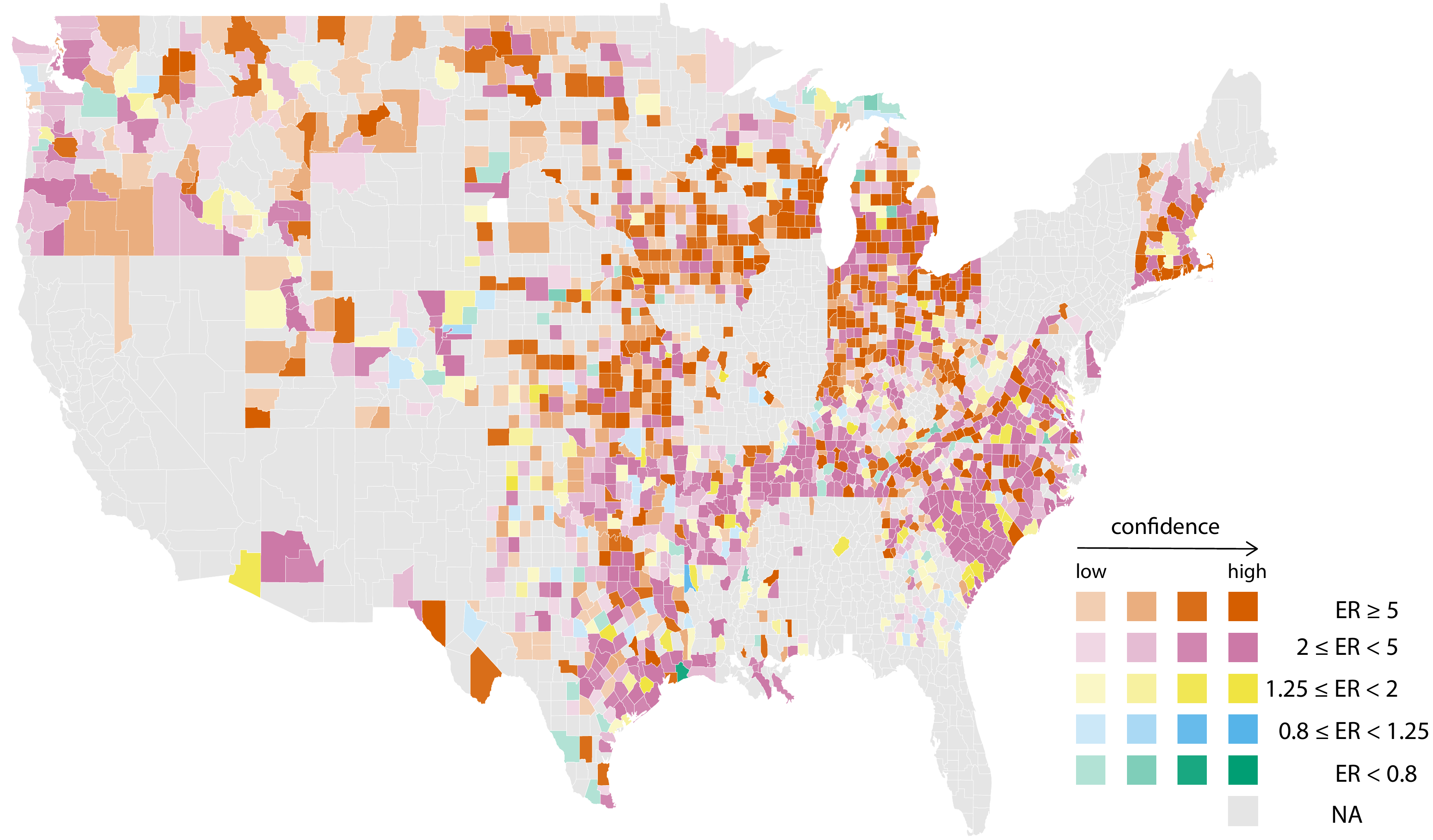}
    \caption{Maps of enforcement ratio for Marijuana incidents \textbf{including hispanic individuals} across counties in the United States, using UCR and NSDUH data. The colors indicate different levels of the ratio, while the opacity level is inversely proportional to the relative standard error. Null values, which correspond to counties in which the agencies did not report data for the period considered, are colored in grey.}
    \label{SIfig:hispanic_map}
\end{figure}

\clearpage

\subsection{Daylight}
\label{SISec:day}

We investigate whether enforcement ratios differ significantly during light hours to the enforcement ratios calculated at night. This is calculated for each county using astral python package\footnote{https://github.com/sffjunkie/astral}, which general solar position calculations to determine sunrise and sunset times.

The enforcement rate for each daylight model was then calculated based on the number of incidents in that time period, for example:

$$
E_{r, \text{Dawn $\rightarrow$ Dusk}} = \frac{I_{\text{r, Dawn $\rightarrow$ Dusk}}}{C_r}
$$

\begin{figure}[!htb]

\begin{minipage}{.5\linewidth}
\centering
\subfloat{\label{daylight:a}\includegraphics[width=\linewidth]{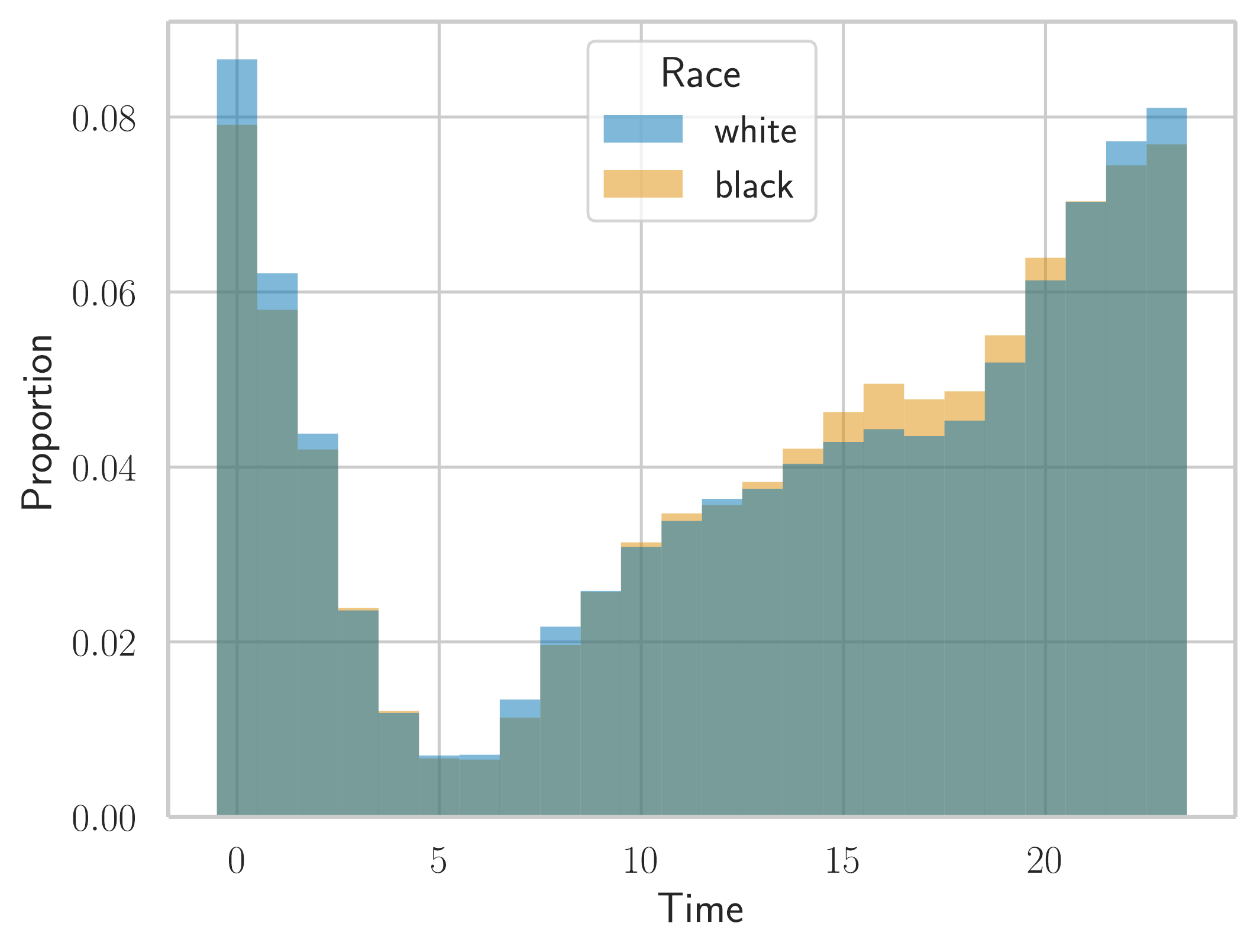}}
\end{minipage}%
\begin{minipage}{.5\linewidth}
\centering
\subfloat{\label{daylight:b}\includegraphics[width=\linewidth]{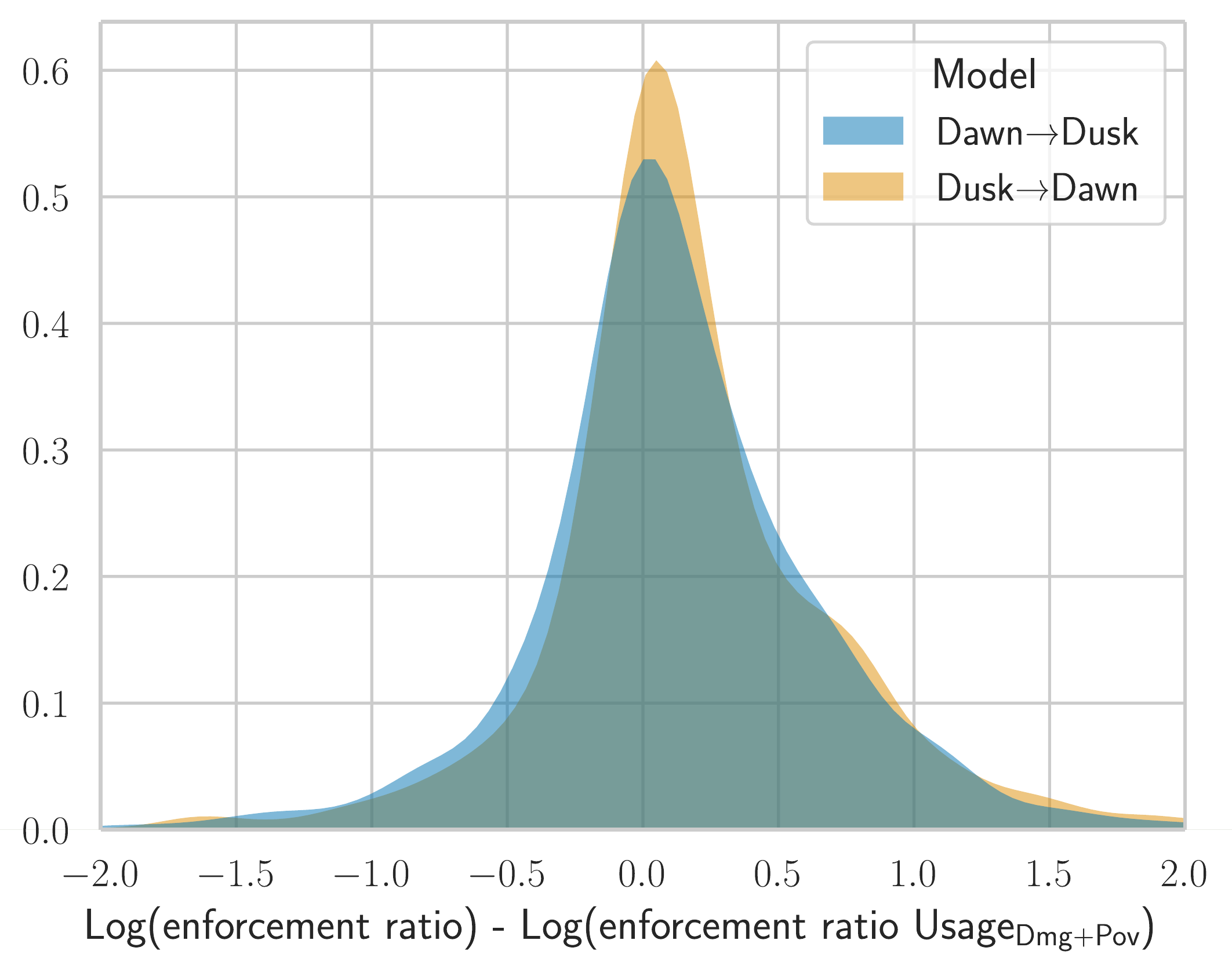}}
\end{minipage}\par\medskip
\caption{\\\textbf{(a)} The distribution of Marijuana incident times for black and white individuals \\\textbf{(b)} The distribution of enforcement ratio differences between the two daylight models and the base model \texttt{Use}\textsubscript{Dmg+Pov}}
\label{fig:daylight}
\end{figure}

\clearpage

\subsection{Arrest Map}

Not all incidents recorded in the NIBRS dataset lead to arrests. Below is a map of the enforcement ratio calculated when incidents that did not lead to an arrest are removed.

\begin{figure}[!htbp]
    \centering
    \includegraphics[width=\textwidth]{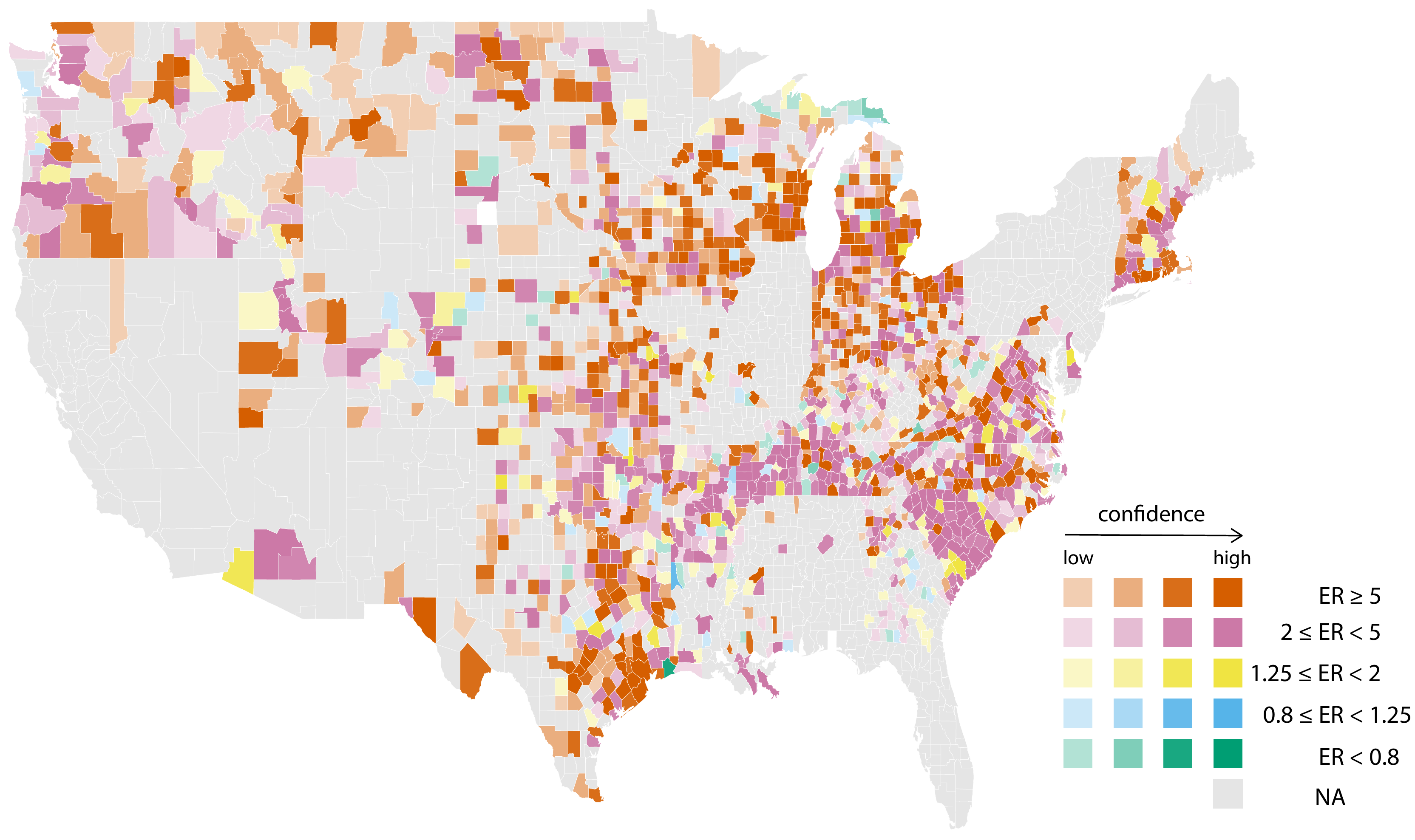}
    \caption{Maps of enforcement ratio for Marijuana incidents \textbf{that lead to an arrest} across counties in the United States, using UCR and NSDUH data. The colors indicate different levels of the ratio, while the opacity level is inversely proportional to the relative standard error. Null values, which correspond to counties in which the agencies did not report data for the period considered, are colored in grey.}
    \label{SIfig:arrest_map}
\end{figure}
\newpage

\section{NIBRS Coverage}

Not all police agencies report to the NIBRS, meaning only a proportion of a counties population is covered by the data collected. The map below shows the proportion of each counties population that is covered by the NIBRS data. This does not take into account agencies serving a population that another agency also serves; the coverage presented may be an overestimate due to this overlap.

\begin{figure}[!htbp]
    \centering
    \includegraphics[width=\textwidth]{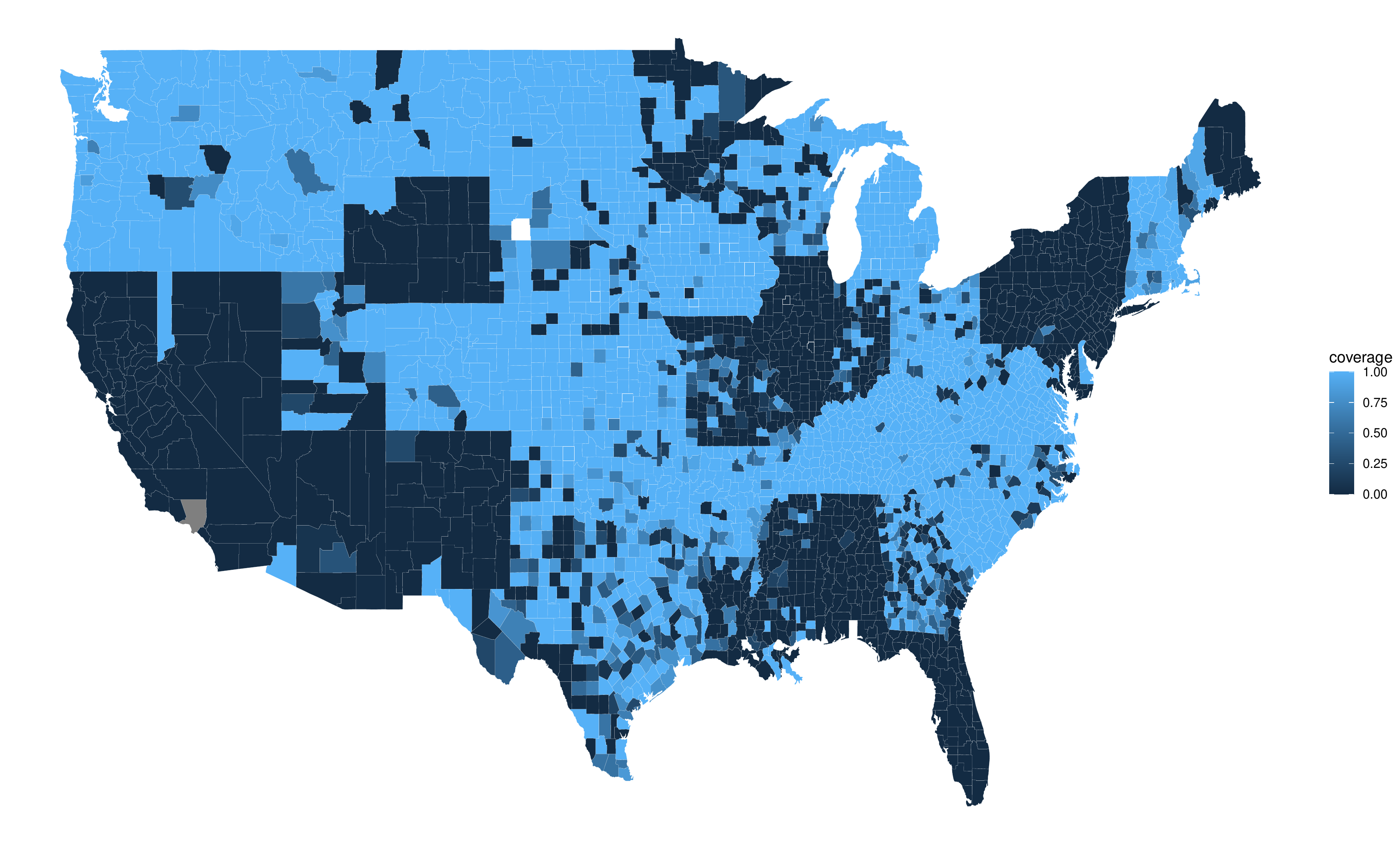}
    \caption{Map of the US with colour indicating the population covered by NIBRS reporting per county. The lighter the colour, the higher the reporting per county. Note that some states and counties do not report at all.}
    \label{SIfig:coverage_map}
\end{figure}
\newpage

\section{Census data}
\label{SISec:Census}
As the census only takes place once every 10 years, the 2011-2019 data are a projection from the previous census that took place in 2010. The projection is calculated using a cohort-component method, where factors effecting population are separately estimated, then summed to obtain the population final estimate. The components can be seen in the following:

\begin{align*}
\text{Population}_{2019} = \text{Population}_{2010} + \text{Births}_{2010-2019} \\
- \text{Deaths}_{2010-2019} + \text{Migration}_{2010-2019}
\end{align*}

Birth and death estimates are obtained from the National Center for Health Statistics (NCHS) and the Federal-State Cooperative for Population Estimates (FSCPE). The NCHS data, sourced from birth and death certificates, comes with a two year lag. An adjustment is made using data obtained from FSCPE, which report more frequently on geographic events which could have an impact on birth or death rates.

Migration statistics are obtained from four data sources; the Inland Revenue Service (IRS), the Social Security Administration (SSA), Medicare registrations, and the Census Bureau's (CB) Demographic Characteristics Files (DCF). The DCF is an internally developed dataset which allocates race and Hispanic origin to individuals who have migrated. It is developed using \say{person-level records from the decennial census, administrative records, and a set of imputation techniques when reported race and Hispanic origin are
not available}. 

National and county-level estimates are calculated separately. Due to \say{slightly different data sources and methodology}, the summation of the county-level estimates is not equal to the national-level estimates. To correct for this, the CB calculate a uniform rake factor, which are then applied to the county level estimates.

$$
\text{Rake} = \frac{\text{National Population}}{\sum^{c}_{c \in \text{counties}} \text{Population}_{c} }
$$

Note that historically, the projection is accurate, with a 3.1\% absolute difference between the estimates and ground-truth for the 2000-2010 period.

\section{Relationship between incidents and usage}

\begin{table}[ht]
\centering
\caption{Univariate regression coefficients that estimate the number of incidents.}
\begin{tabular}{llll}
  \hline
 Race & P(Usage$|$Race=r) & P(Purchasing$|$Race=r) & P(Purchasing Outside$|$Race=r) \\ 
  \hline
Black | White & 0.00011 (5.4e-05) * & 0.00025 (0.00011) * & 0.00025 (0.00011) * \\ 
  Black & 0.00077 (0.00026) ** & 0.0015 (0.00061) * & 0.0018 (8e-04) * \\ 
  White & 4e-05 (4.3e-05)   & 0.00021 (0.00018)   & 0.00034 (0.00035)   \\ 
   \hline
\multicolumn{4}{l}{\parbox[t][][t]{\textwidth}{\vspace{1em}\footnotesize{\textit{Notes:} Univariate regression coefficients estimating the number of incidents per person on a county level, using the probability of usage/purchase/outside purchase in 2019 (taken from NSDUH). This is done for black incidents, white incidents, and the union of each.  Significant relationships are found between usage and the number of incidents for black individuals and the union, but not for white in isolation. This pattern repeats for purchasing and purchasing outside.}}}
\end{tabular}
\end{table}

\end{document}